\documentclass{aa}

\usepackage{txfonts,graphicx,subfigure,natbib,aalongtable,afterpage}
\bibpunct{(}{)}{;}{a}{}{,}
\begin{document}

\title{Global characteristics of GRBs observed with \textit{INTEGRAL} and the inferred large
  population of low-luminosity GRBs\thanks{Based on observations
  with INTEGRAL, an ESA project with instruments and science data
  centre funded by ESA member states (especially the PI countries:
  Denmark, France, Germany, Italy, Switzerland, Spain), Czech Republic
  and Poland, and with the participation of Russia and the USA.}}
  \author{S.~Foley\inst{1} \and S.~McGlynn\inst{1} \and
  L.~Hanlon\inst{1} \and S.~McBreen\inst{2}
  \and B.~McBreen\inst{1}}

\offprints{S.~Foley, \email{sfoley@bermuda.ucd.ie}}

\institute{UCD School of Physics, University College Dublin,
Dublin 4, Ireland \and Max-Planck-Institut f\"{u}r extraterrestrische Physik,
85748 Garching, Germany} 

\date{Received/Accepted}

\abstract
{\textit{INTEGRAL} has two sensitive gamma-ray instruments that have detected and localised 47
  gamma-ray bursts (GRBs) from its launch in October 2002 up to July 2007.}
{We present the spectral, spatial, and temporal properties of the bursts in the \textit{INTEGRAL} GRB catalogue
  using data from the imager, IBIS, and spectrometer, SPI.}
{Spectral properties of the GRBs are determined using power-law and, where appropriate, Band model and
  quasithermal model fits
 to the prompt emission. 
Spectral lags, i.e. the time delay in the arrival of low-energy $\gamma$-rays with respect
to high-energy $\gamma$-rays, are measured for 31 of the GRBs.}
{The photon index
  distribution of power-law fits to the prompt emission spectra is presented and is consistent with that obtained by \textit{Swift}. The peak flux distribution shows that \textit{INTEGRAL} detects
  proportionally more weak GRBs than \textit{Swift} because of its higher  sensitivity in a smaller field of view. 
The all-sky rate of GRBs above $\sim0.15\,\rm{ph}\,\rm{cm}^{-2}\,\rm{s}^{-1}$ is $\sim1400\,\rm{yr}^{-1}$ in the fully coded field of view of IBIS. Two groups are identified
  in the spectral lag distribution of \textit{INTEGRAL} GRBs, one with short lags $<0.75\,\rm{s}$ (between 25--50\,keV and
  50--300\,keV) and one with long lags $>0.75\,\rm{s}$. 
Most of the long-lag GRBs are inferred to have low redshifts because of their long spectral lags, their tendency to have low peak energies, and their faint optical and X-ray afterglows. They are mainly observed in the direction of the supergalactic plane with a quadrupole moment of $Q=-0.225\pm0.090$ and hence reflect the local large-scale structure of the Universe.}
  {The spectral, spatial, and temporal properties of the 47 GRBs in the \textit{INTEGRAL} catalogue are presented and compared with the results from other missions. The rate of long-lag GRBs with inferred low luminosity is $\sim$25\% of Type Ib/c supernovae. Some of these bursts could be produced by the collapse of a massive star without a supernova. Alternatively, they could result from a different progenitor, such as the merger of two white dwarfs or a white dwarf with a neutron star or black hole, possibly in the cluster environment without a host galaxy.}


\titlerunning{Global characteristics of GRBs observed with \textit{INTEGRAL}}

\maketitle
\section{Introduction\label{intro}}

The prompt emission of gamma-ray bursts provides
valuable insight into the mechanisms from which these
extremely explosive events originate. Their short durations
and highly variable temporal structures provide constraints on the
physics of the central engine powering the burst. A number of GRB reviews have been
published
\citep[e.g.][]{zhang:2004,piran:2005,meszaros:2006}. In recent years, the advent
of missions such as the Compton Gamma-Ray Observatory
(CGRO)~\citep{fishman:1995}, along with the improved imaging capabilities of missions such as
\textit{BeppoSAX}~\citep{boella:1997}, HETE~II~\citep{sakamoto:2005}, \textit{INTEGRAL}~\citep{winkler:2003},
and \textit{Swift}~\citep{gehrels:2004}, has led to the precise localisations of GRBs
and enabled rapid multi-wavelength follow-up observations. The X-ray, optical and
radio afterglow detections are listed in Table~\ref{table:ags} for a total of 423 GRBs well localised by these
missions between July
1996 and
July 2007, showing in particular the observed
number of afterglows based on \textit{INTEGRAL} GRB detections. The data are taken from the webpage maintained by Jochen
Greiner\footnote{http://www.mpe.mpg.de/$\sim$jcg/grbgen.html.}.

\small
\begin{table}
\begin{center}
\caption{Afterglow detections for GRBs localised by recent $\gamma$-ray missions
between July 1996 and July 2007.}
\label{table:ags}
\begin{tabular}{@{}l c c c c @{}}
\hline\hline
& \textbf{\textit{BeppoSAX}} & \textbf{HETE II} & \textbf{\textit{INTEGRAL}} &
\textbf{\textit{Swift}} \\
\hline
\textbf{GRBs} & 55 & 79 & \textbf{47} & 242 \\
\textbf{X-ray} & 31 & 19 & \textbf{17} & 209 \\
\textbf{Optical} & 17 & 30 & \textbf{16} & 123 \\
\textbf{Radio} & 11 & 8 & \textbf{8} & 17 \\
\hline
\end{tabular}
\end{center}
\end{table}

\normalsize

There are two main
$\gamma$-ray instruments on board \textit{INTEGRAL}, namely
IBIS~\citep{ubertini:2003} and SPI~\citep{vedrenne:2003},
optimised for high-resolution imaging and spectroscopy of the $\gamma$-ray sky, respectively. The IBIS instrument is comprised of two separate layers of
  dectectors, ISGRI in the 15\,keV--1\,MeV energy range~\citep{lebrun:2003}, and PICsIT
  in the 180\,keV--10\,MeV energy range~\citep{labanti:2003}. IBIS/ISGRI has 16384 CdTe detectors, located 3.4\,m from a tungsten mask which projects a shadowgram on the detector plane. Maps of the sky are reconstructed by decoding the shadowgram with the mask pattern. IBIS has a fully coded field
  of view (FCFoV) of $9^{\circ}\times\,9^{\circ}$ and a partially coded field of
  view (PCFoV) of 
  $19^{\circ}\times\,19^{\circ}$ at 50\% coding and $29^{\circ}\times\,29^{\circ}$  at zero coding. SPI consists of 19
  hexagonal germanium (Ge) detectors covering the energy range
  20\,keV--8\,MeV with high energy resolution of 2.5\,keV at 1.3\,MeV. A
  coded mask is located 1.71\,m above the detector plane for imaging
  purposes, giving a $16^{\circ}$ corner-to-corner FCFoV and a PCFoV of
  $34^{\circ}$.  The SPI and IBIS instruments
are supported by
an optical camera (OMC,~\citet{mas-hesse:2003}) and an X-ray monitor (JEM-X,~\citet{lund:2003}). 

The \textit{INTEGRAL} Burst Alert System
(IBAS\footnote{http://ibas.iasf-milano.inaf.it/IBAS\_Results.html},~\citet{mereghetti:2003c}) is an automatic ground-based
system for the accurate localisation of GRBs and the rapid
distribution of GRB coordinates, providing, on average,
0.8~GRBs per month with an error radius of $\sim$3
arcminutes. \textit{INTEGRAL} has detected 46
long-duration GRBs ($T_{90}\gtrsim2$\,s) and 1 short-duration GRB
($T_{90}\lesssim2$\,s) between October 2002 and July 2007.  \textit{INTEGRAL} bursts of particular interest include the low-luminosity
GRB\,031203~\citep{sazonov:2004}, the very intense
GRB\,041219a~\citep{mcbreen:2006}, a number of X-ray rich
GRBs such as GRB\,040223~\citep{mcglynn:2005,filliatre:2006},
GRB\,040403~\citep{mereghetti:2005} and
GRB\,040624~\citep{filliatre:2006}, and the short-duration GRB\,070707~\citep{mcglynn:070707}.
In addition,~\citet{marcinkowski:2006} have detected a bright, hard  GRB outside the field of view using the ISGRI Compton mode. Spectroscopic redshifts have been determined for
four \textit{INTEGRAL} GRBs, i.e. GRB\,031203  at
z=0.1055~\citep{prochaska:2004}; GRB\,050223 at z=0.584~\citep{pellizza:2006}; GRB\,050525a at
z=0.606~\citep{foley:2005} and GRB\,050502a at
z=3.793~\citep{prochaska:2005}. Non-spectroscopic redshifts have been inferred for GRB\,040812 ($0.3\,<\,z\,<\,0.7$,~\citet{davanzo:2006}) and GRB\,040827~($0.5\,<\,z\,<\,1.7$,~\citet{deluca2005}). The low efficiency for measuring redshifts is partially due to the fact that \textit{INTEGRAL} spends a large amount of observing time pointing towards the galactic plane.

Gamma-ray burst continuum spectra are in most cases well
described by a smoothly broken power law in the 30\,keV--2\,MeV
energy range~\citep{band:1993}. 
The $\gamma$-ray spectral shape as
predicted for optically thin synchrotron emission are two asymptotically
broken power laws but many GRBs are not consistent with this model and
it requires modification to fit the observed
spectra~\citep[e.g.][]{lloyd:2002}. 
It has been proposed that GRB spectra may contain a thermal
component~\citep[e.g.][]{ghirlanda:2003,ryde:2005,kaneko:2006,mcbreen:2006}. The interpretation
of quasithermal emission as opposed to synchrotron emission
can provide an explanation for the observed spectral
characteristics within a more physical framework~\citep[e.g.][]{rees2005,ryde:2006}.  

The time profiles of GRBs often exhibit a complex
and unpredictable nature, displaying considerable diversity
both in terms of structure and duration~\citep{mcbreen:2001,quilligan:2002}. This makes them difficult to
classify on the basis of temporal structure alone. 
 One notable feature of GRB time profiles is the tendency for emission
 in a high-energy band to lead the arrival of photons in a low-energy
 band~\citep[e.g.][]{cheng:1995,wu:2000,norris:2000,bolmont:2006,mcbreen:2006,hakkila:2007,hakkila:2008}. The energy-dependent lag allows the temporal and spectral properties of the GRB prompt $\gamma$-ray
emission to be combined in a single measurement. The typical lag values 
measured for long-duration GRBs detected by the Burst and Transient Source Experiment (BATSE) between the
25--50\,keV and 100--300\,keV channels
concentrate $\sim100$\,ms~\citep{norris:2000}.  An
anti-correlation between spectral lag and isotropic
peak luminosity was first observed by \citet{norris:2000}, using 6 BATSE bursts
with measured redshifts. A similar trend is observed
between lag and luminosity for a number of \textit{Swift}
GRBs of known redshift~\citep{gehrels:2006}. However, there exist notable
outliers, in particular the ultra-low luminosity
bursts GRB\,980425, GRB\,031203 and GRB\,060218, associated with the supernovae
SN\,1998bw, SN\,2003lw and SN\,2006aj, respectively.
Short bursts
($T_{90}<2$\,s) have very small or
negligible lags~\citep{yi:2006,norris:2006,zhang:2006b} and relatively low peak
luminosities and so do not lie on the
correlation~\citep{gehrels:2006}. On this basis, the spectral lag has been suggested by
  \citet{Donaghy:2006} as one of the criteria to determine whether a
  burst is long or short. Pulse width and spectral lag are
strongly related, with wider pulses tending to have longer spectral
lags~\citep{norris:2006}. Relative spectral lags, defined as the ratio of spectral
lag to pulse width, have been found to have normal distributions
centering on $\sim100$\,ms for long bursts~\citep{zhang:2006a} and $\sim14$\,ms for
short GRBs~\citep{zhang:2006b}. The
lag-luminosity  and $\rm{E}_{peak}$-$\rm{E}_{iso}$~\citep{amati:2007} relationships can be used as distance indicators for GRBs~\citep{schaefer:2007}, provided the role of selection effects is understood and quantified~\citep[e.g.][]{butler:2008}.

The physical basis underlying spectral lags is not yet well
understood~\citep{schaefer:2004}. The observed lag of a burst is a direct consequence of its
spectral evolution because the peak of the $\nu\,F_{\nu}$ spectrum,
$E_{peak}$, decays with time~\citep{kocevski:2003,hafizi:2007}. The internal shock model allows for three possible sources
of temporal variations in GRB pulses: cooling, hydrodynamics and
geometric angular effects. Cooling is unable to fully account for the time
lag since the
synchrotron timescale is much
shorter than the lag timescale~\citep{wu:2000}. It has been proposed
that the lag-luminosity relation may arise kinematically, based on the
viewing angle at which the GRB jet is observed~\citep{salmonson:2000}. In this interpretation, a high-luminosity
GRB with short spectral lag corresponds to a
jet with a small viewing angle, while a low-luminosity GRB with long
spectral lag corresponds to a jet with a large viewing angle~\citep{ioka:2001}. A correlation has also been
observed between spectral lag (or luminosity) and jet-break time,
thereby connecting the prompt and afterglow phases of GRBs. This may
be understood in terms of a model in which the Lorentz factor decreases
away from the axis of the GRB jet~\citep{salmonson:2002}. 
The connection between spectral lag and the timescales involved in the hydrodynamic processes and
  radiative mechanisms of the burst has been discussed
  by~\citet{daigne:2003}.

A subpopulation of local, faint, long-lag GRBs has been suggested
by~\citet{norris:2002} from a study of BATSE bursts, which implies that
events with low peak fluxes ($F_{Peak}$ (50-300\,keV) $\sim
0.25$\,ph\,cm$^{-2}$\,s$^{-1}$) should be predominantly long-lag
GRBs. The sensitivity of IBIS is such that bursts fainter than the BATSE limit can
  be well localised. In this paper we present the spectral, spatial and temporal lag properties of the complete sample of the 47 GRBs detected in the field of view
of IBIS and SPI up to July 2007. Section~\ref{acs} describes the capabilities
of SPI's anti-coincidence shield as a GRB detector. The spectral and lag analyses are described in
Sect.~\ref{analysis} and results are presented in
Sect.~\ref{results}. Sections~\ref{discussion} and~\ref{conclusion} discuss the
significance of these results, which imply a large population of long-lag GRBs with low luminosities. The cosmological parameters adopted throughout the paper are $H_0$ = 70\,km\,s$^{-1}$\,Mpc$^{-1}$, $\Omega_m = 0.3$, $\Omega_{vac} = 0.7$. All errors are quoted at the 1$\sigma$ confidence level.

\section{GRBs detected with SPI's Anti-Coincidence Shield\label{acs}}

\begin{figure*}

\mbox{ \subfigure{\includegraphics[width=0.45\columnwidth,
height=0.35\textheight,angle=270]{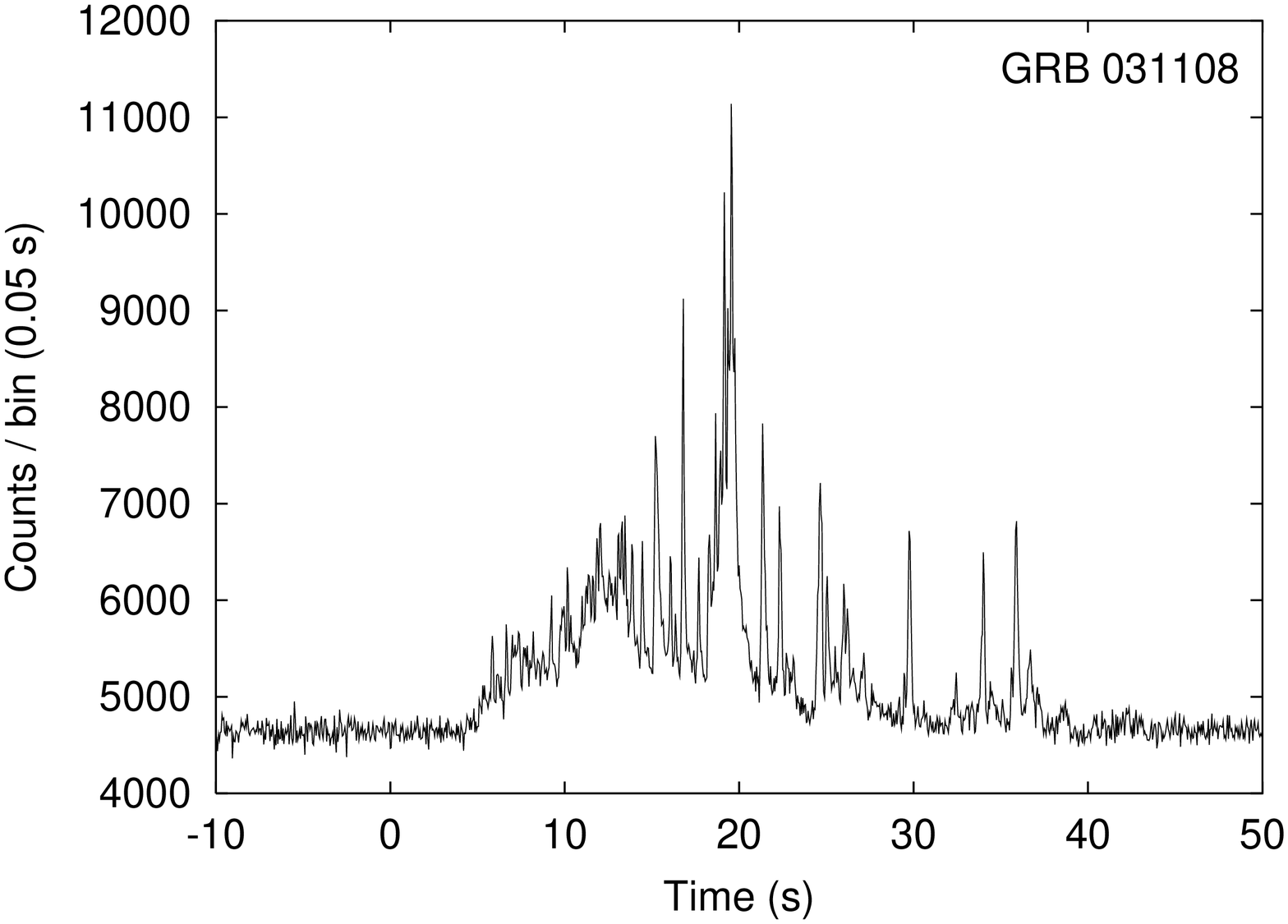}}
\subfigure{\includegraphics[width=0.45\columnwidth,
height=0.35\textheight,angle=270]{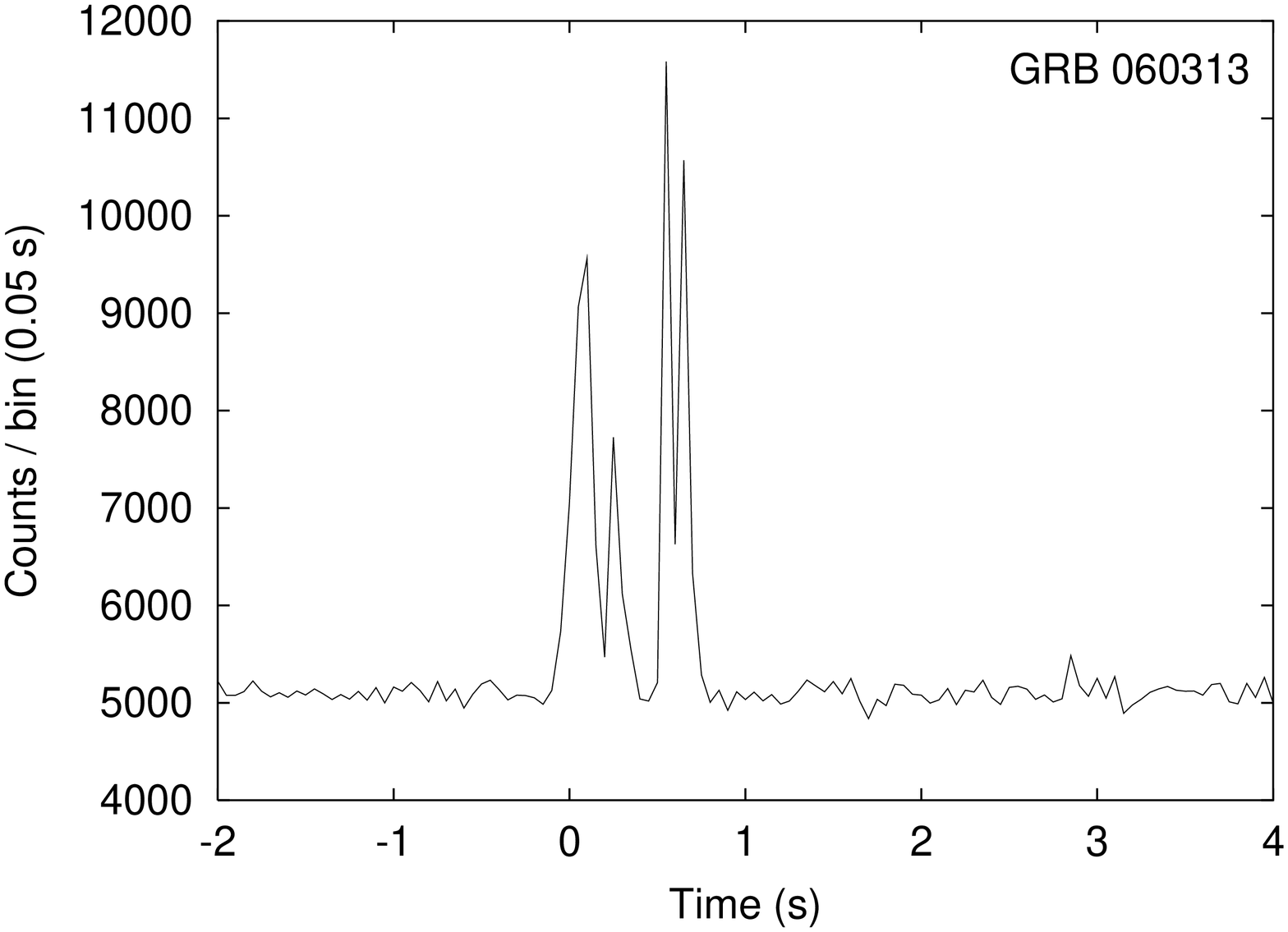}} }

\mbox{ \subfigure{\includegraphics[width=0.45\columnwidth,
height=0.35\textheight, angle=270]{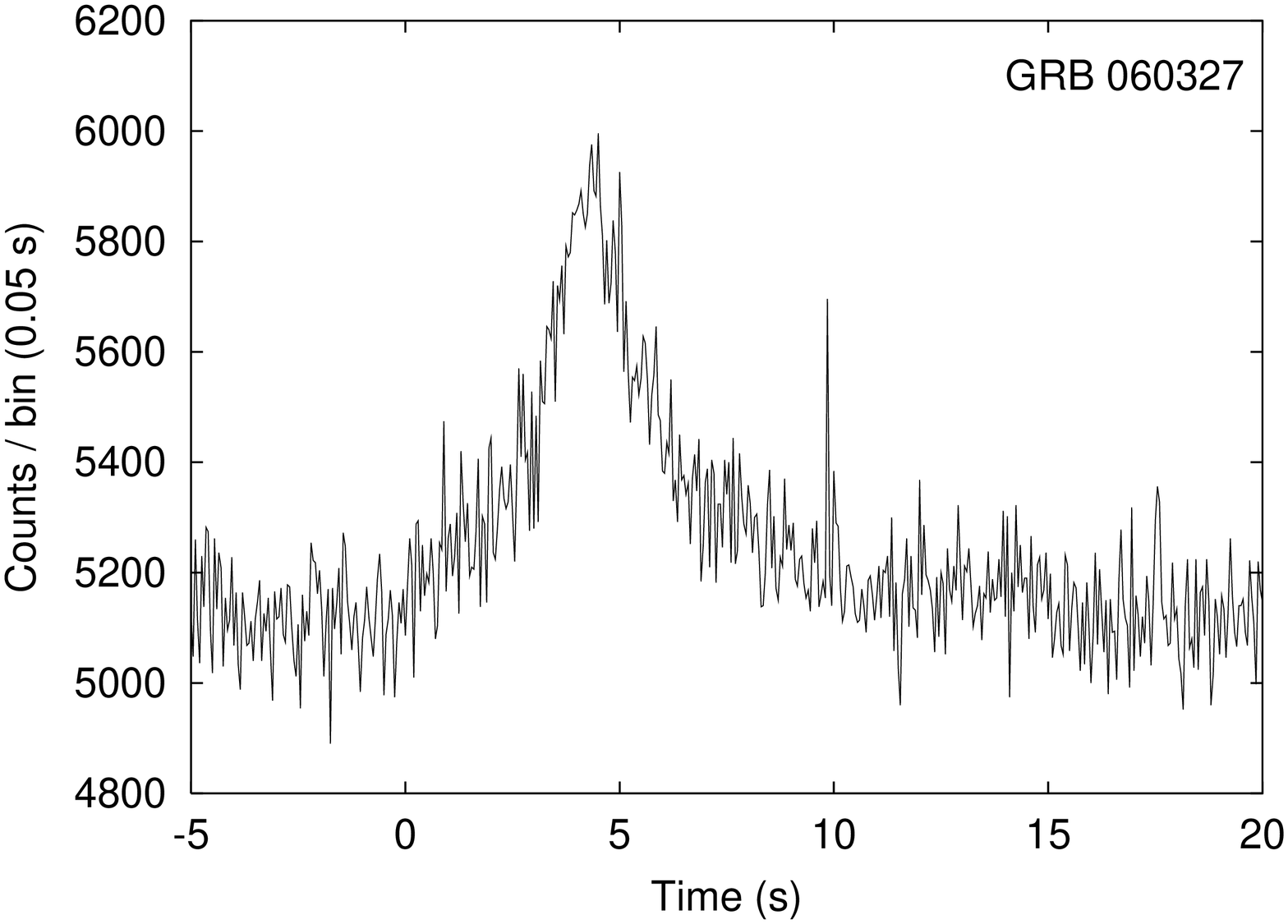}}
\subfigure{\includegraphics[width=0.45\columnwidth,
height=0.35\textheight, angle=270]{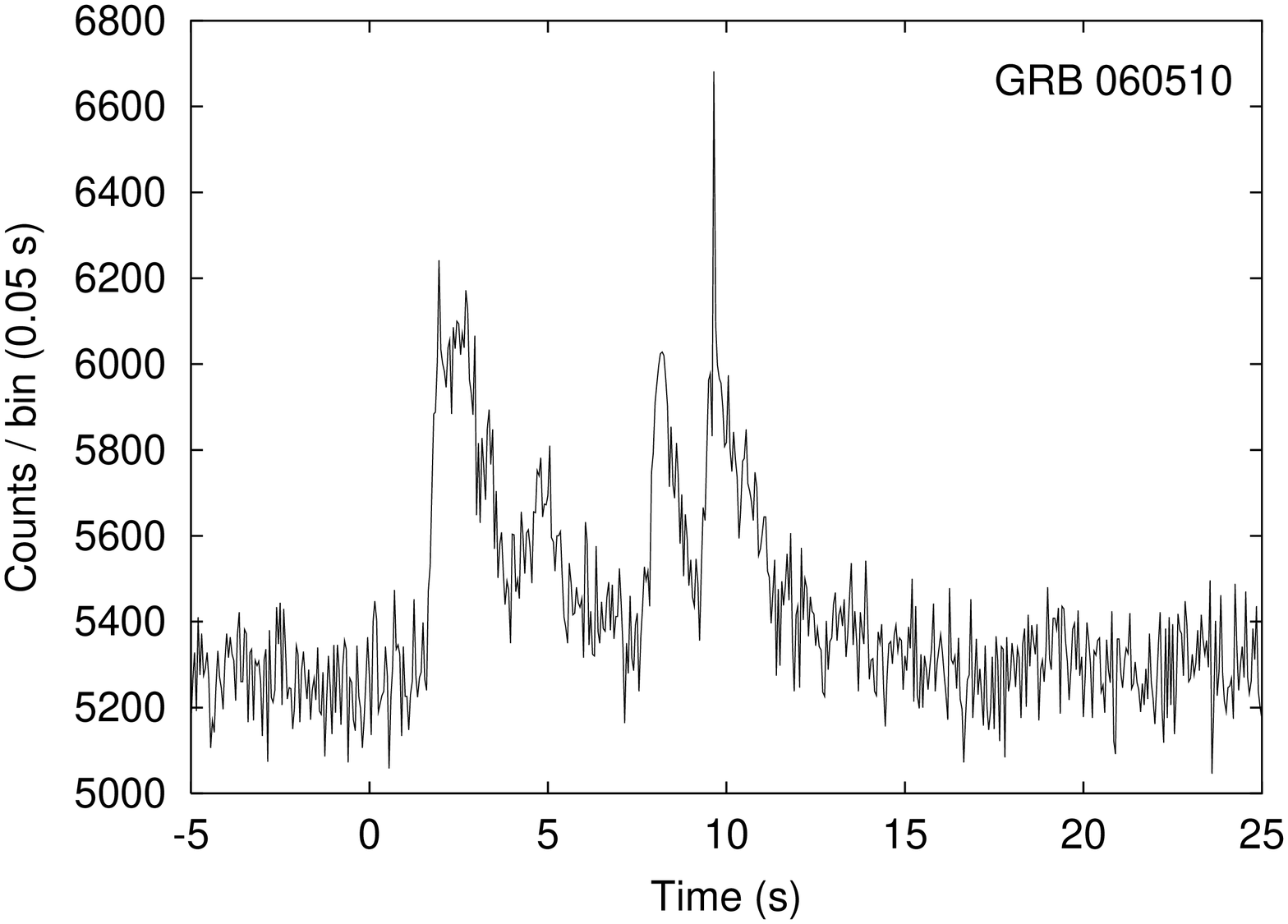}} }

\caption{A selection of GRB lightcurves detected by the
  Anti-Coincidence Shield at photon energies $>$ 80\,keV.}
\label{fig:acs}

\end{figure*}

In addition to SPI and IBIS, the Anti-Coincidence Shield (ACS) surrounding the SPI detectors works as a highly-sensitive GRB detector above $\sim80$\,keV but
 lacks spatial and spectral
 information~\citep{vonkienlin:2003a}.  The ACS consists of 91
 BGO crystals with a total mass of 512\,kg surrounding SPI. It
 has a maximum sensitivity to GRBs at $\sim90\,^{\circ}$ from
 the pointing direction and provides lightcurves in
 50\,ms intervals. The ACS detects GRBs at a rate of
 $\sim$1 every 2--3 days.  A selection of GRBs detected with
 SPI-ACS is shown in Fig.~\ref{fig:acs} using data taken from the publicly available catalogue of SPI-ACS
 GRBs\footnote{http://www.mpe.mpg.de/gamma/science/grb/1ACSburst.html}. Temporal analysis of
 a more complete sample is presented in~\citet{rau:2005}. The ACS is
 used as part of the interplanetary network~\citep{hurley:2006}.

\section{Analysis of GRBs observed with IBIS and SPI\label{analysis}}

\subsection{Spectral Analysis}

Spectral analysis of the GRBs was performed using \textit{INTEGRAL}'s
Online Software Analysis version 5.1  available from the
\textit{INTEGRAL Science Data Centre}
\footnote{http://isdc.unige.ch/}. The T$_{90}$ duration (i.e.,
the time during which 5\% to 95\% of the GRB counts are
recorded) was determined for each GRB in the 20--200\,keV energy range and the spectrum was
generated for that time interval. For cases in which the GRB
had a multi-peaked time profile, the GRB was divided into its
constituent pulses and a spectrum was generated for
each pulse to investigate evolution of the spectral
parameters during the burst.

Each GRB was fit by a simple power-law model, the Band
model and a quasithermal (combined power-law + blackbody)
model. Weak GRBs are best fit by a single
power-law model because the limited statistics are insufficient to
constrain any additional parameters. For brighter GRBs, the Band
model or combined power-law + blackbody fits usually result in an
improved $\chi^{2}$ value. The peak flux was measured over the
brightest 1\,s time interval in the 20--200\,keV energy range. The fluence of each GRB
was determined for the T$_{90}$ interval in the 20--200\,keV
energy range for the IBIS spectra, and the 20--200\,keV and
20\,keV--8\,MeV energy ranges for the SPI spectra.

\subsection{Spectral Lag Analysis}

In order to measure the lag,
background-subtracted lightcurves were extracted in three
energy bands comparable to  those used with BATSE, namely 25--50\,keV (Channel 1), 50--100\,keV
(Channel 2) and 100--300\,keV (Channel 3). The lag, $\tau$, between two energy
channels was determined by computing the
cross-correlation function (CCF) between the two lightcurves
as a function of temporal lag as described by
\citet{band:1997} and ~\citet{norris:2000}. Assuming the time profiles  in both energy
channels display sufficient similarity, the peak in the CCF
then corresponds to the time lag of the GRB between the two
energy channels in  question. The lag was determined between Channels 1 and 2, $\tau_{2,1}$, Channels 1
and 3, $\tau_{3,1}$, and to account for those cases in which the signal
level in Channel 3 was insufficient to determine an accurate
lag, the counts in Channels 2 and 3 were combined and
correlated with Channel 1 to give $\tau_{2+3,1}$. In this paper, GRBs with $\tau_{2+3,1}>0.75$\,s are defined as long-lag. 

\begin{figure}[ht]
\centering
\subfigure{\label{fig:lag070615a}\includegraphics[width=0.50\columnwidth, height=0.35\textheight,angle=270]{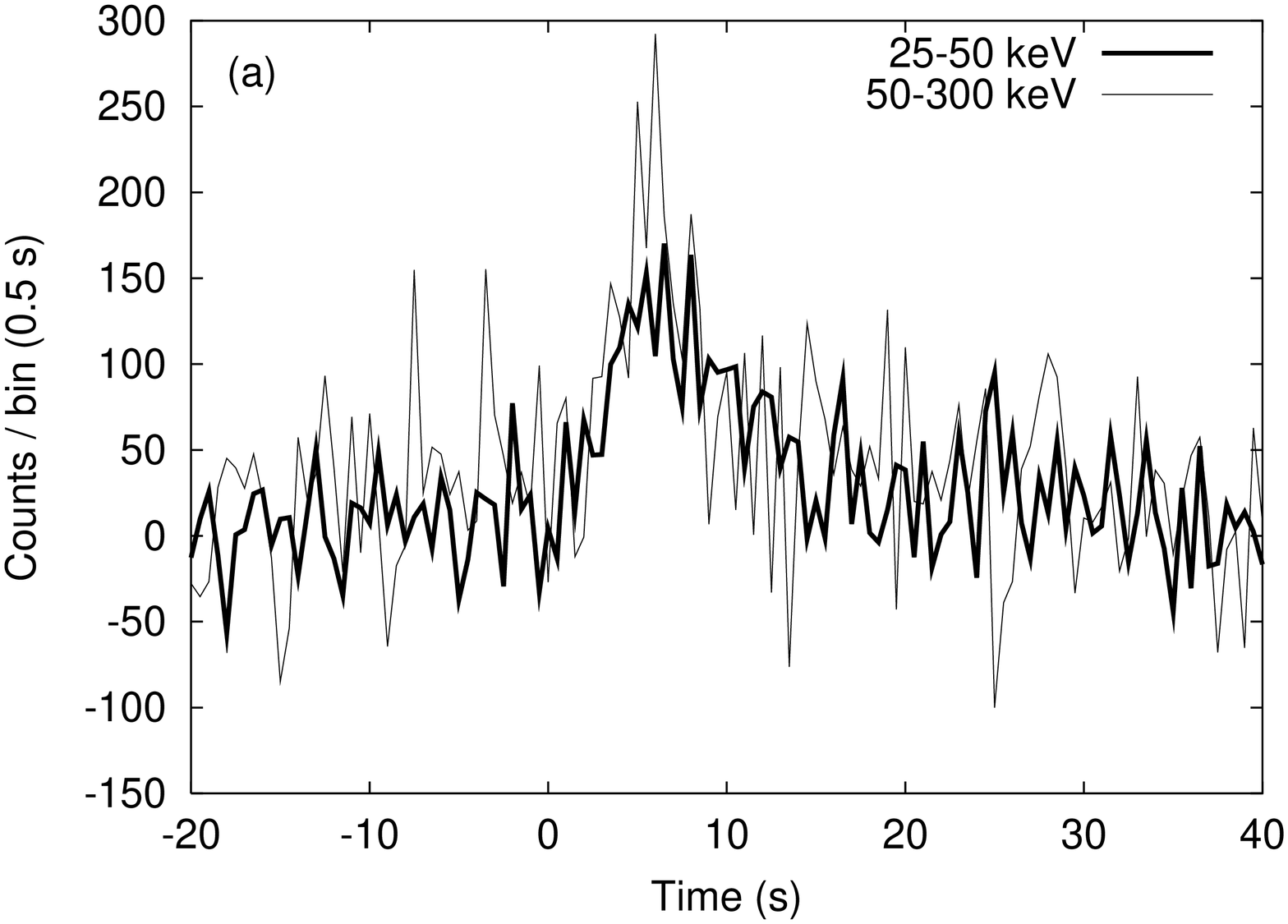}}
\subfigure{\label{fig:lag070615b}\includegraphics[width=0.50\columnwidth, height=0.35\textheight,angle=270]{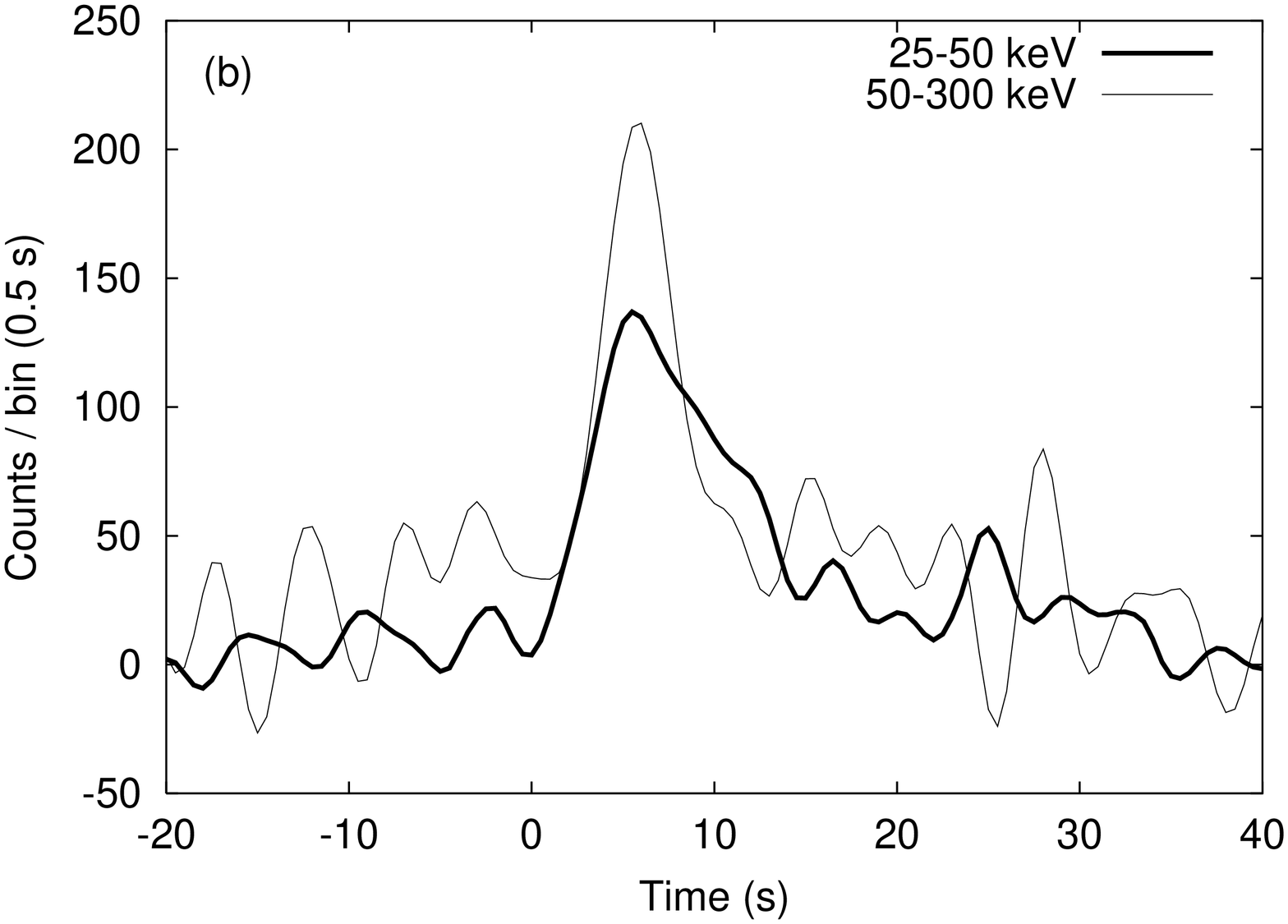}}
\subfigure{\label{fig:lag070615c}\includegraphics[width=0.50\columnwidth, height=0.35\textheight,angle=270]{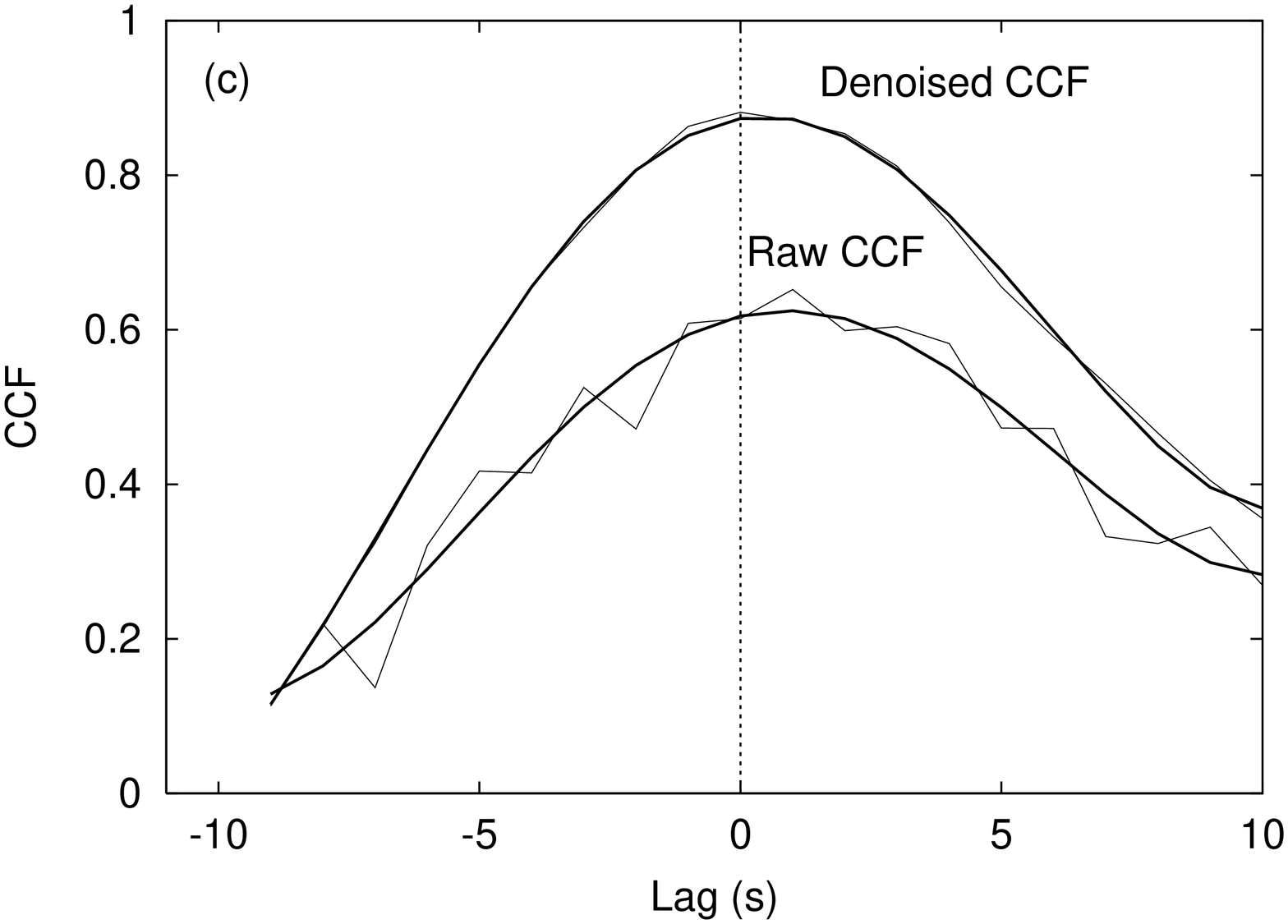}}
\caption{(a) Raw and (b) denoised lightcurves of GRB\,070615 in the
  25-50\,keV (dark line) and 50-300\,keV (light line) energy
  bands. (c)~Cross correlation functions and polynomial fits giving
  a lag of $0.40^{+0.15}_{-0.25}$
  and $0.40^{+0.15}_{-0.20}$ for the raw and denoised lightcurve data,
  respectively. The dashed line represents a lag of 0\,s.}
\label{fig:lag070615}
\end{figure}

The reliability of the cross-correlation technique was limited
primarily by the signal to noise ratio of the data. The faint
nature of many of the GRBs detected by \textit{INTEGRAL} can result in a noisy CCF
with an ambiguous peak. A denoising technique was
used to smooth the lightcurve while retaining the
structure of the burst. This technique
involved filtering the signal to remove the high frequency components in order to produce a denoised
lightcurve \citep{quilligan:2002}. The wavelet analysis was
carried out on the weakest GRBs with
$F_{peak}$~(20--200\,keV)~$\lesssim0.6\,\rm{ph}\,\rm{cm}^{-2}\,\rm{s}^{-1}$
using the wavelet toolbox in MATLAB. As an example, the lightcurves of GRB\,070615 in the 25--50\,keV and 50--300\,keV
energy bands are shown in Fig.~\ref{fig:lag070615a} before and Fig.~\ref{fig:lag070615b} after denoising. The
CCFs resulting from the raw data and denoised data are plotted in
Fig.~\ref{fig:lag070615c}. The denoised data results in a smoother CCF
peak and a more
significant correlation, while retaining the
position of the peak at a similar lag value to that measured with
the raw data. Six \textit{INTEGRAL} GRBs were too weak for a
  reliable lag to be
determined using either the raw or denoised data.

The CCF was fit with a fourth order polynomial in
order to account for the asymmetry of the CCF~\citep{norris:2000}. 
The peak of the polynomial fit to the CCF was then taken to be the true lag
value. For each GRB, an average spectral
lag over the total burst duration was determined. For the cases in
which separate pulses could be clearly distinguished by eye,
spectral lags were measured for the individual pulses to investigate
the evolution of spectral lag during the burst.  In
each case, the spectral lag was determined for regions of the
lightcurve above 10\%, 30\% and 50\% of the peak count rate and over a
number of different lag ranges to ensure that consistent results were
obtained. The optimum
lag range was taken to be that for which the CCF was concave
down but short enough that the CCF peak was well fit.  Statistical errors
were calculated using a bootstrap method as described in \citet{norris:2000}. This involves adding
Poissonian noise based on the observed counts to the
lightcurves in the different energy channels and  re-computing
the CCF in 100 realisations for each burst. The 50th ranked
value is then the mean lag and the 16th and  84th ranked
values represent $\pm1\sigma$. The lightcurve data was over-resolved by
a factor of 10 in order to compute the errors at a time resolution
less than the natural binning of the raw data. 

\section{Results\label{results}}

\begin{figure}[ht]
\subfigure{\label{fig:exp_gal}\includegraphics[width=0.9\columnwidth,
height=0.2\textheight]{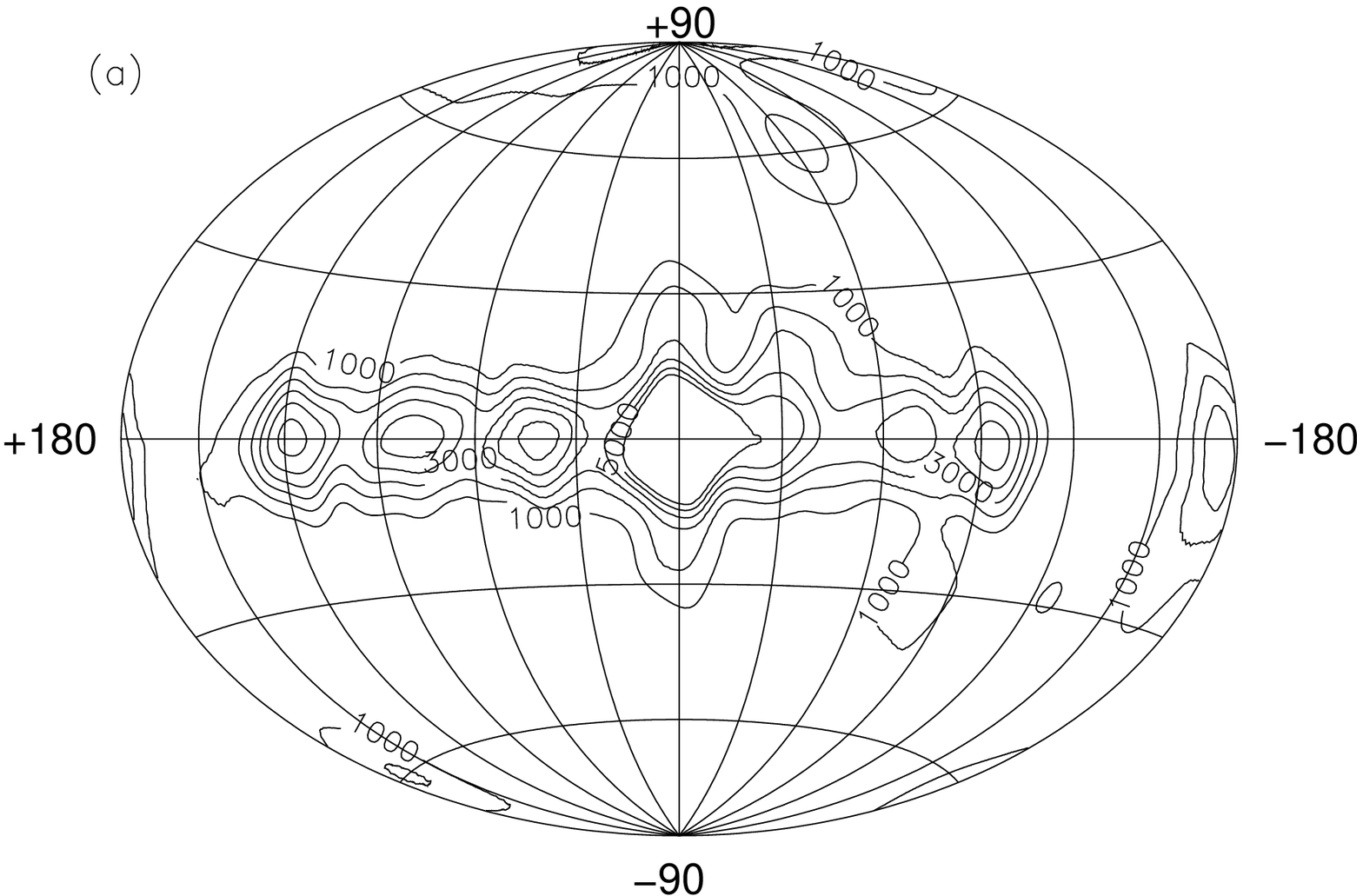}}
\subfigure{\includegraphics[width=0.9\columnwidth, 
height=0.2\textheight]{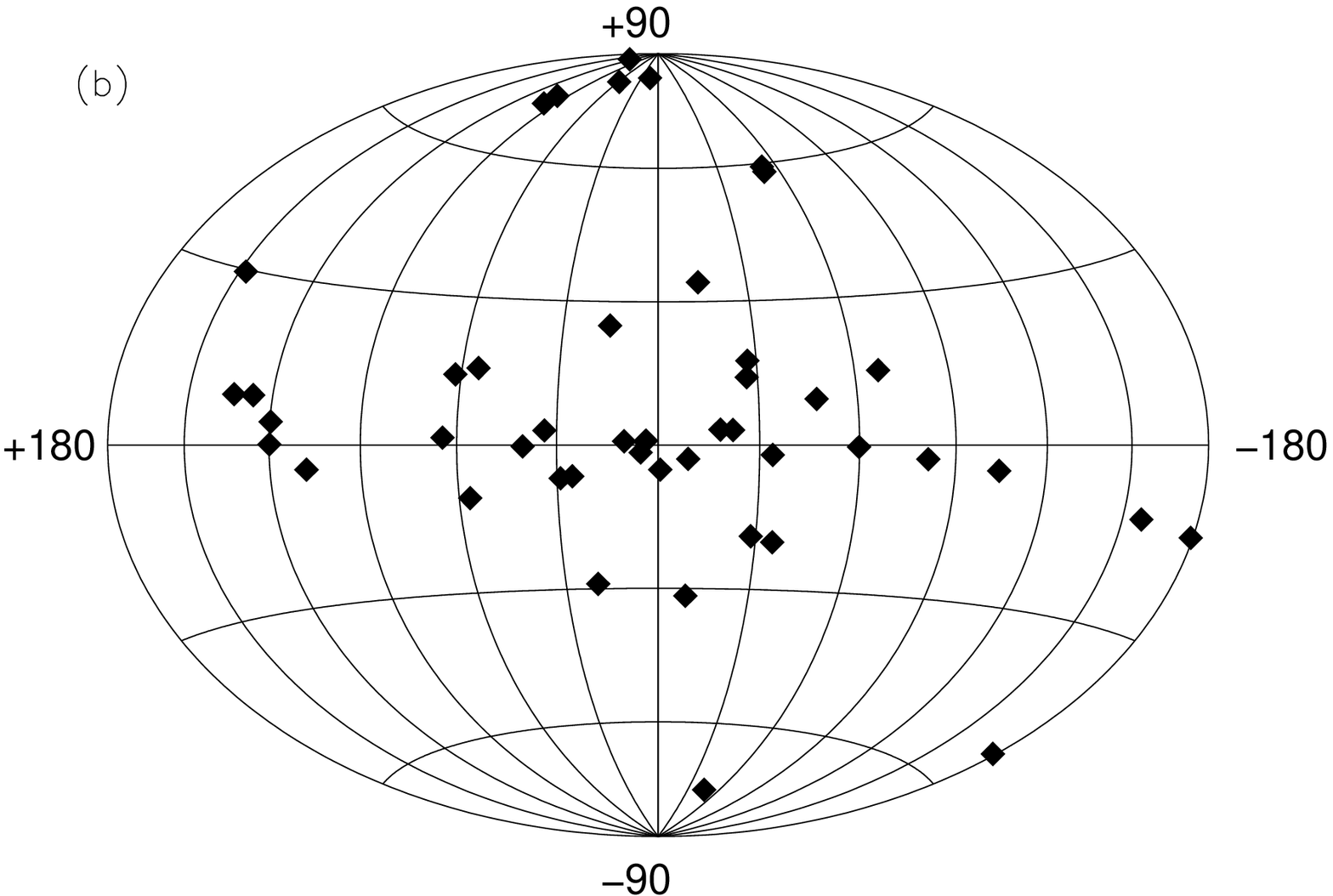}}
\caption{(a) \textit{INTEGRAL} exposure map in galactic coordinates from October 2002 up to
  July 2007 (contours in
  units of kiloseconds), showing the concentration of exposure in the
  direction of the galactic plane (Erik Kuulkers, private communication). (b)
  Spatial distribution of 47 \textit{INTEGRAL} GRBs detected between
  October 2002 and July 2007 in galactic coordinates.}
\label{fig:spatial}
\end{figure}

The exposure map and spatial distribution of the 47 GRBs observed with IBIS are shown in galactic coordinates in Fig.~\ref{fig:spatial} for the period from October 2002 to July 2007. 
The burst distribution is
significantly concentrated towards the galactic plane, reflecting the
direction in which the satellite is pointed. \textit{INTEGRAL} spent 64\% of its observing time in the half of the sky at galactic latitudes between $\pm30^\circ$.

\begin{figure}[ht]
\centering
\resizebox{\hsize}{!}{\includegraphics[angle=270]{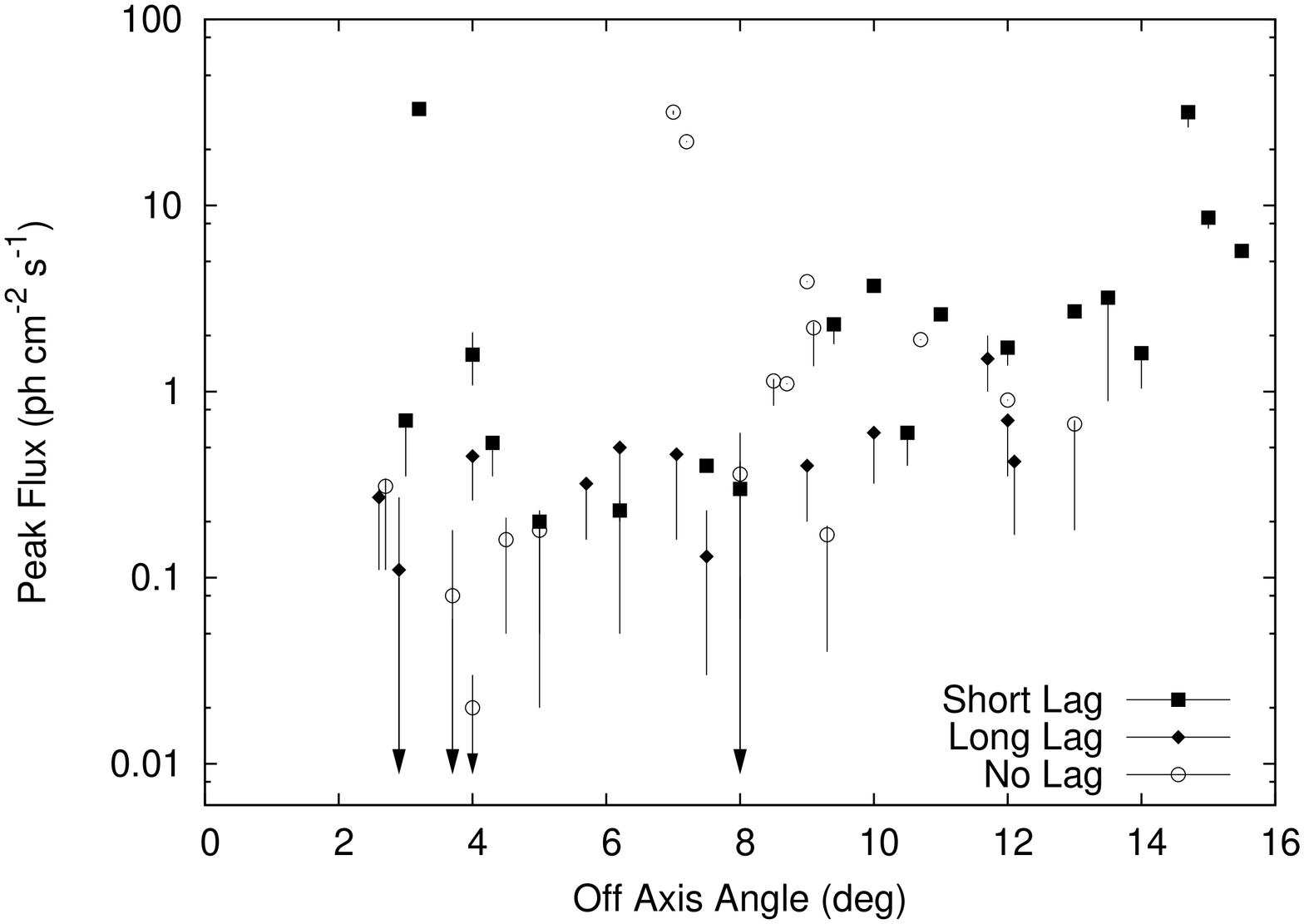}}
\caption{The off-axis angle distribution of the GRBs in the IBIS FoV as a function of peak flux (20--200\,keV). The FCFoV is $9^{\circ}\,\times\,9^{\circ}$ and the boundary of the PCFoV is between $4.5^{\circ}$ and $6.2^{\circ}$, depending on the azimuthal angle of the GRB position. The long-lag GRBs are marked with filled diamonds. There are 11 GRBs within the FCFoV and 4 have long lags. IBIS is less sensitive to $\gamma$-ray sources outside the FCFoV as reflected by the decrease in the number of faint GRBs.}
\label{fig:offaxis}
\end{figure}

\begin{figure}[ht]
\centering
\includegraphics[width=0.3\textheight,angle=270]{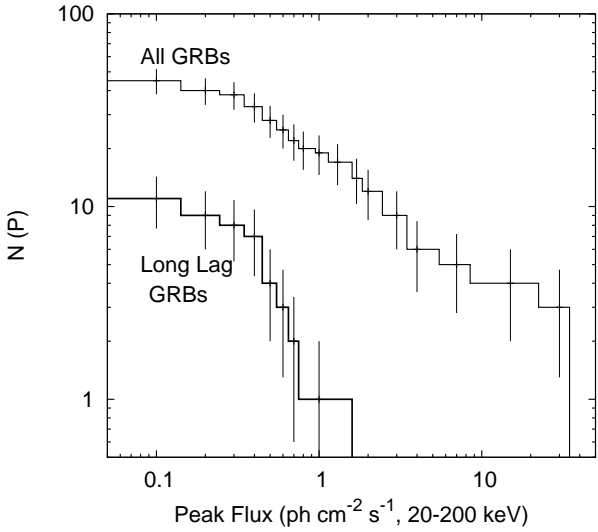}
\caption{The cumulative log\,N-log\,P distribution of the 47 GRBs detected by IBIS, with peak flux P measured between 20--200\,keV. The distribution is biased by the lower sensitivity of IBIS at large off-axis angles (Fig.~\ref{fig:offaxis}). The small subset of 11 long-lag GRBs is shown separately.}
\label{fig:logn_logp}
\end{figure}

\begin{figure}[ht]
\resizebox{\hsize}{!}{\includegraphics{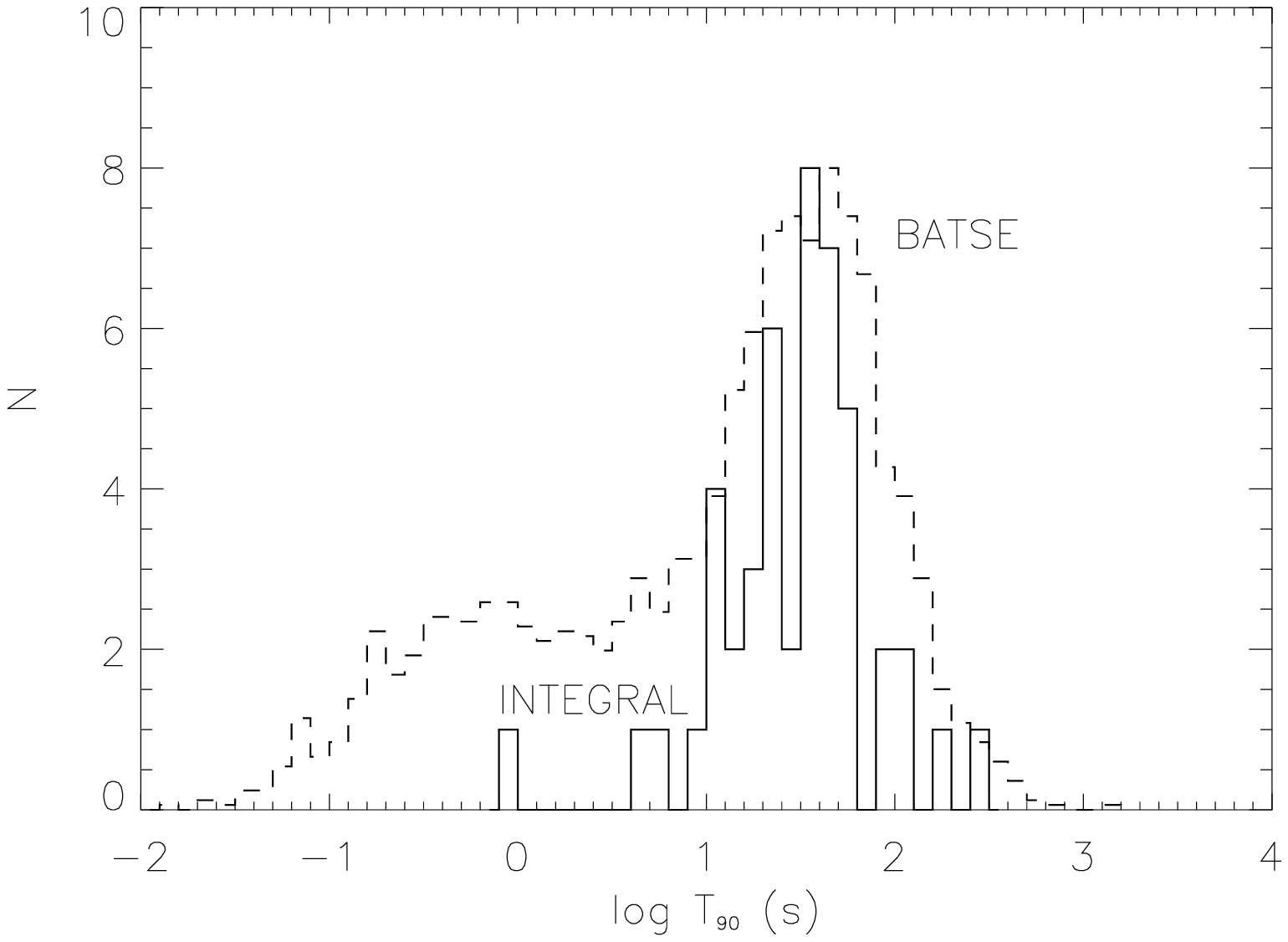}}
\caption{$T_{90}$ distribution of \textit{INTEGRAL} GRBs (solid line) in
  comparison to that of BATSE (dashed line). The BATSE distribution is normalised to
  the peak of the \textit{INTEGRAL} distribution for clarity. The BATSE
  data for 2041 GRBs is taken from the Current Catalog at 
  http://www.batse.msfc.nasa.gov/batse/grb/catalog/current.}
\label{fig:t90s}
\end{figure}

The properties of the 47 \textit{INTEGRAL} GRBs are presented in
Tables~\ref{table:grbs} and~\ref{table:band}. The coordinates and afterglow information are taken
  from the webpage maintained by Jochen Greiner. The typical
  size of the \textit{INTEGRAL} error box is 3$\arcmin$.  GRB\,060428c did not trigger IBAS but was discovered during subsequent analysis of \textit{INTEGRAL} archival data~\citep{grebenev:2007}. The spectral lag for each burst is presented in Table~\ref{table:speclags}, with 11 GRBs having long lags ($\tau_{2+3,1}>0.75$\,s). The off-axis angle distribution of the GRBs in the IBIS FoV is given in Fig.~\ref{fig:offaxis}. The bursts have preferentially higher peak fluxes at larger angles outside the FCFoV of IBIS due to the reduced sensitivity at lower coding levels. The log\,N-log\,P distribution is given in Fig.~\ref{fig:logn_logp} for all IBIS GRBs and separately for the small subsample of 11 long-lag GRBs. The long-lag GRBs appear to form a separate population at low values of P.  The $T_{90}$ distribution of \textit{INTEGRAL} GRBs is shown in
Fig.~\ref{fig:t90s} and compared with the bimodal distribution for BATSE GRBs~\citep{kouveliotou:1993}. 
There is reasonable agreement between the two distributions,
  especially when the small number of \textit{INTEGRAL} GRBs is taken
  into account. The IBIS lightcurves of 43 \textit{INTEGRAL} bursts are given in the Appendix in the 25--50\,keV and 50--300\,keV energy ranges. In general, the faint bursts have smooth, long-duration profiles with only one or two weak pulses.

\subsection{Spectral Properties of \textit{INTEGRAL} GRBs}

\begin{figure}[ht]
\centering
\subfigure{\label{fig:060114spec}\includegraphics[width=0.5\columnwidth,height=0.35\textheight,angle=270]{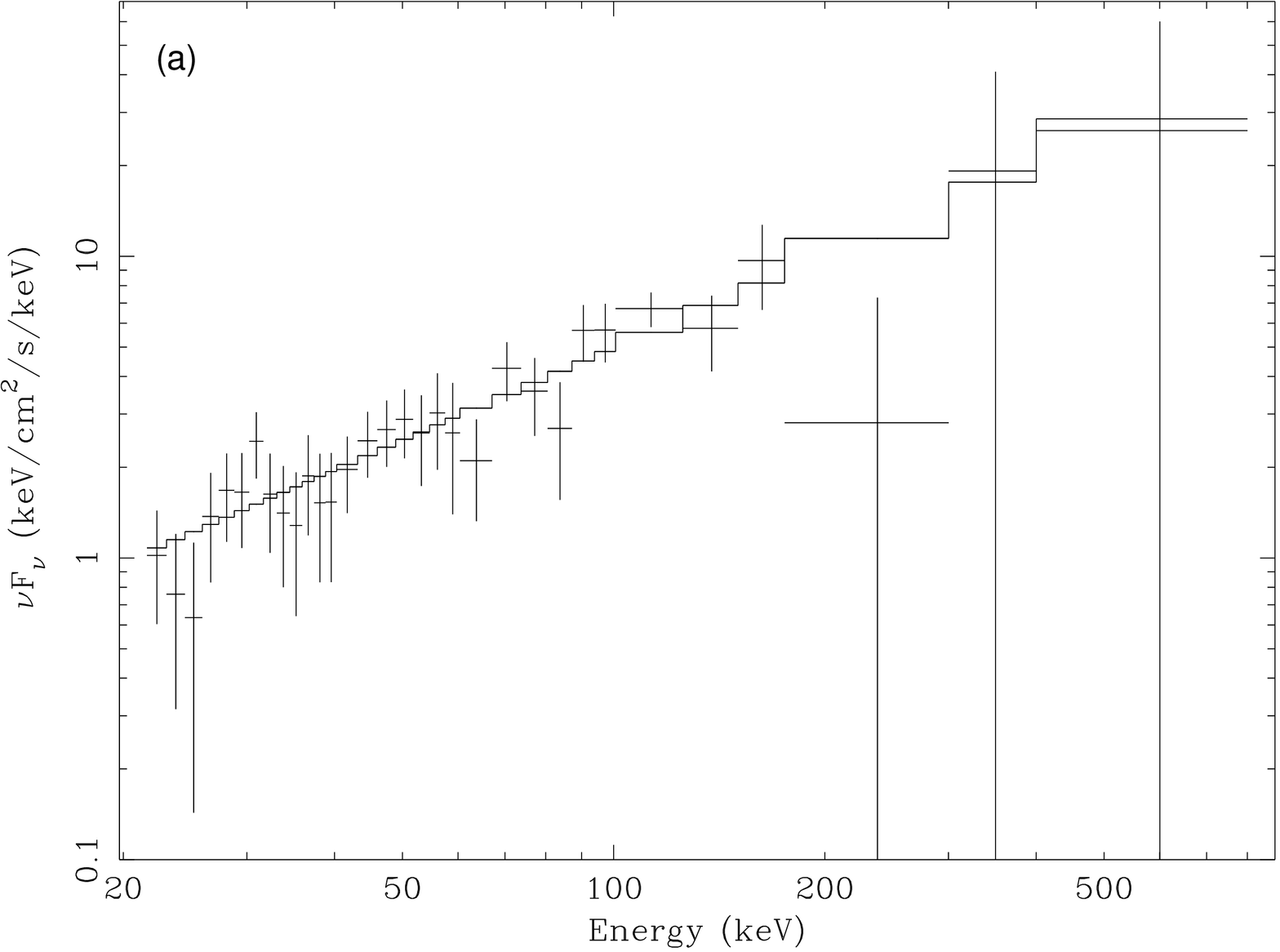}}
\subfigure{\label{fig:061025spec}\includegraphics[width=0.5\columnwidth,height=0.35\textheight,
   angle=270]{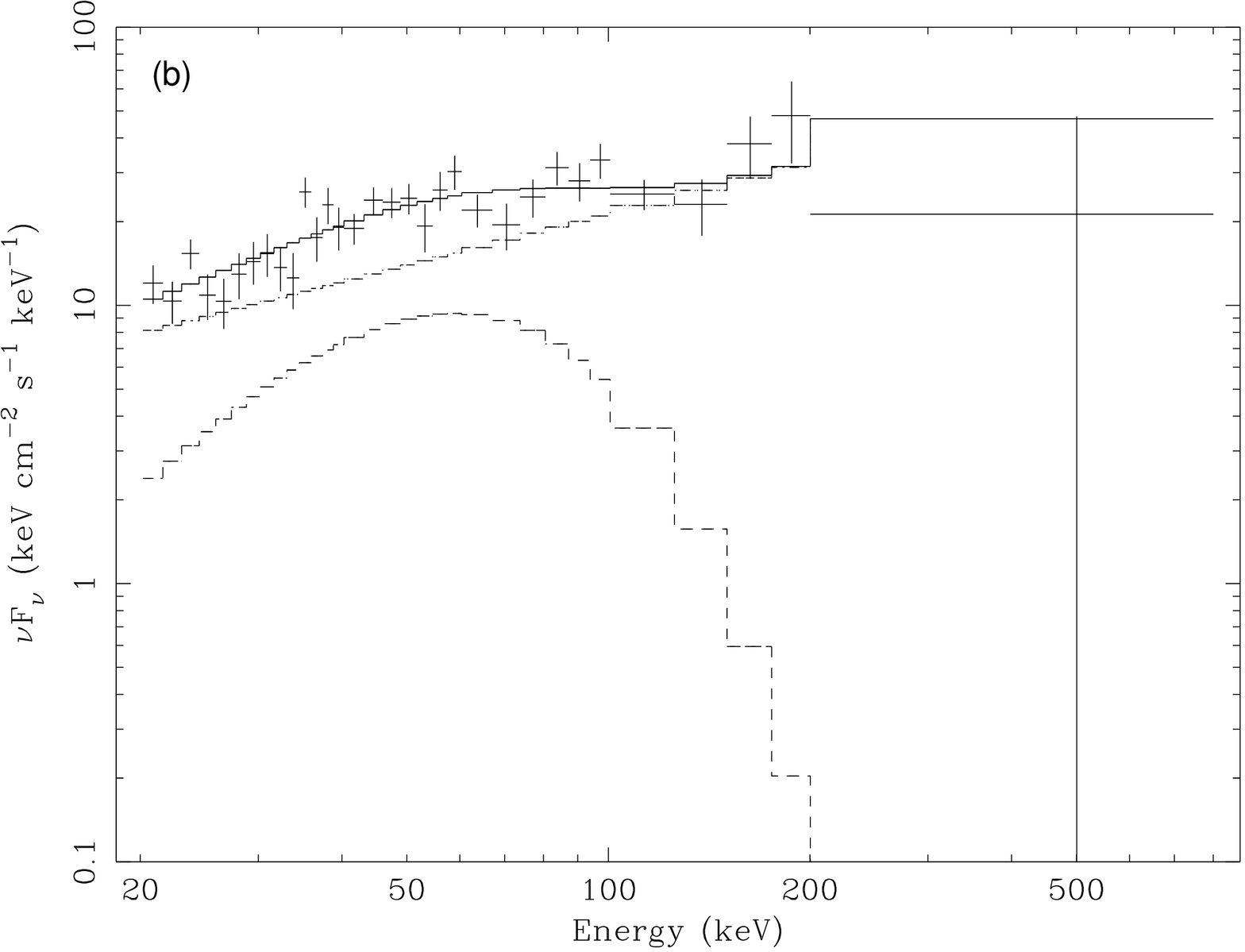}}
\subfigure{\label{fig:061122spec}\includegraphics[width=0.5\columnwidth,height=0.35\textheight,angle=270]{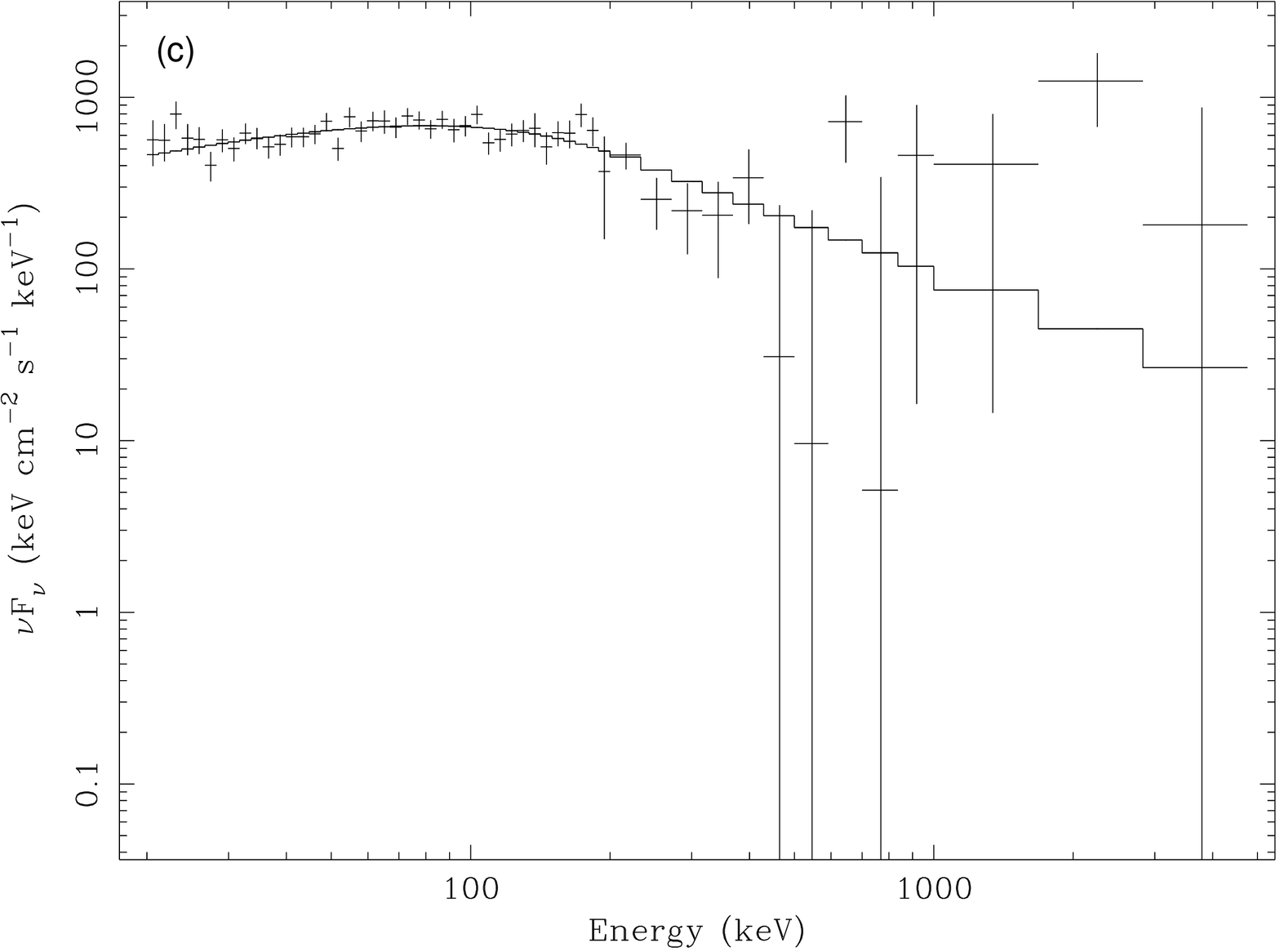}}
\caption{(a) IBIS ${\nu}$F$_{\nu}$ power-law spectrum of
  GRB\,060114 in the 20--800\,keV energy range. (b) IBIS ${\nu}$F$_{\nu}$ spectrum of GRB\,061025,
  fit by the quasithermal (blackbody + power-law) model in the
  20--800\,keV energy range. (c) SPI
 ${\nu}$F$_{\nu}$  spectrum of GRB\,050525a, from 20\,keV--8\,MeV, fit with the Band model.} 
\label{fig:spectra}
\end{figure}

The spectral information
given in Table~\ref{table:grbs} is obtained from the IBIS detector,
with the exception of GRB\,041219a, GRB\,050525a, GRB\,060901 and GRB\,061122 for
which the peak flux and fluence values are measured with SPI
in the 20--200\,keV energy range due to telemetry gaps in the IBIS data~\citep{mcbreen:2006,mcglynn:inprep}. 
Limited spectral information is
available for GRB\,021125 due to significant telemetry loss
while the satellite was in calibration mode~\citep{malaguti:2003}. Data for
GRB\,060930 and GRB\,070311 are currently not publicly available and values shown are taken
from~\citet{gotz:2006} and \citet{mereghetti:2007}, respectively. Peak flux values
are integrated over the brightest 1 second time interval in the
20--200\,keV energy band and fluence values are measured
over the  same energy range. Photon indices are given for a power-law fit to
the spectra in the 20--200\,keV energy range. Spectral
parameters obtained from IBIS and SPI spectra for which the Band
model and quasithermal model are fit are listed in
Table~\ref{table:band}. The values obtained from a combined IBIS and SPI spectral fit to GRB\,060428c over the 20--400\,keV energy range are taken from~\citet{grebenev:2007}. GRB\,030131 and GRB\,050502a were detected during a satellite slew so SPI
spectral analysis was not possible for these bursts. A sample of \textit{INTEGRAL} ${\nu}$F$_{\nu}$
spectra is shown in Fig.~\ref{fig:spectra}.

\begin{figure}[ht]
\resizebox{\hsize}{!}{\includegraphics{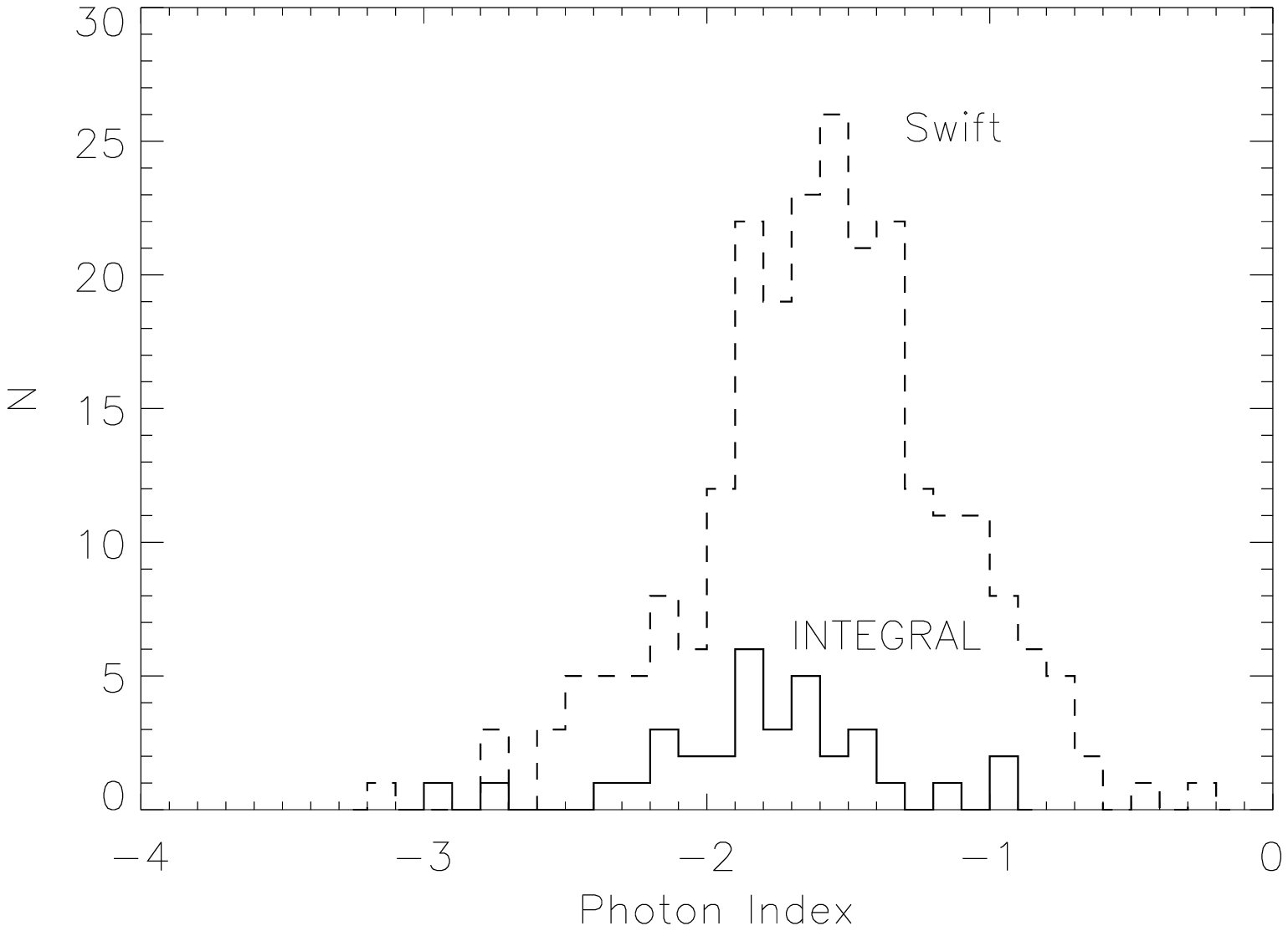}}
\caption{Power-law photon index distribution for \textit{INTEGRAL}
  GRBs detected between October 2002 and
July 2007
  (solid line) and \textit{Swift} GRBs detected between November 2004
and July 2007 (dashed line), in the 20--200\,keV and 15--150\,keV energy bands, respectively. The \textit{Swift} data for 238 GRBs is taken from
http://swift.gsfc.nasa.gov/docs/swift/archive/grb\_table.html. }
\label{fig:phot_ind}
\end{figure}

The distribution of photon indices is shown in Fig.~\ref{fig:phot_ind} for
\textit{INTEGRAL} and \textit{Swift} GRBs for which a power-law model
was fit to the
spectral data in the 20--200\,keV and 15--150\,keV energy ranges, respectively. 
In comparison to \textit{Swift}, \textit{INTEGRAL} detects proportionally more soft GRBs with
  steeper power-law photon indices.

\begin{figure}[ht]

\resizebox{\hsize}{!}{\includegraphics{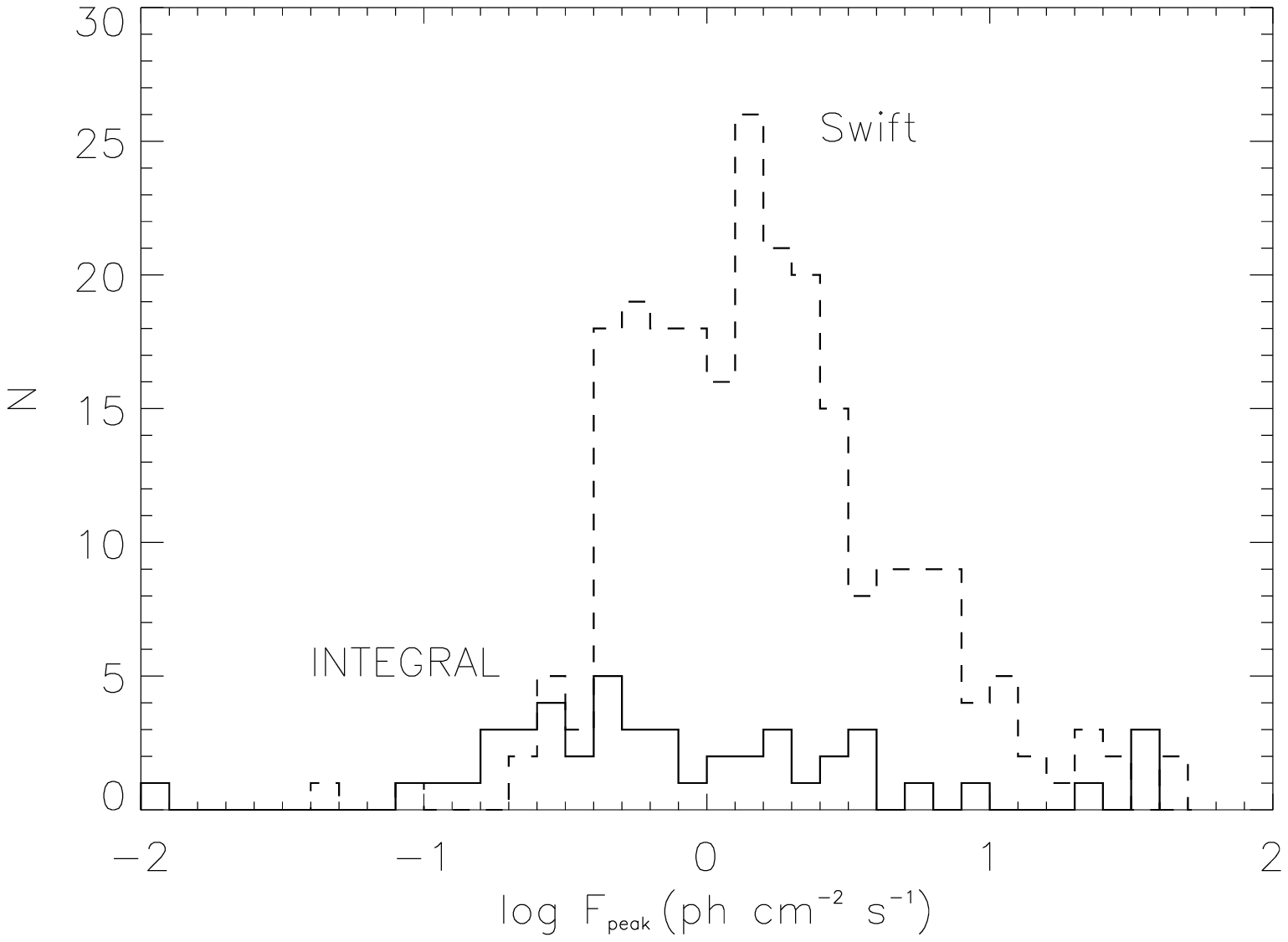}}
\caption{Peak flux distribution for GRBs detected by
  \textit{INTEGRAL} (20--200\,keV, solid line) and \textit{Swift}
  (15--150\,keV, dashed line). The \textit{Swift} data for 237 GRBs is taken from
http://swift.gsfc.nasa.gov/docs/swift/archive/grb\_table.html.}
\label{fig:fpeak}
\end{figure}

Fig.~\ref{fig:fpeak} compares the peak flux (20--200\,keV) distribution of
the GRBs observed by IBIS (solid line) to
the peak flux (15--150\,keV) distribution of the GRBs detected by the BAT
instrument on \textit{Swift}. IBIS detects proportionally more
weak GRBs than \textit{Swift} because of its better sensitivity within a FoV that is smaller by a factor of $\sim12$.

 \subsection{Spectral Lags of \textit{INTEGRAL} GRBs}

\begin{figure}[ht]
\centering
\resizebox{\hsize}{!}{\includegraphics{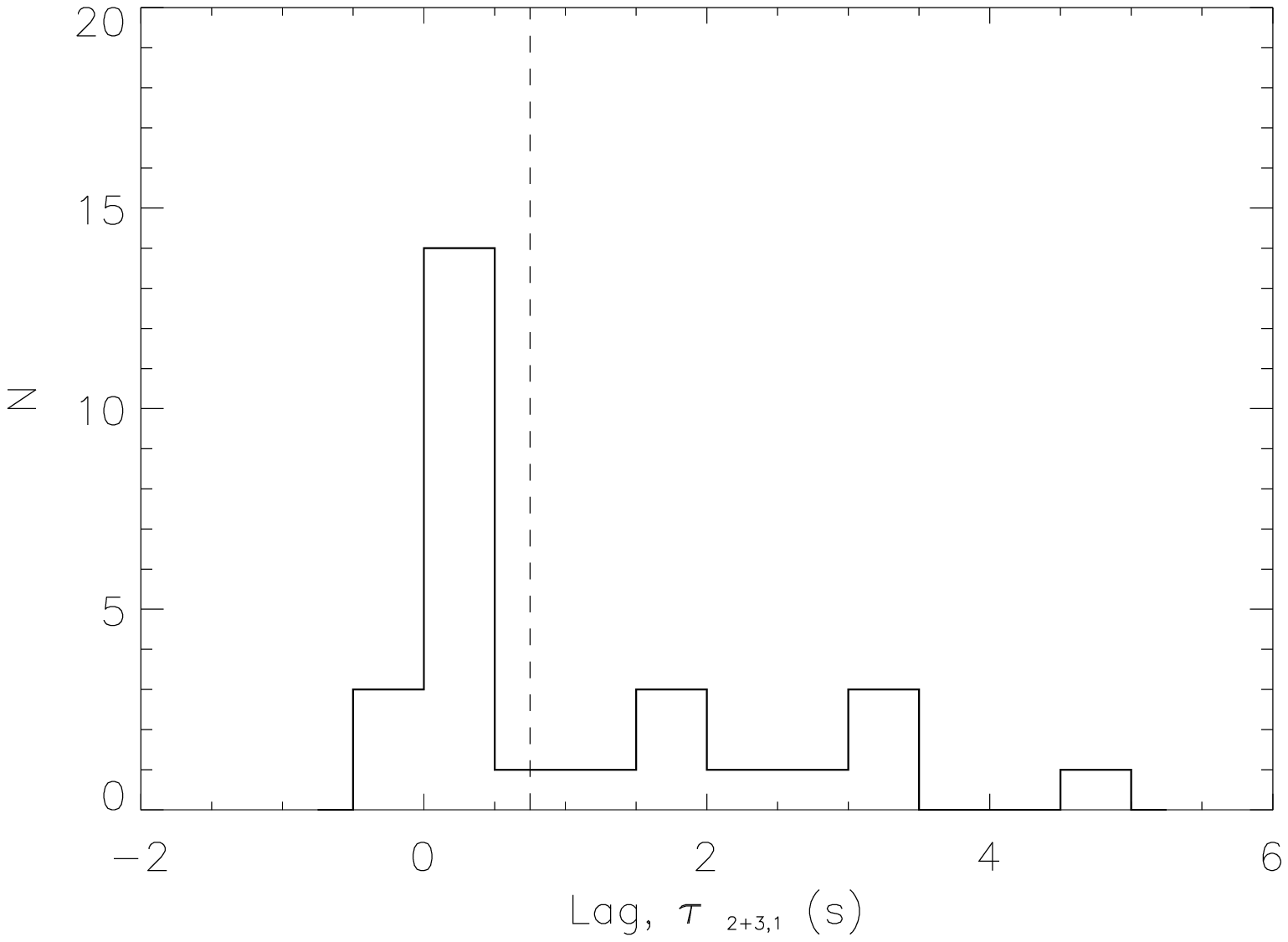}}
\caption{Spectral lag distribution for the 28 \textit{INTEGRAL} GRBs
  for which a lag could be measured between 25--50\,keV and 50--300\,keV
  ($\tau_{2+3,1}$). The distribution is separated by the dashed line
  into short-lag and long-lag GRBs at $\tau_{2+3,1}=0.75\,\rm{s}$.}
\label{fig:lag_hist}
\end{figure}

The spectral lag results are presented in
Table~\ref{table:speclags}. A positive spectral lag follows the usual convention of high energy emission preceding low
energy emission. The time intervals which were correlated to
determine the spectral lags are marked with vertical lines on the lightcurves in the Appendix.
Within these intervals, only those counts above 10\% of the peak
  count rate
were correlated to give the lag values in Table~\ref{table:speclags}. The number distribution of 
spectral lags measured over the full burst duration
is given in Fig.~\ref{fig:lag_hist} for the 28 long-duration
GRBs with a measured lag between  Channel~1 (25--50\,keV) and Channels 2 \& 3 (50--300\,keV). No statistically
significant negative spectral lags are found. Negative lags, which violate the typical hard-soft evolution of GRBs, have been observed in
a small minority of cases~\citep[e.g.][]{chen:2005} and may be more prevalent in short
bursts~\citep{gupta:2002,yi:2006}. A long tail extending to $\sim5$\,s is observed in the lag
distribution in Fig.~\ref{fig:lag_hist} and a clear separation between
short and long lag is drawn at $\tau_{2+3,1}\sim0.75$\,s. Thus, long-lag bursts have $\tau_{2+3,1}>0.75$\,s and those with $\tau_{2+3,1}<0.75$\,s are
referred to as short-lag GRBs. The 11 long-lag GRBs are identified in Table~\ref{table:speclags}, with 4 in the FCFoV, a further 4 in the PCFoV to the 50\% coding level and the remaining 3 GRBs at larger off-axis angles (Fig.~\ref{fig:offaxis}).

\begin{figure}[ht]
\centering
\resizebox{\hsize}{!}{\includegraphics[width=0.3\textheight,angle=270]{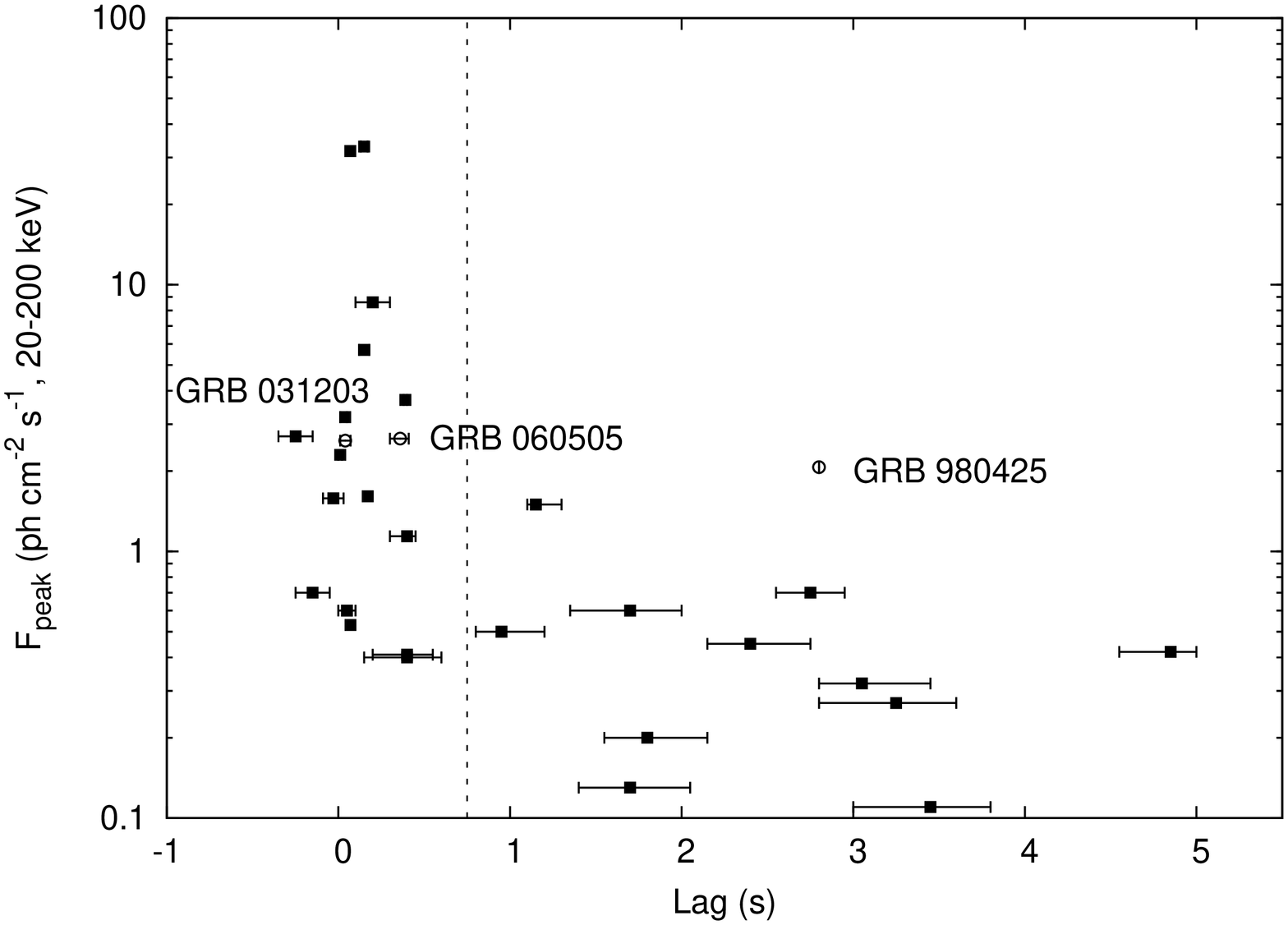}}
\caption{Spectral lag distribution of \textit{INTEGRAL} GRBs
  as a function of peak flux (20--200\,keV). The SN bursts GRB\,980425 and GRB\,031203 are identified and represented by open circles, as is GRB\,060505 which does not have an associated SN. XRF\,060218 has a peak flux of $0.6\,\rm{ph}\,\rm{cm}^{-2}\,\rm{s}^{-1}$ and is not included in the figure because of its very long lag of $61\pm26$\,s. GRBs with the
  longest lags tend to have low peak fluxes, whereas GRBs with short
  lags have both low and high peak fluxes. The dashed line indicates
  the separation between long and short-lag GRBs at $\tau_{2+3,1}=0.75$\,s.}
\label{fig:fp_lag}
\end{figure}

The spectral lag distribution of \textit{INTEGRAL} GRBs as
a function of peak flux is shown in Fig.~\ref{fig:fp_lag}.  The BeppoSAX SN burst GRB\,980425~\citep{galama:1998} is also shown in Fig.~\ref{fig:fp_lag}. The SN burst XRF\,060218, with a peak flux of $0.6\,\rm{ph}\,\rm{cm}^{-2}\,\rm{s}^{-1}$, is not included in the figure because it has an extremely long lag of $61\pm26\,\rm{s}$~\citep{liang:2006}. The other low luminosity burst shown in Fig.~\ref{fig:fp_lag} is GRB060505, which has no associated SN~\citep{fynbo:2006,mcbreen:inprep}. The figure shows that both bright and faint GRBs have short spectral lags, but there is an obvious absence of bright long-lag GRBs. Therefore GRBs with long spectral lags tend to be weak bursts with low peak flux. This trend is in good agreement with that
observed using BATSE GRBs~\citep{norris:2002}, where the proportion of long-lag GRBs is negligible
among bright BATSE bursts and increases to around 50\% at the
trigger threshold. Using 1429
BATSE GRBs, \citet{norris:2002} identified three GRB groups that consist of bright
short-lag bursts, weak short-lag bursts and weak long-lag bursts. These groups are also
clearly identifiable for the \textit{INTEGRAL} sample in
Fig.~\ref{fig:fp_lag}. 

\begin{figure}[ht]
\centering
\resizebox{\hsize}{!}{\includegraphics[width=0.5\textheight,angle=270]{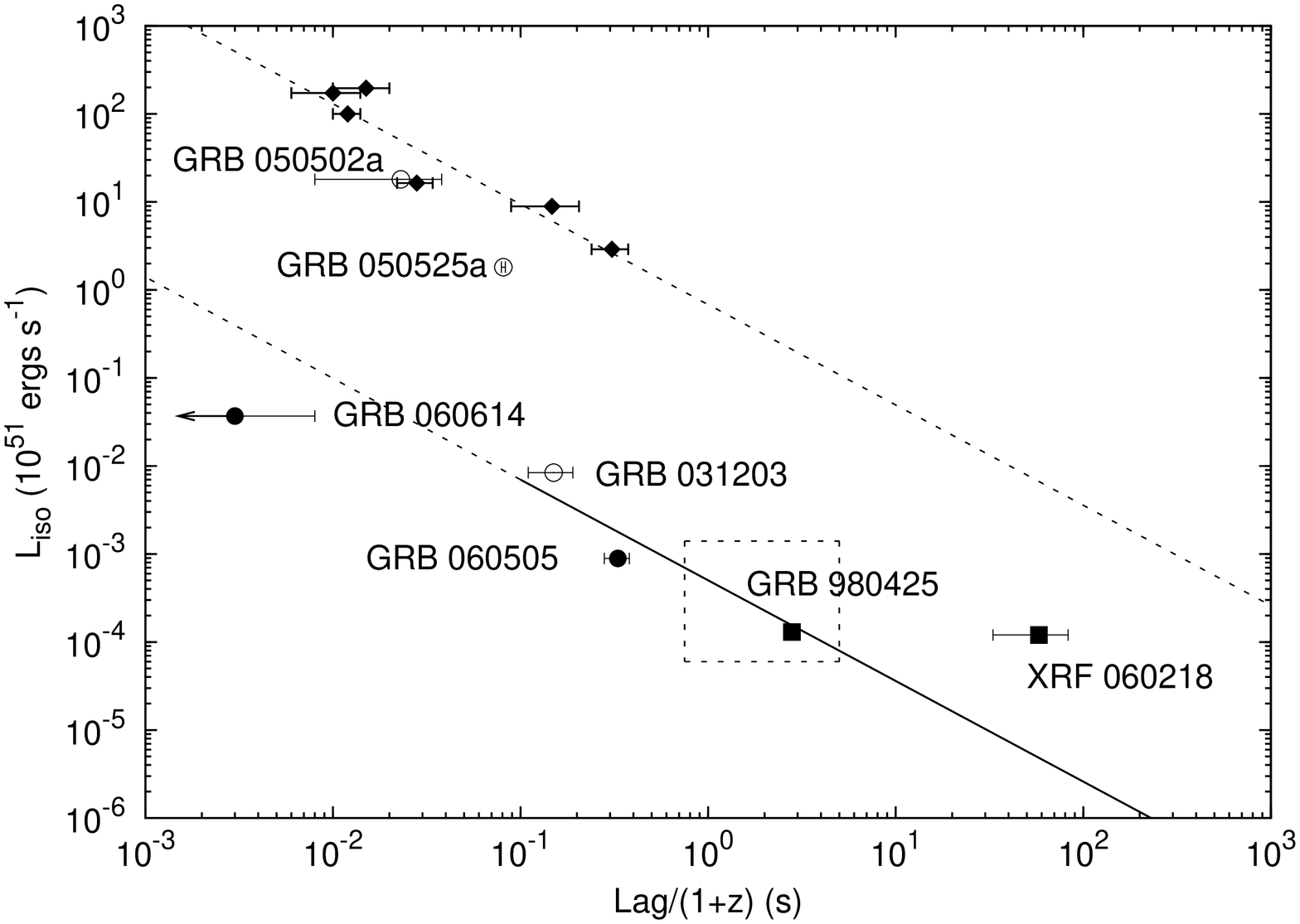}}
\caption{Isotropic peak luminosity
  ($\times10^{51}\,\rm{ergs}\,\rm{s}^{-1}$, 50--300\,keV) as a function of spectral
  lag measured between 25--50\,keV and 100--300\,keV. The two exceptions are GRB\,060505 and GRB\,060614 for which the lag was measured between 25--50\,keV and 50--100\,keV and 15--25\,keV and 50--100\,keV, respectively. The dashed line
  is the anti-correlation obtained for 6 bursts (diamonds) at known redshift~\citep{norris:2000}. The 3 \textit{INTEGRAL} GRBs with a redshift and measured lag are represented by open circles, including the SN burst GRB\,031203. The other low-luminosity SN bursts GRB\,980425 and XRF\,060218 are identified and represented by squares and the two GRBs without SNe by filled circles. The long-lag GRBs lie within the dashed box if they are at the adopted distance of 250\,Mpc, with peak luminosities ranging from $6\times10^{46}$ to $1.4\times10^{48}\,\rm{ergs}\,\rm{s}^{-1}$
 and total luminosities from $10^{48}$ to $10^{50}\,\rm{ergs}$. 
The solid line is the proposed lag-luminosity relationship for the low-luminosity GRBs and extrapolates to short GRBs at low values of the lag, e.g. GRB\,060614, which lies in the region of the plot occupied by short GRBs. No k-correction was applied to the data.}
\label{fig:lag_lum}
\end{figure}

The isotropic peak luminosity as a
function of spectral lag is shown in Fig.~\ref{fig:lag_lum} and includes the 3 \textit{INTEGRAL} GRBs with known
redshift for which a lag was measured. The low-luminosity bursts GRB\,980425, XRF\,060218 and GRB\,060505 are
also plotted in Fig.~\ref{fig:lag_lum}. The dashed line is the anti-correlation between lag and luminosity proposed
by~\citet{norris:2000}. The bright \textit{INTEGRAL} bursts
GRB\,050502a and GRB\,050525a have peak luminosities of
$1.8\times10^{52}$\,erg\,s$^{-1}$ and $1.8\times10^{51}$\,erg\,s$^{-1}$ and
spectral lags of $0.11^{+0.07}_{-0.06}$\,s and $0.130^{+0.003}_{-0.002}$\,s,
respectively and follow the trend of the relation. At $z=0.106$, GRB\,031203 has a peak luminosity of
$8.4\times10^{48}$\,erg\,s$^{-1}$ and spectral lag of $0.17^{+0.03}_{-0.04}$\,s, causing it to
fall significantly below the
correlation~\citep{sazonov:2004,gehrels:2006}. The HETE burst GRB\,030528 has a long lag similar to that of
GRB\,980425 but a relatively high luminosity and is consistent with
the lag-luminosity relation~\citep{gehrels:2006}. A long spectral lag is
therefore not an exclusive determinant for a low-luminosity GRB.  

For the SN burst GRB\,031203, spectral lags of
$0.24\pm0.12$\,s between 20-50\,keV and 100-200\,keV~\citep{sazonov:2004} 
and $0.30\pm0.20$\,s between 15-50\,keV and
50-150\,keV~\citep{shrader:2006} have been reported and are consistent with the value
 $\tau_{3,1}=0.17^{+0.03}_{-0.04}$\,s in Table~\ref{table:speclags} and Fig.~\ref{fig:fp_lag}. A spectral lag for GRB\,040403 of $0.6\pm0.1\,\rm{s}$ between 15--40\,keV and
40--200\,keV has previously been reported~\citep{mereghetti:2005},
consistent with the value of $\tau_{2+3,1}=0.95^{+0.25}_{-0.15}$\,s
in Table~\ref{table:speclags}, taking into account the
different energy ranges.  The spectral lag of GRB\,050525a
was found to be $\tau_{3,1}=0.124\pm0.006$\,s~\citep{norris:2005}, in agreement with
the value of $\tau_{3,1}=0.130^{+0.003}_{-0.002}$\,s obtained in Table~\ref{table:speclags}. The spectral lag for the short
burst, GRB\,070707 is $5\pm5$\,ms ($\tau_{2+3,1}$), consistent with
the negligible lag values
expected for short GRBs~\citep{norris:2006,zhang:2006b}.  
For most of the GRBs with distinct separate pulses, the spectral lag does not
evolve significantly during the burst, i.e. GRB\,030320, GRB\,040106,
GRB\,040422,
GRB\,040812, GRB\,050522 and GRB\,050525a. Some evolution is
  evident in GRB\,050918 for which the lag varies from
  $0.50^{+0.05}_{-0.04}$\,s for the first pulse of the burst, to a
  slightly negative lag of $-0.14\pm0.01$\,s for the second pulse. For GRB\,041219a, the lag
of the precursor is relatively long and is much shorter for the main
emission pulses, as discussed in~\citet{mcbreen:2006}. Spectral lag
evolution is often seen in multi-peaked GRBs~\citep[e.g.][]{hakkila:2004,chen:2005,Ryde_Kov:2005,hakkila:2008}, and poses obvious
difficulties for the use of the lag as a luminosity indicator.

Spectral lags are measured between wavelet-smoothed lightcurves for
a number of GRBs with $F_{peak}\,$(20--200\,keV)\,$\lesssim0.6\,\rm{ph}\,\rm{cm}^{-2}\,\rm{s}^{-1}$ and are
denoted by $\dagger$ in
Table~\ref{table:speclags}. Table~\ref{table:wavelet} gives the lags
determined with and without denoising for these GRBs. The
  denoising method allows
the lag to be measured in some cases for very weak GRBs where it
cannot be determined using the  raw data and in most cases the
lag is better constrained with smaller errors.   

Spectral lag measurements were not possible for 16 GRBs for the
following reasons:
\begin{enumerate}
\item{Significant telemetry gaps occur in the IBIS data for
  GRB\,021125 (Fig.~\ref{fig:lcs1}\,(a)), GRB\,030131, GRB\,030227
  (Fig.~\ref{fig:lcs1}\,(c)) and GRB\,061122 (Fig.~\ref{fig:lcs4}\,(d)).}
\item{GRB\,041015 (Fig.~\ref{fig:lcs2}\,(f)), GRB\,050129
  (Fig.~\ref{fig:lcs2}\,(i)), GRB\,050922a (Fig.~\ref{fig:lcs3}\,(g)),
  GRB\,060114 (Fig.~\ref{fig:lcs3}\,(j)), GRB\,060204a (Fig.~\ref{fig:lcs3}\,(l)) and
  GRB\,060912b (Fig.~\ref{fig:lcs4}\,(b)) are too weak, even when denoised lightcurves are used
  for the correlation.}
\item{GRB\,050223
  (Fig.~\ref{fig:lcs2}\,(j)), GRB\,050714a (Fig.~\ref{fig:lcs3}\,(e))
  and GRB\,051105b (Fig.~\ref{fig:lcs3}\,(h)) have significant
  temporal variations which results in incorrect pulses being
  correlated and so significant CCF peaks were not found for these bursts.}
\item{IBIS data are currently unavailable to this team for GRB\,060428c, GRB\,060930 and GRB\,070311.} 
\end{enumerate}

There are special circumstances for some GRBs and these include:
\begin{enumerate}
\item{Spectral lags for
GRB\,021219 (Fig.~\ref{fig:lcs1}\,(b)) and GRB\,050502a
(Fig.~\ref{fig:lcs2}\,(k)) are measured over the region of the burst
before telemetry saturation occurred.}
\item{For GRB\,030529 only the first peak is correlated  because
telemetry gaps occur in the later data (Fig.~\ref{fig:lcs1}\,(f)).}
\item{GRB\,040903 (Fig.~\ref{fig:lcs2}\,(e)) and
GRB\,050522 (Fig.~\ref{fig:lcs3}\,(b)) are not detected above 50\,keV so the lag is
determined  between 15--25\,keV and 25--50\,keV (Channel
2).}
\item{GRB\,041219a (Fig.~\ref{fig:lcs2}\,(h))~\citep{fenimore:2004} and GRB\,050525a
  (Fig.~\ref{fig:lcs3}\,(c)) were simultaneously observed by \textit{INTEGRAL} and \textit{Swift}. There are telemetry gaps in the IBIS data for both
  bursts and SPI data types do not permit energy-resolved
  lightcurves to be extracted. The spectral
lags are therefore determined using data from the BAT instrument on \textit{Swift} in the same
energy ranges.}
\end{enumerate}

\section{Discussion\label{discussion}}

\subsection{\textit{INTEGRAL} GRBs of particular note\label{particulars}}

GRB\,031203 is the third nearest GRB at
$z=0.1055$~\citep{prochaska:2004}
and is notable for its
unambiguous association with the supernova
SN\,2003lw~\citep{malesani:2004}. It is a confirmed low-luminosity 
GRB with an isotropic energy of $\sim4\times10^{49}\,\rm{erg}$ (20-200\,keV)~\citep{sazonov:2004} and a lag of $0.17^{+0.04}_{-0.03}$\,s between 25--50 and 50--300\,keV. The
IBAS localisation of GRB\,031203 enabled \textit{XMM-Newton} to begin
follow-up observations 6 hours later~\citep{santos:2003}. The X-ray observations
showed concentric ring-like structures centred on the GRB
location, making this the first detection of a GRB X-ray
halo, caused by X-ray scattering from dust columns in our galaxy along
the line of sight to the GRB~\citep{vaughan:2004,tiengo:2006}. Dust
scattering X-ray halos have recently been observed for two \textit{Swift}
GRBs~\citep{vianello:2007}. The X-ray observations of GRB\,031203 inferred a very high
soft X-ray flux for this burst, implying that this GRB may have been an
X-ray Flash (XRF)~\citep{watson:2004,watson:2006,sazonov:2006}.

GRB\,041219a is the brightest burst localised by
\textit{INTEGRAL} \citep{mcbreen:2006}. The
peak flux  of 43\,ph\,cm$^{-2}$\,s$^{-1}$ (20\,keV--8\,MeV,
1\,s integration) is greater than that for $\sim$98\% of all
bursts  and the total duration of $\sim$520\,s is longer than
all but a small number of bursts. The SPI instrument was used to measure GRB polarisation
through multiple scattering events in its 19 Ge
detectors~\citep{kalemci:2004}, since the scatter angle
distribution depends on the polarisation of the incoming
photons~\citep{lei:1997}. A search for linear polarisation in
the most intense pulse (66 seconds) in GRB~041219a and sub-intervals was
performed. \citet{kalemci:2007} and~\citet{mcglynn:2007} have shown
that there is evidence for a high degree of polarisation, but at a low significance level.

\subsection{Afterglows of \textit{INTEGRAL} GRBs}

X-ray observations were performed for 2 long-lag GRBs with \textit{XMM-Newton}~\citep{GCN2533,GCN2688} and 5 long-lag GRBs with \textit{Swift}~\citep{GCN3426,GCN3564,GCN4347,GCN4624,GCN6369}. There were also X-ray observations of 8 of the GRBs with no lag measurement. The median X-ray flux for the long-lag GRBs is $\sim4\,\times$ 10$^{-13}$ erg cm$^{-2}$ s$^{-1}$ between 5 to 10 hours after the burst, with an upper limit in the two cases of non-detection of $\sim4\,\times$ 10$^{-14}$\,erg\,cm$^{-2}$\,s$^{-1}$, indicating that the long-lag weak GRBs tend to have a weak X-ray afterglow component. In contrast, the other \textit{INTEGRAL} GRBs with a measured afterglow have a typical flux of $\sim9\times10^{-13}$\,erg\,cm$^{-2}$\,s$^{-1}$ between 2 and 12 hours after the burst.

Of $\sim$300 GRBs localised by \textit{Swift} up to November 2007, $\sim$60\% have optical or near-IR afterglows and only $\sim$30\% have measured redshifts~\citep{coward:2007}, 
even though deep observations down to
$\sim$21--22\,magnitudes are carried out for most events within 24 hours of
the burst. Dark GRBs without a detectable afterglow may therefore make up a significant proportion of
the GRB population.  

Almost 70\% (31/46) of the GRBs observed by \textit{INTEGRAL} do not
have a detected optical counterpart, including 9 of the 11 long-lag GRBs. The optical observations revealed faint afterglows for GRB\,040323 and GRB\,040827, and near-IR afterglows for GRB\,040223 and GRB\,040624~\citep{filliatre:2006}. A non-spectroscopic redshift in the range $0.5<z<1.7$ was obtained for GRB\,040827~\citep{deluca2005}. The IBIS error box of GRB\,060114 contains galaxies from the cluster A1651 (z=0.087) and the optical afterglow was fainter than $R=19$ just 1.9 min after the GRB~\citep{guidorzi:2006}. Only one radio afterglow has been detected for the long-lag GRBs~\citep{GCN4350}, compared to a total of 8/46 for all of the \textit{INTEGRAL} GRBs. 

GRB\,040223 and GRB\,040624~\citep{filliatre:2006} provide good
examples of GRBs with dark or faint optical
afterglows. GRB\,040223 was observed close to the galactic
plane, so NIR observations were carried out to overcome the
high dust obscuration. Observations were undertaken at the NTT
of ESO, 17 hours after the GRB and no afterglow was
found. GRB\,040624 was located far from the galactic plane at
high latitude where the optical extinction is
negligible. Afterglow observations were carried out 13 hours
after the burst using the VLT and TNG. Magnitude limits were
obtained in the optical that are fainter than the very faint
end of the distribution of the magnitudes of a compilation of
39 promptly observed counterparts. The position of GRB\,040624 is less than 5\,$\arcmin$ from a galaxy in the cluster A1651. A search for a supernova was carried out up to a month after the GRB but none was found to a faint limit of $R > 22.6$~\citep{GCN2632}. 

The lack of a detected optical afterglow may be due to a number of
factors. Possible explanations include dust obscuration, a low-density environment, an intrinsically faint
afterglow in the optical, a rapidly decaying afterglow or the burst
occuring at a high redshift~\citep[e.g.][]{Jakobsson,Rol}. Dust
obscuration may be due to a burst environment with a
high gas column density~\citep[e.g.][]{lamb:2001, AJCT2} or dust in the
host galaxy along the line of sight to the GRB ($\sim10$\% of dark
events,~\citet{Piro}).  Low-density GRB environments can also produce a
very faint optical afterglow but the association between GRBs and
core-collapse SNe does not favour this scenario~\citep{Taylor}. Some
GRBs have intrinsically faint afterglows~\citep{Fynbo}. GRBs at high
redshift can only account for $\sim10$\% of these dark bursts~\citep[e.g.][]{gorosabel:2004,AJCT1}. Using early observations of \textit{Swift} GRBs,~\citet{roming:2006} found that $\sim$25\% of the sample were extincted by galactic dust,  $\sim$25\% by absorption in the local burst environment and $\sim$30\% were most probably affected by Ly-$\alpha$ absorption at high redshift.

\subsection{\textit{INTEGRAL} Spectra\label{spectra}}

Most GRB continuum spectra can be fit by the Band model~\citep{band:1993}, an
 empirical function comprising two smoothly broken power laws. The distributions of the low energy power-law photon index and
 high energy power-law photon index are distributed around
 values of $\alpha=-1$ and $\beta=-2.2$, respectively, for time-resolved
 spectra of 156 GRBs
 detected by BATSE~\citep{preece:2000}. The majority of GRBs have low energy power-law spectral
 indices in the range $-2\,<\,\alpha\,<\,0$~\citep[e.g.][]{preece:2000}. 
Of the 10 \textit{INTEGRAL} GRBs fit by the Band model
  (Table~\ref{table:band}), all have low energy spectral indices consistent with synchrotron
 emission, i.e. $-3/2\,<\,\alpha\,<\,-2/3$.

A thermal component of the prompt emission spectra has been proposed
 by several authors~\citep[e.g.][]{ghirlanda:2003, kaneko:2006, ryde:2005,bosnjak:2006,mcbreen:2006}. 
 In a study of BATSE GRBs with very hard
 spectra,~\citet{ghirlanda:2003} found that the time-resolved spectra
 were not adequately described by non-thermal emission models and that the
 early parts of the bursts were well fit by a blackbody
 component. \citet{ryde:2005} has shown that GRB
 spectra may be composed of a thermal and a power-law
 component. \citet{mcbreen:2006} found that for quasithermal model fits to the precursor and main emission
 of the \textit{INTEGRAL} burst GRB\,041219a, the blackbody component is more dominant in
 the precursor of the burst, while the main burst emission is
 well fit by both the Band and quasithermal
 models. The blackbody temperature decreases from the precursor of
 GRB\,041219a to the
 main burst emission as expected from~\citet{ryde:2005}. The
 10 bright GRBs in Table~\ref{table:band} are equally well fit by the
 Band model and blackbody + power-law model fits to the average 
  prompt emission spectra, with similar
  $\chi^{2}$ values.

Many of the \textit{INTEGRAL} GRBs in Table~\ref{table:grbs} have
steep power-law spectra, which are significantly outside the usual
range of the Band model low-energy spectral index of $-3/2\,<\alpha\,<\,-2/3$. For these
GRBs, the power-law index is similar to the high energy photon
index above the break energy, with typical values of $\beta$ between
$-2$ and $-2.5$~\citep{preece:2000}. If the steep power-law indices are
assumed to be $\beta$, the break energy, $E_{0}$, must be at or
below the sensitivity threshold of IBIS, $\sim20$\,keV. There is therefore an excess of counts in
the X-ray/soft $\gamma$-ray region, implying that they are
X-ray rich.  The peak
energy is given by $E_{peak}=(2+\alpha)\,E_{0}$ and has a value of
$E_{peak}=E_{0}/2$ for a typical value of $\alpha=-1.5$. These GRBs have a low $E_{peak}$ and will have a low luminosity if they are at low redshift~\citep{amati:2007}.

\textit{INTEGRAL} has detected a number of X-ray rich GRBs. X-ray flashes, X-ray rich GRBs and classical GRBs appear
to possess a continuum of spectral properties and it is
probable that they have a similar
origin~\citep{sakamoto:2005}. \textit{INTEGRAL} X-ray rich bursts tend to be weak and their time
profiles consist of long, slow pulses (see Appendix). Examples include
GRB\,040223 (Fig.~\ref{fig:lcs1}\,(i)), GRB\,040624
(Fig.~\ref{fig:lcs2}\,(a)),
GRB\,050626 (Fig.~\ref{fig:lcs3}\,(d)) and GRB\,060130 (Fig.~\ref{fig:lcs3}\,(k)).

\subsection{The rate of \textit{INTEGRAL} GRBs}
GRBs detected at off-axis angles outside the FCFoV of IBIS have a higher peak flux (Fig.~\ref{fig:offaxis}) because of the reduction in collecting area. For this reason, we estimate the rate of GRBs both in the FCFoV and in the PCFoV at greater than 50\% coding. There are 11 GRBs within the FCFoV of 0.025\,sr of IBIS in 4 years of observation time, yielding an all-sky rate of $\sim1400\,\rm{yr}^{-1}$ above the threshold of $\sim0.15\,\rm{ph}\,\rm{cm}^{-2}\,\rm{s}^{-1}$ in the energy range 20--200\,keV. The 4 long-lag GRBs in the FCFoV give a rate of $\sim500\,\rm{yr}^{-1}$. The 33 GRBs within the PCFoV to the 50\% coding level yield an all-sky rate of $\sim930\,\rm{yr}^{-1}$ and $\sim230\,\rm{yr}^{-1}$ for the 8 long-lag GRBs above the higher threshold of $\sim0.25\,\rm{ph}\,\rm{cm}^{-2}\,\rm{s}^{-1}$. The long-lag GRBs contribute significantly at faint flux levels and appear to form a separate population in the log\,N-log\,P distribution (Fig.~\ref{fig:logn_logp}). The rate of GRBs is in good agreement with the values obtained from the more sensitve analysis of BATSE archival data~\citep{kommers:2001,stern:2001}.

\subsection{The population of long-lag GRBs with low luminosity}
The lack of redshift determinations for the weak GRBs detected by IBIS prohibits progress by using individual GRBs. However there are a number of redshift indicators that can be used on weak GRBs. In this case the best redshift indicator is the spectral lag which combines the spectral and temporal properties of the prompt GRB emission. GRBs have a long lag when a typical value of 0.1\,s is redshifted by a large factor or alternatively is an intrinsic property of a low-luminosity GRB such as GRB\,980425 and XRF\,060218. The rate of $z>5$ GRBs in IBIS has been modelled~\citep{salvaterra:2007,gorosabel:2004,guetta:2007b,lapi:2008} and is unlikely to be more than 1 or 2 GRBs in 5 years of observations. The long lag is therefore taken to be an intrinsic property of most of the long-lag GRBs indicating their low luminosity and we investigate the consequences.

The median properties of long and
short lag \textit{INTEGRAL} GRBs are given in Table~\ref{table:medians} for the 28 bursts with a measured
value of $\tau_{2+3,1}$. Approximately 40\% of
\textit{INTEGRAL} GRBs with a measured lag belong to the
long-lag category. The median peak flux for long-lag
GRBs is a factor of $\sim5$ lower than for GRBs with short lags. 

\begin{figure}[ht]
\centering
\subfigure{\includegraphics[width=0.9\columnwidth,
height=0.2\textheight]{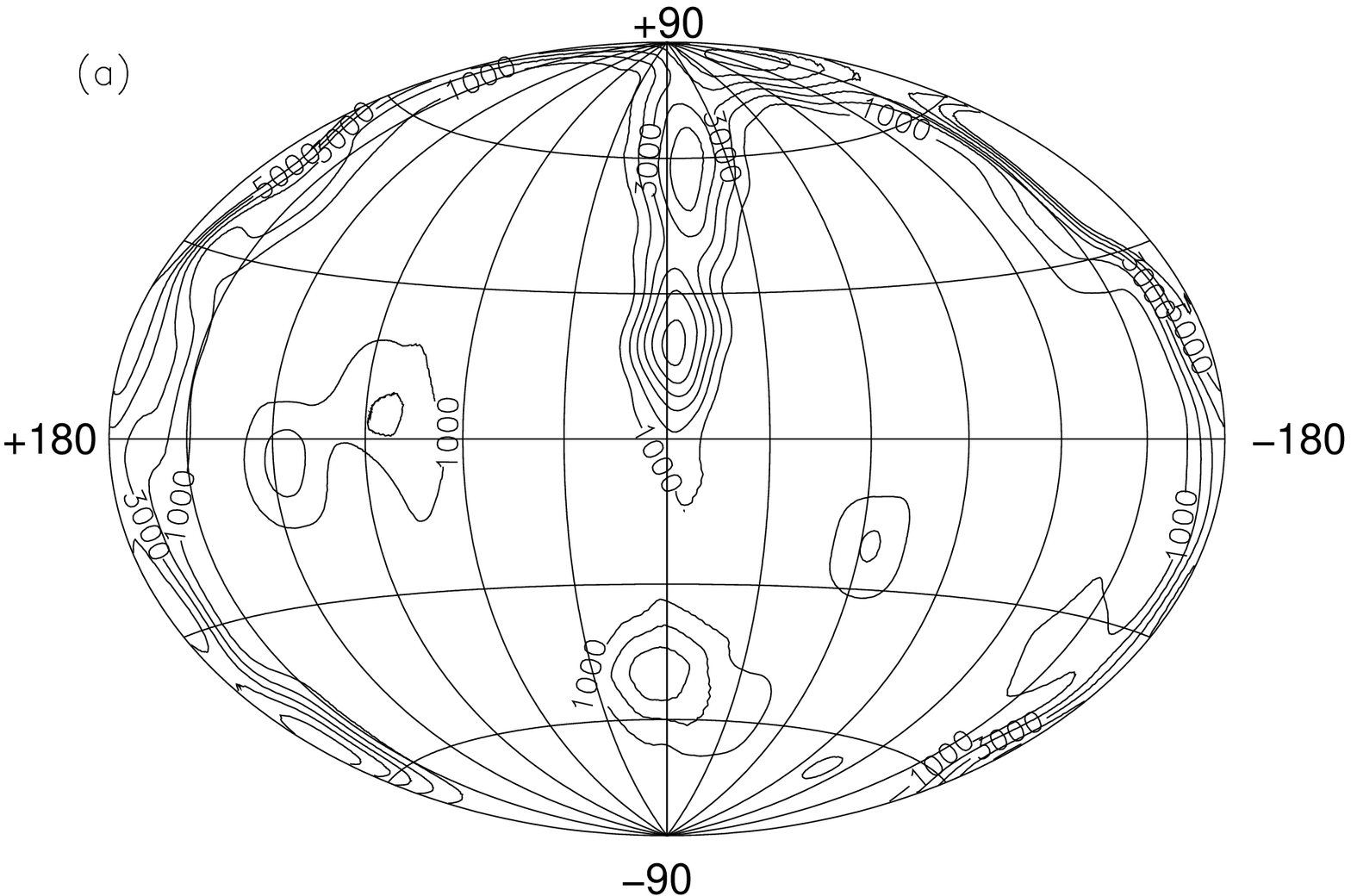}}
\subfigure{\includegraphics[width=0.9\columnwidth,
height=0.2\textheight]{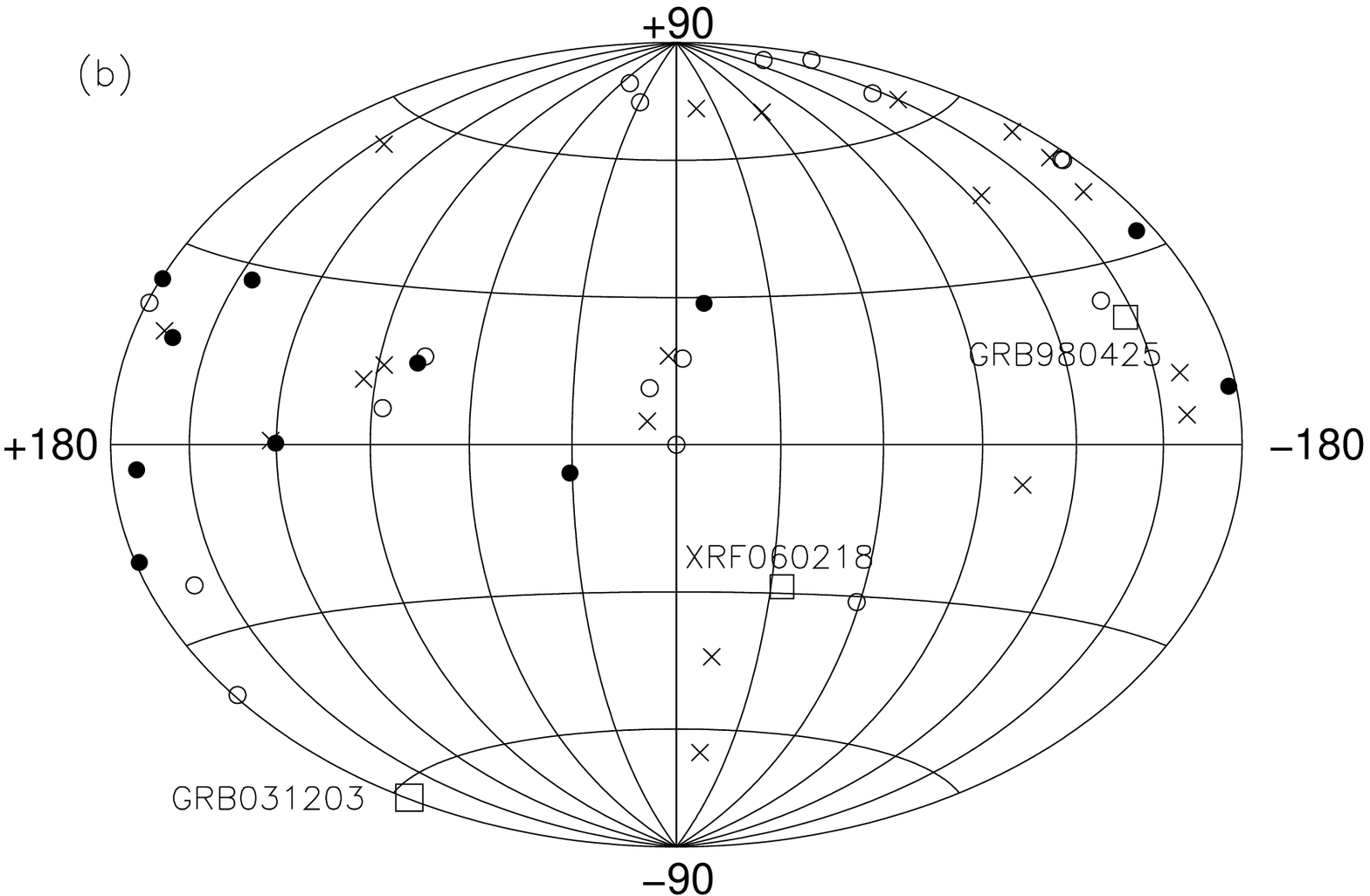}}
\caption{(a) \textit{INTEGRAL} exposure map in supergalactic
  co-ordinates up to July 2007 (contours in units of kiloseconds). (b) The distribution of \textit{INTEGRAL} GRBs
  in supergalactic co-ordinates; the open circles represent
  short-lag GRBs ($\tau_{2+3,1}<0.75$\,s)  and
  filled-in circles those GRBs with long lags
  ($\tau_{2+3,1}>0.75$\,s). The 16 GRBs for which a lag could
  not be determined are denoted by an 'x', as are the XRFs which do
  not have a measured lag between 25--50\,keV and 50--300\,keV, and
  the short burst GRB\,070707. Ten out of the 11 long-lag GRBs are within $\pm30^\circ$
  of the supergalactic plane.}
\label{fig:lags_sg}
\end{figure}

The distribution of the \textit{INTEGRAL} GRBs in supergalactic coordinates is shown in
Fig.~\ref{fig:lags_sg}. All of the \textit{INTEGRAL} GRBs are divided almost equally between the
half of the sky above and below $\pm30^{\circ}$, in agreement with the
exposure map which has $\sim52$\% of the exposure time
within $\pm30^{\circ}$ of the supergalactic plane. However, 10 of the 11 long-lag GRBs are concentrated at supergalactic
latitudes between $\pm30^{\circ}$. The quadrupole moment~\citep{hartmann:1996} has a value of $Q=0.007\pm0.043$ for all \textit{INTEGRAL} GRBs and $Q=-0.225\pm0.090$ for the long-lag GRBs. The quadrupole moment of the 47 bursts is consistent with zero and an isotropic distribution. The non-zero moment of the long-lag bursts indicates an anisotropy in the distribution of these GRBs with respect to the supergalactic plane. The binomial probability that this is a chance occurrence is $7\times10^{-3}$. GRB\,980425 and XRF\,060218 have
  long lags and lie within $\pm30^{\circ}$ of the supergalactic plane,
  while GRB\,031203 has a relatively short lag and lies at a high
  supergalactic latitude. The long-lag GRBs observed with BATSE are also significantly concentrated in the direction $\pm30^{\circ}$ of the supergalactic plane with a quadrupole moment $Q = -0.097\pm 0.038$~\citep{norris:2002}.  The combined results of more than 14 years of observations with IBIS and BATSE lead us to 
conclude that long-lag GRBs trace the features of the nearby large-scale structure of the Universe as revealed with superclusters, galaxy surveys~\citep{lahav:2000,stoughton:2002} and the very high energy cosmic rays~\citep{abraham:2007}.  This result is a further indication that most long-lag GRBs are nearby and have low luminosity.  The local supercluster seems to be appended to a web of filaments and sheets, rather than an isolated pancake structure, with superclusters evident to $\sim$400\,Mpc.  It has been pointed out that weak BATSE GRBs appear to be correlated with galaxies out to distances of $\sim$155\,Mpc with the limit determined by galaxy surveys~\citep{chapman:2007}.

A nearby population of long-lag, low-luminosity GRBs has previously been
proposed based on the detections of  GRB\,980425, GRB\,031203 and
XRF\,060218~\citep[e.g][]{guetta:2004,daigne:2007}. GRB\,980425 has an isotropic-equivalent
$\gamma$-ray energy release of $\sim10^{48}$\,ergs, approximately 3 orders of magnitude lower
than that of ``standard'' GRBs, assuming
its association with SN\,1998bw~\citep{galama:1998} and redshift of $z=0.0085$~\citep{tinney:1998}. 
XRF\,060218, the second closest GRB localised to date,
was detected at a redshift of $z=0.033$~\citep{mirabal:2006}. This burst had an extremely
long duration ($T_{90}\sim2100$\,s) , a very low isotropic energy of $\sim6\times10^{49}$\,ergs and is
classified as an XRF~\citep{campana:2006}. The distance to the long-lag GRBs can also be constrained by association and comparison with the two low luminosity bursts GRB\,980425 ($\tau\sim2.8\,\rm{s}$) and XRF\,060218 ($\tau\sim60\,\rm{s}$).  The distances to GRB\,980425~\citep{galama:1998} and XRF\,060218~\citep{soderberg:2006} are 36\,Mpc and 145\,Mpc, respectively, and these bursts would have been detected in the FCFoV of IBIS to 135\,Mpc and 290\,Mpc.  The association of the long-lag GRBs with the known low luminosity GRBs and with the supergalactic plane implies that they are at similar distances. GRB\,060505 has no SN to faint limits and has a smaller lag of $\sim$0.4\,s and a distance of 404\,Mpc~\citep{fynbo:2006,mcbreen:inprep}. 

 The region marked with a box on the lag-luminosity plot (Fig.~\ref{fig:lag_lum}) contains the long-lag GRBs if they belong to the low luminosity population at an adopted distance of 250 Mpc.  The box contains the prototype low luminosity GRB\,980425 and is bracketed on one side by GRB\,060218 and on the other by GRB\,060505 ($\tau\sim0.4\,\rm{s}$) and GRB\,031203 ($\tau\sim0.17\,\rm{s}$) and at short lag by GRB\,060614 which lies in the region occupied by short GRBs~\citep{gehrels:2006}. With a long lag of
$\sim2.8$\,s, GRB\,980425 qualitatively follows
the lag-luminosity trend but falls significantly below the relation in Fig.~\ref{fig:lag_lum}~\citep{norris:2002},
as do GRB\,031203~\citep{sazonov:2004} and GRB\,060218~\citep{liang:2006}.
The solid line in Fig.~\ref{fig:lag_lum} is the proposed lag-luminosity relationship for low-luminosity GRBs and is parallel to but shifted from the corresponding fit for the long cosmological GRBs by a factor of $\sim10^{3}$.

Both GRB\,980425 and GRB\,031203
violate the Amati correlation~\citep{amati:2007} between isotropic energy, $E_{iso}$,
and peak energy, $E_{peak}$~\citep{ghisellini:2006}.  
GRB\,060218 however is consistent with
the the Amati relation, fuelling the debate that GRB\,980425 and
GRB\,031203 may just be apparent outliers and their intrinsic
properties may be consistent with the relations. Invoking a jet geometry, the low observed luminosities
for these GRBs may be due to such factors as wider jet opening angle,
variations in viewing angle such that the more off-axis a burst is
viewed the lower its luminosity or profiled jets with the Lorentz
factor decreasing off-axis~\citep{yamazaki:2003,ramirez-ruiz:2005}. 

It has also been proposed that GRBs\,980425 and 031203 are intrinsically
sub-energetic events seen on-axis, based on afterglow
observations~\citep{soderberg:2004} and within the constraints of the
internal shock model~\citep{daigne:2007}. Such inherently
sub-luminous events would be permitted by the internal shock model,
where the burst is produced by a mildly relativistic outflow. If
intrinsically low-luminosity, the rate of bursts such as GRB\,980425 may
be far higher than that of ``standard'' cosmological GRBs but
only detectable locally. 

The possibility that low luminosity GRBs could be part of the same population as cosmological GRBs or form a separate sub-energetic population with a much higher rate has been considered~\citep{soderberg:2006,cobb:2006,chapman:2007,guetta:2007,liang:2007,coward:2007,le:2007,virgili:2008}.  The large number of long-lag GRBs detected with IBIS favours the latter conclusion and indicates that low redshift GRBs are dominated by the low luminosity class.

 The collapsar model can account for low-luminosity GRBs if they have low
Lorentz factors and fail to produce highly-relativistic jets due to
baryon loading~\citep{woosley:1999}. It has also been proposed that the nature of the central engine
may be different, e.g., a magnetar progenitor as opposed to a black
hole~\citep[e.g.][]{soderberg:2006,toma:2007}. Some authors suggest that the
$\gamma$-rays are produced in a supernova shock breakout~\citep[e.g.][]{matzner:1999,ghisellini:2007}. A very luminous X-ray outburst was observed from the core collapse SN in NGC\,2770~\citep{soderberg:2008} that may be the shock breakout as recently observed from a red supergiant in a galaxy in the COSMOS field~\citep{schawinski:2008} or an X-ray flash~\citep{xu:2008,li:2008}.

We evaluate the rate of GRBs over the whole sky using the 8 long-lag GRBs in the PCFoV (Table~\ref{table:speclags}) at 50\% coding (0.1\,sr) over an exposure time of 4 years, adopting a distance of 250 Mpc and assuming that 2 of the 8 GRBs are not at low redshift. We obtain $2640\,\rm{Gpc}^{-3}\,\rm{yr}^{-1}$ which is $\sim$25\% of the local rate of Type Ib/c SNe~\citep{soderberg:2006}.  The major uncertainty in this estimate is the distance, where a change of only a factor of 2 increases or decreases this number by 8 to $21,100\,\rm{Gpc}^{-3}\,\rm{yr}^{-1}$ and $330\,\rm{Gpc}^{-3}\,\rm{yr}^{-1}$, respectively. 

The rate of low-luminosity GRBs at the adopted distance of 250 Mpc exceeds the upper limit of 3\% or $<300\,\rm{Gpc}^{-3}\,\rm{yr}^{-1}$ of Type Ib/c SN producing GRBs, which was derived assuming that all low luminosity GRBs would produce a SN and be as radio bright as the SN GRBs~\citep{soderberg:2006}. However, the low luminosity GRB\,060605 has no associated SN to faint limits and is evidence for a quiet end for some massive stars~\citep{fynbo:2006,dado:2006,mcbreen:inprep}.  A GRB may occur without a corresponding SN being observed if the $^{56}$Ni does not have sufficient energy to escape the black hole or if the progenitor star has a low angular momentum.  The association of low luminosity GRBs with the supergalactic plane is not proof that they are associated with clusters of galaxies but indicates that clusters may play a role. It is interesting to note that the rate of Type Ia SNe is higher in elliptical galaxies in clusters than in field ellipticals by a factor of $\sim$3~\citep{mannucci:2007}.  This effect is due to galaxy-galaxy interactions in clusters~\citep{boselli:2006} either producing a small amount of young stars or affecting the evolution and properties of binary systems. In the latter case, there should also be an increase in the merger rate of white dwarfs or a white dwarf with a neutron star or black hole.  A merger involving a white dwarf~\citep{fryer:1999,middleditch:2004,levan:2006,king:2007} should produce a long GRB that is likely to be fainter than the formation of a black hole in cosmological GRBs.  There will be no supernova in the merger of a white dwarf with a neutron star or black hole, and probably a faint afterglow.  In addition, the merger could take place in the intercluster region without a host galaxy if the binary is ripped from its host in the merger interaction involving the cluster galaxies~\citep{niino:2008}. 

\section{Conclusions\label{conclusion}}

\textit{INTEGRAL} observations of gamma-ray bursts have
yielded many interesting results and offer significant insight
into the prompt $\gamma$-ray emission.   IBAS successfully
provides accurate, fast localisations of GRBs to the community
at a rate of $\sim0.8$ GRBs per month, enabling
multi-wavelength afterglow observations to be carried out by
other space-based missions and ground-based
telescopes. 

We have presented the spectral, spatial, and temporal lag analysis of
the 47 \textit{INTEGRAL} GRBs up to July 2007. Most weak
\textit{INTEGRAL} GRBs are well fit with a power law,
while bright GRBs can be fit by both the Band and quasithermal
models. Approximately 40\% of the GRBs for which a lag was measured have long
spectral lags ($>0.75$\,s between 25--50 and 50--300\,keV). The long-lag GRBs are characterised by low peak flux,
long slow pulses and are concentrated towards the supergalactic
plane, reflecting the nearby large-scale structure of the Universe.

\textit{INTEGRAL} therefore detects a large proportion of
faint, long-lag GRBs that are inferred to be local. The sensitivity of IBIS is such that it can detect very faint
GRBs, allowing \textit{INTEGRAL} to probe the population of
low-luminosity GRBs with long lags. This population appears to be
distinct from that of high-luminosity GRBs and dominates the
local GRB population.

\begin{acknowledgements}

The authors thank Darach Watson and Rob Preece for useful discussions and comments, Andreas von Kienlin for providing the GRB\,060901 data and Erik Kuulkers for the \textit{INTEGRAL} exposure map data. SMB acknowledges the support of the European Union through a Marie Curie Intra-European Fellowship within the Sixth Framework Program. 

\end{acknowledgements}


\clearpage

\begin{table*}[h]
\begin{center}
\begin{small}
\label{table:grbs}
\caption{GRBs detected by
  \textit{INTEGRAL}. The columns refer to (from left to
  right): GRB; right ascension; declination; $T_{90}$ duration; peak
  flux; fluence; photon index; $\chi^2$/degrees of freedom
  (\textit{d.o.f.}); afterglow detections in radio,
  R, infrared, IR, optical, O and X-ray, X. Peak fluxes, fluences and
  photon indices are given in the 20--200\,keV energy range.} 
\begin{minipage}{\textwidth}
\begin{tabular}[l]{@{}l l r r c c c c c c@{}}
\hline\hline GRB & RA & DEC & $T_{90}$  &
\multicolumn{2}{c}{Peak Flux} & Fluence & Photon Index &
$\chi^{2}$ &
Afterglow \\ 
& & & \textit{s} & $ph\,cm^{-2}\,s^{-1}$ &
$erg\,cm^{-2}\,s^{-1}$ & $erg\,cm^{-2}$ & & /~\textit{d.o.f.} &\\ 
\hline
GRB\,021125\,\footnote{\citet{malaguti:2003},
  $^{b}$\,\citet{mereghetti:2003a}, $^{c}$\,\citet{gotz:2003},
  $^{d}$\,\citet{mereghetti:2003b}, $^{e}$\,\citet{vonkienlin:2003b},
  $^{f}$\,\citet{beckmann:2003}, $^{g}$\,\citet{sazonov:2004},
  $^{h}$\,\citet{moran:2005},
  $^{i}$\,\citet{mcglynn:2005}, $^{j}$\,\citet{mereghetti:2005},
  $^{k}$\,\citet{filliatre:2005}, $^{l}$\,\citet{filliatre:2006},
  $^{m}$\,\citet{mcbreen:2006}, $^{n}$\,\citet{grebenev:2007},
  $^{o}$\,A. von Kienlin (priv. comm.)

$^\ast$~25--500\,keV

$^\star$~GRBs with known redshift; GRB\,031203 at z=0.106~\citep{prochaska:2004}; GRB\,050223 at z=0.584~\citep{pellizza:2006}; GRB\,050502a
 at z=3.793~\citep{prochaska:2005}; GRB\,050525a~\citep{foley:2005} at z=0.606.

$^\dagger$~Data unavailable for GRB\,060930 (values taken
  from~\citet{gotz:2006}) and GRB\,070311 (values taken from~\citet{mereghetti:2007}).} & 19:47:57 & 28:23:35 & $24$ & 22 &
--- & $4.8\times10^{-5}$\large{$^\ast$} & $-2.2$ & --- & ---  \\  
GRB\,021219\,$^{b}$ & 18:50:27 & 31:57:17 & 5.5 & 3.7 &
$3.5\times10^{-7}$ & $9.0\times10^{-7}$ & $-2.00\pm0.10$ & --- & R \\  
GRB\,030131\,$^{c}$ & 13:28:21 & 30:40:43 & 124 &
1.9 & $1.7\times10^{-7}$ & $7.0\times10^{-6}$ &
Table~\ref{table:band} & --- & O  \\
GRB\,030227\,$^{d}$ & 04:57:34 & 20:28:16 & 33 & 1.1 &
$9.6\times10^{-8}$ & $7.5\times10^{-7}$ & $-1.85\pm0.20$ & --- & X,O \\ 
GRB\,030320\,$^{e}$ & 17:51:36 & -25:18:52 & 48 & 5.7 &
$5.4\times10^{-7}$ & $1.1\times10^{-5}$ & $-1.69\pm0.08$ & --- & ---  \\ 
GRB\,030501\,$^{f}$ & 19:05:33 & 06:15:57 & 40 & 2.7 &
$2.5\times10^{-7}$ & $3.0\times10^{-6}$ & $-1.75\pm0.10$ & --- & ---  \\ 
GRB\,030529 & 09:40:30 & -56:20:31 & 20 & 0.4 &
$3.7\times10^{-8}$ & $4.0\times10^{-7}$ & $-2.07\pm0.30$ & 2.9/5 & --- \\ 
GRB\,031203\,$^{g}$\,$^{\star}$ & 08:02:32 & -39:50:47 & 39 &
2.6 & $2.5\times10^{-7}$ & $2.0\times10^{-6}$ & $-1.63\pm0.06$
& --- & X,O,R  \\ 
GRB\,040106\,$^{h}$ & 11:52:18 &
-46:47:15 & 47 & 0.6 & $6.5\times10^{-8}$ & $8.2\times10^{-7}$
& $-1.72\pm0.15$ & --- & X,O,R \\ 
GRB\,040223\,$^{i}$ & 16:39:31 & -41:55:47 &
258 & 0.2 & $1.6\times10^{-8}$ & $4.4\times10^{-7}$ &
$-2.30\pm0.19$ &20.1/20 & X \\ 
GRB\,040323 & 13:53:49 & -52:20:45 & 14 &
1.5 & $2.0\times10^{-7}$ & $3.0\times10^{-6}$ & $-1.44\pm0.18$
&10.6/8 & O \\ 
GRB\,040403\,$^{j}$ & 07:40:54 & 68:12:55 & 21 & 0.5 &
$4.3\times10^{-8}$ & $5.0\times10^{-7}$ & $-1.90\pm0.15$ & --- & --- \\ 
GRB\,040422\,$^{k}$ & 18:42:01 & 01:59:04 & 4 & 2.3 &
$1.8\times10^{-7}$ & $3.4\times10^{-7}$ &
Table~\ref{table:band} & --- & IR \\
GRB\,040624\,$^{l}$ & 13:00:08 & -03:34:08 & 62 & 0.7 & $4.0\times10^{-8}$
& $7.0\times10^{-6}$ & $-2.11\pm0.20$ &28.1/33 & --- \\ 
GRB\,040730 &
15:53:14 & -56:28:15 & 51 & 0.32 & $3.4\times10^{-8}$ &
$6.6\times10^{-7}$ & $-1.44\pm0.14$ & 34.3/33& --- \\ 
GRB\,040812 &
16:26:05 & -44:42:32 & 16 & 0.7 & $5.0\times10^{-8}$ &
$2.3\times10^{-7}$ & $-2.34\pm0.29$ & 19.5/26& X,R \\ 
GRB\,040827 &
15:17:00 & -16:08:21 & 35 & 0.42 & $6.0\times10^{-8}$ &
$1.2\times10^{-6}$ & $-1.83\pm0.18$ & 32.5/29& X,O \\ 
XRF\,040903 &
18:03:22 & -25:15:23 & 15 & 0.23 & $2.0\times10^{-8}$ &
$1.4\times10^{-7}$ & $-2.94\pm0.44$ & 16.6/20& --- \\ 
GRB\,041015 &
00:18:37 & 66:51:37 & 33 & 0.17 & $4.0\times10^{-8}$ &
$6.0\times10^{-7}$ & $-0.95\pm0.28$ & 17.4/21 & --- \\ 
GRB\,041218 &
01:39:06 & 71:20:05 & 40 & 3.19 & $2.8\times10^{-7}$ &
$5.5\times10^{-6}$ & Table~\ref{table:band} & --- & O \\
GRB\,041219a\,$^{m}$ & 00:24:26 & 62:50:06 & 186 & 33 &
$3.6\times10^{-6}$ & $1.6\times10^{-4}$ &
Table~\ref{table:band} & --- & O,R \\
GRB\,050129 & 16:51:12 & -03:04:44 & 40 & 0.36 &
$3.0\times10^{-8}$ & $4.5\times10^{-7}$ & $-1.91\pm0.31$ & 21.6/26& ---
\\ 
GRB\,050223$^{\star}$ & 18:05:36 & -62:28:26 & 39 & 0.67 &
$6.0\times10^{-8}$ & $8.2\times10^{-7}$ & $-1.84\pm0.25$ &21.3/30 & X  \\
GRB\,050502a$^{\star}$ & 13:29:45 & 42:40:27 & 20 & 1.58
& $2.0\times10^{-7}$ & $>1.4\times10^{-6}$ &
Table~\ref{table:band} & --- & O,IR \\
GRB\,050504 & 13:24:00 & 40:41:45 & 58 & 0.45 &
$7.4\times10^{-8}$ & $1.3\times10^{-6}$ & $-1.32\pm0.10$ & 50.8/29& X \\
GRB\,050520 & 12:50:03 & 30:27:02 & 80 & 0.53 & $4.0\times10^{-8}$
& $2.4\times10^{-6}$ & $-1.61\pm0.07$ & 26.1/30 & X,R \\
XRF\,050522 & 13:20:35 & 24:47:30 & 17 & 0.2 &
$1.3\times10^{-8}$ & $7.0\times10^{-8}$ & $-2.72\pm0.67$ &16.2/17 & --- \\ 
GRB\,050525a$^{\star}$ & 18:32:33 & 26:20:23 & 12 &
31.7 &
$3.1\times10^{-6}$ & $2.0\times10^{-5}$ &
Table~\ref{table:band} & --- & O,R \\
GRB\,050626 & 12:26:58 & -63:08:03 & 50 & 0.27 &
$2.0\times10^{-8}$ & $7.6\times10^{-7}$ & $-2.20\pm0.14$ & 27.6/30& --- \\  
GRB\,050714a & 02:54:21 & 69:07:34 & 34 & 0.18 &
$1.5\times10^{-8}$ & $4.7\times10^{-7}$ & $-2.06\pm0.20$ & 21.8/28 & --- \\ 
GRB\,050918 & 17:50:26 & -25:24:51 & 115 & 1.61 & $1.6\times10^{-7}$
& $3.1\times10^{-6}$ & Table~\ref{table:band} & --- & --- \\ 
GRB\,050922a & 18:04:37 &
-32:01:24 & 8 & 0.01 & $5.3\times10^{-9}$ & $5.3\times10^{-8}$ & $-1.85\pm1.01$ & 0.6/2 & --- \\ 
GRB\,051105b &
00:37:51 & -40:28:52 & 16 & 0.31 & $3.0\times10^{-8}$ &
$2.9\times10^{-7}$ & $-1.83\pm0.23$ &25.7/27 & --- \\ 
GRB\,051211b &
23:02:45 & 55:04:44 & 60 & 0.6 & $6.1\times10^{-8}$ &
$2.0\times10^{-6}$ & $-1.62\pm0.10$ & 38.4/28 & X,R \\ 
GRB\,060114 & 13:01:07 & -04:44:53 & 63 & 0.31 &
$3.0\times10^{-8}$
& $9.0\times10^{-7}$ & $-0.98\pm0.16$ & 18.9/30 & --- \\ 
GRB\,060130 &
15:16:54 & -36:54:43 & 29 & 0.11 & $7.7\times10^{-9}$ &
$2.8\times10^{-7}$ & $-1.51\pm0.40$ & 27.7/22 & --- \\ 
GRB\,060204a &
15:28:56 & -39:26:38 & 34 & 0.16 & $2.1\times10^{-8}$ &
$6.6\times10^{-7}$ & $-1.43\pm0.27$ &20.3/26 & X \\ 
GRB\,060428c\,$^{n}$ & 19:00:52 & -09:33:00 & 12 & 3.9 & $3.6\times10^{-7}$ & $2.3\times10^{-6}$ & Table~\ref{table:band} & --- & --- \\
GRB\,060901\,$^{o}$ &
19:08:38 & -06:38:22 & 20 & 8.6 & $9.0\times10^{-7}$ &
$8.7\times10^{-6}$ & Table~\ref{table:band} & --- & X,O \\ 
GRB\,060912b &
18:04:52 & -19:52:50 & 91 & 0.08 & $8.8\times10^{-9}$ &
$7.1\times10^{-7}$ & $-1.52\pm0.23$ & 18.5/33 & --- \\ 
GRB\,060930$^\dagger$ & 20:18:09 & -23:37:31 & 20 & 0.3 &
$2.2\times10^{-8}$ & $2.5\times10^{-7}$ & --- & --- 
& --- \\ 
GRB\,061025 & 20:03:39 & -48:14:39 & 11 & 1.14 &
$1.3\times10^{-7}$ & $1.1\times10^{-6}$ &
Table~\ref{table:band} & --- & X,O \\ 
GRB\,061122 & 20:15:21 &
15:30:51 & 11 & 31.7 & $3.1\times10^{-6}$ & $2.0\times10^{-5}$ &
Table~\ref{table:band} & --- & X,O \\ 
GRB\,070309 & 17:34:44 & -37:56:40 & 38 & 0.13 & $1.1\times10^{-8}$ & $5.4\times10^{-7}$ &
$-1.73\pm0.27$ & 26.5/26 & X? \\
GRB\,070311$^\dagger$ & 05:50:10 &  03:22:29 & $50$ & $0.9$ & --- &
$2.0\times10^{-6}$ & --- & --- & X,O \\
GRB\,070615 & 02:57:14 & -04:24:28 & 27 & 0.41 & $2.8\times10^{-8}$ & $5.0\times10^{-7}$ &
$-1.62\pm0.28$ & 21.9/27 & --- \\
GRB\,070707 & 17:51:00 & -68:52:52 & 0.8 & 1.72 & $2.1\times10^{-7}$ & $2.1\times10^{-7}$ &
$-1.19\pm0.14$ & 51.2/47 & X,O \\
\end{tabular}
\end{minipage}
\end{small}
\end{center}
\end{table*}

\begin{table*}[h]
\begin{center}
\caption{IBIS and SPI spectral properties of GRBs for which Band
  and quasithermal (PL + BB) models are fit to
spectra. Parameters quoted are low energy power-law index,
$\alpha$, high energy power-law index, $\beta$, and break
energy, $E_{0}$, for Band model fits and temperature, \textit{kT} and power-law index,
$\Gamma$ for quasithermal model fits.}
\label{table:band}
\begin{tabular}{@{}l c | c c c r | c c r @{}}
\hline\hline
& & \multicolumn{4}{c}{Band Model} & \multicolumn{3}{|c}{Quasithermal Model} \\ 
\hline
GRB & & $\alpha$ & $\beta$ & $E_{0}$\,(keV) &
  $\chi^{2}$~/\textit{d.o.f} & \textit{kT} & $\Gamma$ & $\chi^{2}$~/\textit{d.o.f}  \\
\hline
 & & & & & & & & \\ 
GRB\,030131 & \textit{IBIS} & $-1.40\pm0.20$ & $-3.0\pm1.0$
& $70\pm20$ & 22.5/16 & --- & --- & --- \\ 
GRB\,040422 & \textit{IBIS} & $-1.26\pm0.08$ &
$-3$\,{\large{$^\dagger$}} & $55\pm31$ & 42.0/35 & $14\pm2$ & $-2.49\pm0.48$ & 38.4/34\\
& \textit{SPI} & \multicolumn{3}{c}{$-2.17\pm0.28$ (PL)} & 7.5/3 &
--- & --- & --- \\ 
GRB\,041218 & \textit{IBIS} &
$-1.15\pm0.20$ & $-2.08\pm0.16$ & $80\pm40$ & 64.4/35 & $14\pm2$ &
$-1.76\pm0.08$ & 64.1/35 \\
& \textit{SPI} & $-0.97\pm0.47$\footnotemark[1] & $-3$\,{\large{$^\dagger$}} &
$116\pm152$ & 20.0/35 & $26\pm9$ & $-1.82\pm0.38$ & 19.4/34 \\ 
GRB\,041219a & \textit{IBIS} & $-1.67$\,{\large{$^\dagger$}}&
$-1.92\pm0.02$ & $114\pm144$ & 184.0/157 & $30\pm12$ & $-2.03\pm0.07$ &
109.6/87 \\
& \textit{SPI} & $-1.48\pm0.08$ & $-1.92\pm0.13$ & $365.9\pm192$ &
51.9/30 & $29\pm3$ & $-1.77\pm0.03$ & 17.6/18 \\
GRB\,050502a & \textit{IBIS}
& $-0.89\pm0.25$ & $-3$\,{\large{$^\dagger$}} & $83\pm44$ & 34.7/38 &
$17\pm2$ & $-1.64\pm0.11$ & 33.2/41 \\
GRB\,050525a & \textit{IBIS} & $-0.94\pm0.45$ & 
$-3$\,{\large{$^\dagger$}} & $65\pm62$ & 59.2/31 & $18\pm4$ & $-2.05\pm0.58$ &60.1/31\\
& \textit{SPI} & $-1.37\pm0.23$ & $-3$\,{\large{$^\dagger$}} & $127\pm45$ & 46.9/51
& $26\pm4$ & $-2.24\pm0.15$ & 46.0/50 \\ 
GRB\,050918 & \textit{IBIS} & $-1.42\pm0.51$ & $-3$\,{\large{$^\dagger$}} &
$117\pm1484$ & 26.4/29 & $21\pm5$ & $-2.44\pm1.22$ & 23.8/28 \\
& \textit{SPI} & $-1.28\pm2.58$ & $-3$\,{\large{$^\dagger$}} & $69\pm890$ & 4.4/7 & $15\pm11$ & $-2.24\pm3.22$ & 3.8/6 \\ 
GRB\,060428c & \textit{IBIS/SPI} & $-0.71\pm0.26$ & $-2.0$ & $54\pm14$ & 32.3/33 & --- & --- & --- \\
GRB\,060901 & \textit{IBIS} & $-1.47\pm0.20$ & $-3$\,{\large{$^\dagger$}}
& $998\pm9002$ & 40.2/30 & $36\pm11$ & $-1.91\pm0.39$ & 34.8/29 \\
& \textit{SPI} & $-0.50\pm0.61$ & $-1.89\pm0.42$ & $85\pm151$ & 9.0/14
& $28\pm11$ & $-1.54\pm0.15$ & 14.0/14 \\ 
GRB\,061025 & \textit{IBIS} &
$-0.85\pm0.43$ & $-3$\,{\large{$^\dagger$}} & $94\pm134$ & 38.3/29 &
$15\pm6$ & $-1.38\pm0.30$ & 35.6/28 \\
& \textit{SPI} & \multicolumn{3}{c}{$-1.57\pm0.25$ (PL)} & 10.4/6
& $17\pm6$ & $-0.75\pm2.65$ & 7.2/5 \\
GRB\,061122 & \textit{IBIS} & $-0.97\pm0.40$ & $-2.00\pm0.12$ &
$56\pm52$ & 47.4/30 & $12\pm3$ & $-1.72\pm0.12$ & 48.6/30 \\ 
& \textit{SPI} & $-0.98\pm0.12$ & $-2.72\pm0.85$ & $166\pm39$ & 66.6/56
& $36\pm3$ & $-1.81\pm0.08$ & 67.8/56 \\
& & & & & & & & \\ 
\hline
$\dagger$ Parameter frozen \\
\end{tabular}
\end{center}
\end{table*}

\renewcommand{\thefootnote}{\alph{footnote}}

\begin{table*}[h]
\begin{minipage}{\textwidth}
\begin{center}
\begin{small}
\label{table:speclags}
\caption{Spectral Lag Measurements for \textit{INTEGRAL}
  GRBs. The columns refer to (from left to right): GRB; supergalactic
  longitude (SGL) and latitude (SGB); burst interval used in
  spectral lag determination relative to trigger time and  as marked on time profiles in
  the Appendix; spectral lags measured between Channel 2
  (50--100\,keV) and Channel 1 (25--50\,keV), $\tau_{2,1}$,
  Channels  2 and 3 combined
  (50--300\,keV) and Channel 1 (25--50\,keV), $\tau_{2+3,1}$ and
  Channel 3 (100--300\,keV) and
  Channel 1 (25--50\,keV), $\tau_{3,1}$.}
\begin{tabular}[c]{@{}l r r c c c c@{}}
\hline\hline
GRB & \textit{SGL} & \textit{SGB} & Lag Interval & \multicolumn{3}{c}{Spectral Lag} \\ 
& & & & $\tau_{2,1}$ (s) & $\tau_{2+3,1}$ (s) & $\tau_{3,1}$ (s) \\
\hline
GRB\,021219 & 29.91 & 73.72 & 29.5--33\,s & $0.33\pm0.02$ & $0.39\pm0.01$ &
$0.63\pm0.02$ \\
GRB\,030320 & $-174.44$ & 46.17 & $-5-50$\,s & $0.08\pm0.03$ & $0.15\pm0.03$ &
$0.28\pm0.03$ \\ 
& & &  5--20\,s & $0.33\pm0.03$ & $0.33\pm0.03$ & $0.28\pm0.03$ \\ 
& & & 37--49\,s & $-0.05\pm0.03$ & $0.05\pm0.03$ &
$0.28^{+0.05}_{-0.03}$  \\
GRB\,030501 & $-137.97$ & 80.21 &  0--30\,s & $-0.30\pm0.10$ & $-0.25\pm0.10$ & --- \\
GRB\,030529\,{\large{$^\dagger$}} & 178.44 & $-38.74$ & $-5$--16\,s & $0.30\pm0.30$ &
$0.40^{+0.20}_{-0.25}$  & --- \\
GRB\,031203 & 178.37 & $-61.73$ & $-5$--20\,s & $-0.05^{+0.04}_{-0.05}$ &
$0.04\pm0.03$ & $0.17^{+0.03}_{-0.04}$ \\
GRB\,040106 & 161.11 & $-22.05$ & $-2$--53\,s & $0.00^{+0.10}_{-0.05}$ &
$0.05\pm0.05$ & --- \\
& & & $-2$--10\,s & $0.15^{+0.05}_{-0.10}$ & $0.05^{+0.10}_{-0.05}$ &
--- \\
& & & 37--53\,s & $-0.35\pm0.10$ & $-0.30\pm0.05$ & --- \\ 
GRB\,040223\,\footnote{Long-lag GRB in PCFoV to 50\% coding. $^{b}$ Long-lag GRB in PCFoV outside 50\% coding. $^{c}$ Long-lag GRB in FCFoV.

$\dagger$ Lag measured using wavelet-smoothed lightcurves.

$\ast$ Short-duration GRB.}\,{\large{$^\dagger$}} & 179.35 & 24.44 & $-15$--15\,s & $1.05^{+0.30}_{-0.35}$ &
$1.80^{+0.35}_{-0.25}$ & --- \\
GRB\,040323\,$^{a}$ & 170.14 & $-3.73$ & 0--20\,s & $0.80^{+0.05}_{-0.10}$ &
$1.15^{+0.15}_{-0.05}$ & $1.80^{+0.10}_{-0.20}$  \\
GRB\,040403\,$^{a}$\,{\large{$^\dagger$}} & 30.70 & $-5.66$ & $-10$--10\,s & $0.60\pm0.20$
& $0.95^{+0.25}_{-0.15}$ & --- \\
GRB\,040422 & $-167.01$ & 75.91 & 0--10\,s & $0.01\pm0.01$ & $0.01\pm0.01$ & ---  \\ 
& & & 2.5--4\,s & $-0.01\pm0.01$ & $-0.01\pm0.01$ & ---   \\
& & & 4--8\,s & $0.04\pm0.01$ & $0.02^{+0.02}_{-0.02}$ & --- \\
GRB\,040624\,$^{b}$\,{\large{$^\dagger$}} & 120.23 & 0.26 & $-5$--50\,s & --- & $2.75\pm0.20$ & --- \\
GRB\,040730\,$^{c}$\,{\large{$^\dagger$}} & $-177.00$ & 8.47 & 15--40\,s  & $2.80\pm0.50$
& $3.05^{+0.40}_{-0.45}$ & $3.30^{+0.45}_{-0.35}$ \\
GRB\,040812 & 179.07 & 20.75 & 4--12\,s & $-0.05^{+0.05}_{-0.10}$ & $-0.15\pm0.10$ &
--- \\
& & & 4--7.7\,s & $0.00\pm0.05$ & $0.00\pm^{+0.10}_{-0.15}$ & ---  \\
& & & 7.7--12\,s & $-0.20^{+0.15}_{-0.10}$ & $-0.30^{+0.15}_{-0.10}$ & --- \\
GRB\,040827\,$^{b}$\,{\large{$^\dagger$}} & 144.61  & 27.49 & 5--30\,s & $3.15^{+0.20}_{-0.30}$
& $4.85^{+0.15}_{-0.30}$ & --- \\
GRB\,040903\,{\large{$^\dagger$}} & $-170.84$ & 47.15 & 0--20\,s &
\multicolumn{3}{c}{$\tau_{1,15-25}=0.10^{+0.25}_{-0.15}$} \\
GRB\,041218 & 7.76 & 11.40 & 20--63\,s & $0.05^{+0.02}_{-0.01}$ &
$0.04^{+0.01}_{-0.03}$ & $-0.03^{+0.04}_{-0.02}$ \\ 
& & & 20--29\,s & $0.24^{+0.06}_{-0.04}$ & $0.15^{+0.05}_{-0.06}$ &
--- \\ 
& & & 31--40\,s & $-0.01^{+0.02}_{-0.03}$ &
$-0.02\pm0.02$ & $-0.01^{+0.04}_{-0.02}$ \\ 
& & & 50--63\,s & $-0.02\pm0.03$ & $-0.06^{+0.04}_{-0.02}$ & --- \\  
GRB\,041219a & $-1.86$ & 17.42 & 261--414\,s & $0.12\pm0.02$ &
$0.15\pm0.02$ & $0.21\pm0.02$ \\
& & & $-1$--7\,s & $0.50^{+0.50}_{-0.08}$ &
$0.65^{+0.05}_{-0.06}$ & $1.00^{+0.10}_{-0.09}$  \\ 
& & & 261--327\,s & $0.06\pm0.02$ & $0.08\pm0.02$ & $0.14\pm0.02$ \\ 
& & & 356--414\,s & $0.10\pm0.02$ & $0.13\pm0.03$ &
$0.18\pm0.03$ \\ 
GRB\,050502a  & 75.87 & 16.88 & $-16.5$--$\,-3.5$\,s & $-0.12^{+0.05}_{-0.08}$ &
$-0.03\pm0.06$ & $0.11^{+0.07}_{-0.06}$ \\ 
GRB\,050504\,$^{c}$ & 77.79 & 15.58 & $-5$--60\,s & $1.35^{+0.30}_{-0.20}$ &
$2.40^{+0.35}_{-0.25}$  & $4.35^{+1.45}_{-0.70}$ \\
GRB\,050520 & 86.68 & 6.85 & 0--80\,s & $0.07\pm0.02$ & $0.07\pm0.02$ & --- \\
GRB\,050522\,{\large{$^\dagger$}} & 93.81 & 12.17 & 0--20\,s &
\multicolumn{3}{c}{$\tau_{1,15-25}=0.70^{+0.15}_{-0.25}$} \\
& & & 2.5--6\,s &
\multicolumn{3}{c}{$\tau_{1,15-25}=0.80^{+0.75}_{-0.50}$} \\
& & & 7--16\,s &
  \multicolumn{3}{c}{$\tau_{1,15-25}=0.70^{+0.15}_{-0.20}$} \\
GRB\,050525a & 51.74 & 78.15 & 0--12\,s & $0.056\pm0.001$ & $0.069\pm0.001$ &
$0.130^{+0.003}_{-0.002}$ \\ 
& & & 0--4\,s & $0.053\pm0.001$ & $0.068\pm0.001$ & $0.131^{+0.003}_{-0.002}$
 \\ 
& & & 4--12\,s & $0.066^{+0.003}_{-0.002}$ &
$0.068\pm0.002$ & $0.096^{+0.003}_{-0.005}$ \\
GRB\,050626\,$^{c}$\,{\large{$^\dagger$}} & 178.46 & $-17.14$ & 50--110\,s &
$3.35^{+0.45}_{-0.40}$ & $3.25^{+0.35}_{-0.45}$ & --- \\
GRB\,050918 & $-174.74$ & 45.97 & 0--325\,s & $0.17^{+0.04}_{-0.03}$ & $0.17^{+0.03}_{-0.02}$
& --- \\
& & & 0--50\,s & $0.43^{+0.04}_{-0.05}$ &
$0.50^{+0.05}_{-0.04}$ & --- \\
& & & 220--350\,s & $-0.06\pm0.01$ & $-0.14\pm0.01$ & --- \\
GRB\,051211b\,$^{b}$ & $-8.74$ & 28.77 & 10--80\,s & $1.95\pm{0.35}$ &
$1.70^{+0.30}_{-0.35}$ & $1.55^{+0.65}_{-0.35}$ \\
GRB\,060130\,$^{c}$\,{\large{$^\dagger$}} & 163.67 & 16.50 & $-10$--45\,s &
$4.05^{+0.15}_{-0.30}$ & $3.45^{+0.35}_{-0.45}$ & --- \\
& & & & & & \\
GRB\,060901 & $-144.72$ & 67.40 & 0--4\,s & $0.2\pm0.1$ & $0.2\pm0.1$ & $0.2\pm0.1$ \\
GRB\,061025 & $-140.43$ & 24.19 & $-10$--10\,s & $0.35^{+0.10}_{-0.05}$ &
$0.40^{+0.05}_{-0.10}$ & $0.40\pm0.05$ \\
GRB\,070309\,$^{a}$\,{\large{$^\dagger$}} & $-172.41$ & 33.13 & $-25$--10\,s & --- &
$1.70^{+0.35}_{-0.30}$ & --- \\
GRB\,070615\,{\large{$^\dagger$}} & $-59.34$ & $-30.58$ & 0--20\,s & $0.70^{+0.25}_{-0.15}$ &
$0.40^{+0.15}_{-0.20}$ & --- \\
GRB\,070707\,{\large{$^\ast$}} & $-159.26$ & 4.64 & 0.4--1.5\,s & $0.025\pm0.005$ &
$0.005\pm0.005$ & $0.020\pm0.005$ \\ 
\end{tabular}
\end{small}
\end{center}
\end{minipage}
\end{table*}

\renewcommand{\thefootnote}{\arabic{footnote}}

\begin{table*}
\begin{center}
\caption{Wavelet-denoised spectral lags of weak GRBs. The lags derived
from raw data are shown for comparison where available. Spectral lags
measured between Channels  2 and 3 combined
  (50--300\,keV) and Channel 1 (25--50\,keV), $\tau_{2+3,1}$, with the
exception of the XRFs for which the lag is measured between 15--25\,keV
and 25--50\,keV (Channel 1), $\tau_{1,15-25}$.}
\label{table:wavelet}
\begin{tabular}{@{}c c c @{}}
\hline\hline
GRB & $\tau_{2+3,1}$ (raw data) (s)  & $\tau_{2+3,1}$ (denoised data) (s) \\
\hline
& & \\
GRB\,030529 & --- & $0.40^{+0.20}_{-0.25}$ \\ 
GRB\,040223 & $1.70^{+0.30}_{-0.25}$ & $1.80^{+0.35}_{-0.25}$ \\
GRB\,040403 & $1.05^{+0.55}_{-0.30}$ & $0.95^{+0.25}_{-0.15}$ \\
GRB\,040624 & $3.10^{+0.90}_{-0.30}$ & $2.75\pm0.20$ \\
GRB\,040730 & $3.35^{+0.80}_{-0.55}$ & $3.05^{+0.40}_{-0.45}$ \\
GRB\,040827 & $3.65\pm0.30$ & $4.85^{+0.15}_{-0.30}$ \\
XRF\,040903 & --- & $\tau_{1,15-25}=0.10^{+0.25}_{-0.15}$ \\
XRF\,050522 & $\tau_{1,15-25}=1.15^{+0.25}_{-0.45}$ & $\tau_{1,15-25}=0.70^{+0.15}_{-0.25}$ \\
GRB\,050626 & $3.25^{+1.25}_{-0.65}$ & $3.25^{+0.35}_{-0.45}$ \\
GRB\,060130 & $3.65^{+1.00}_{-0.75}$ & $3.45^{+0.35}_{-0.45}$ \\
GRB\,070309 & --- & $1.70^{+0.35}_{-0.30}$\\
GRB\,070615 & $0.40^{+0.15}_{-0.25}$ & $0.40^{+0.15}_{-0.20}$ \\
& & \\
\hline
\end{tabular}
\end{center}
\end{table*}

\begin{table*}
\begin{center}
\caption{Median properties of the 28 long-duration GRBs with a measured
  $\tau_{2+3,1}$, categorised into those with short lags
  ($\tau_{2+3,1}<0.75\,\rm{s}$) and long lags
  ($\tau_{2+3,1}>0.75\,\rm{s}$.)}
\label{table:medians}
\begin{tabular}{@{}c c c @{}}
\hline\hline
 & Short-lag GRBs & long-lag GRBs \\
\hline
Fraction of \textit{INTEGRAL} GRBs & 17/28 & 11/28 \\
Median $\tau_{2+3,1}\,\rm{(s)}$ & 0.07 & 1.70 \\
Median $T_{90}\,\rm{(s)}$ & 27 & 50 \\
Median $F_{peak}\,\rm{(erg}\,\rm{cm}^{-2}\,\rm{s}^{-1}\rm{)}$ & $2.0\times10^{-7}$ & $4.0\times10^{-8}$ \\
Median Fluence\,(erg\,cm$^{-2}$) & $2.0\times10^{-6}$ & $7.6\times10^{-7}$ \\ 
\% within $\pm30^\circ$\,SGB & 41\% & 91\% \\
\hline
\end{tabular}
\end{center}
\end{table*}

\clearpage
\appendix
\section{\textit{INTEGRAL} Lightcurves\label{app_lcs}}

The lightcurves of 43 of the 47 \textit{INTEGRAL} GRBs are presented
in the 25--50\,keV
(dark lines) and 50--300\,keV (light lines) energy bands. Exceptions
are the XRFs which were not detected above 50\,keV and for which the
lightcurves are plotted in the 15--25\,keV (dark lines) and
25--50\,keV (light lines) and the very weak GRBs and GRBs with
telemetry gaps for which a lag could not be determined which are given
over the full energy range of 25--300\,keV. The regions
which were used in the spectral lag analysis are denoted by solid
vertical lines and the temporal resolution is that at which the lag
was determined. All lightcurve data is from the IBIS instrument on
board \textit{INTEGRAL} with the exceptions of GRB\,041219a and
GRB\,050525a for which data is taken from the BAT instrument on \textit{Swift} due to
IBIS telemetry saturation. A satellite slew occurred during
GRB\,030131 and the lightcurve is not presented here but is available in~\citet{gotz:2003}. GRB\,060428c was discovered in the \textit{INTEGRAL} archival data and the lightcurve is available in~\citet{grebenev:2007}. IBIS
data is currently unavailable for GRB\,060930 and GRB\,070311.

\begin{figure*}
\centering
\caption{Lightcurves of GRBS observed with \textit{INTEGRAL}.}
\label{fig:lcs1}
\mbox{
\subfigure{\label{021125}\includegraphics[height=0.35\textheight,width=0.18\textwidth,angle=270]{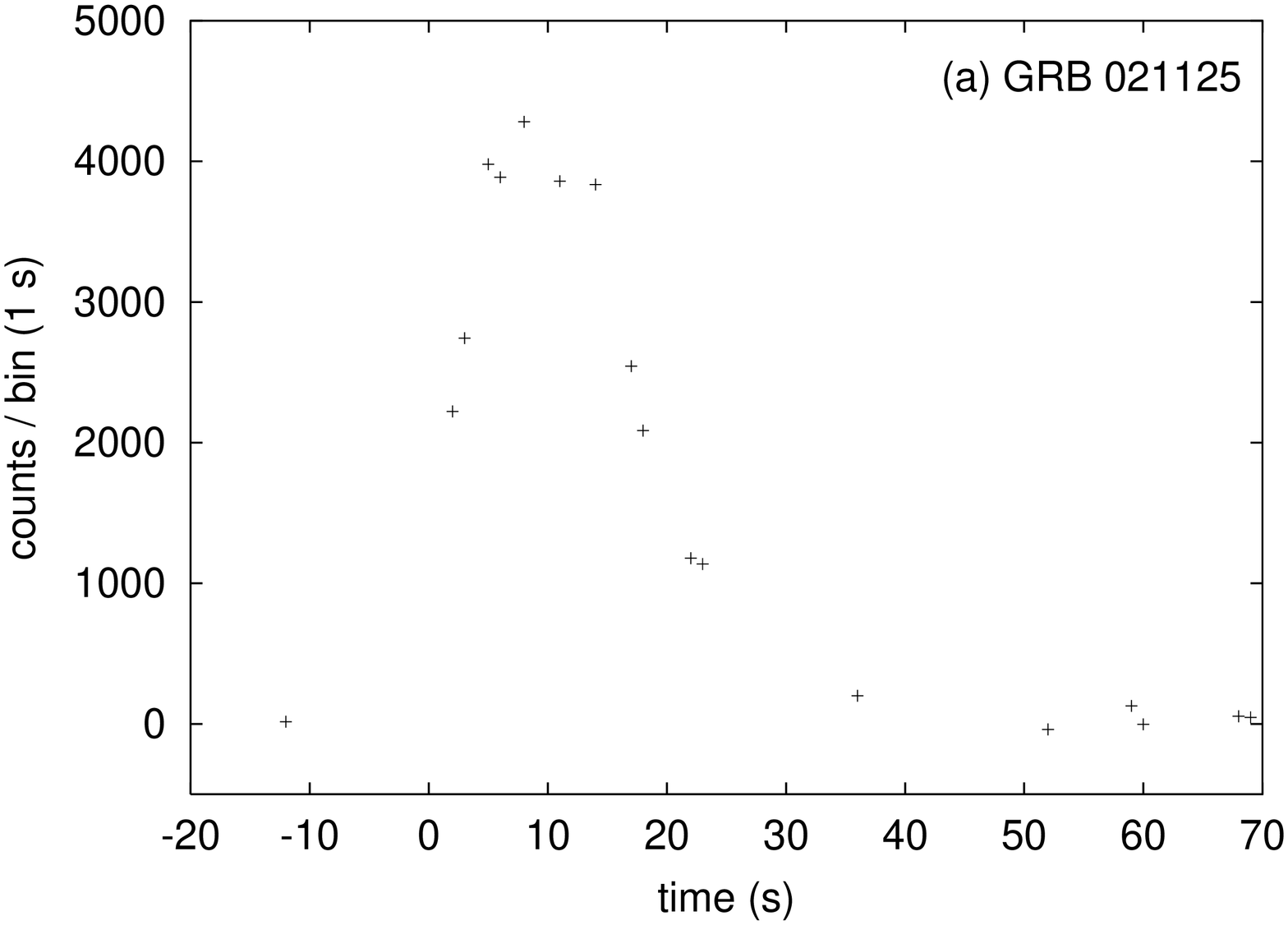}}
\subfigure{\label{021219}\includegraphics[height=0.35\textheight,width=0.18\textwidth,angle=270]{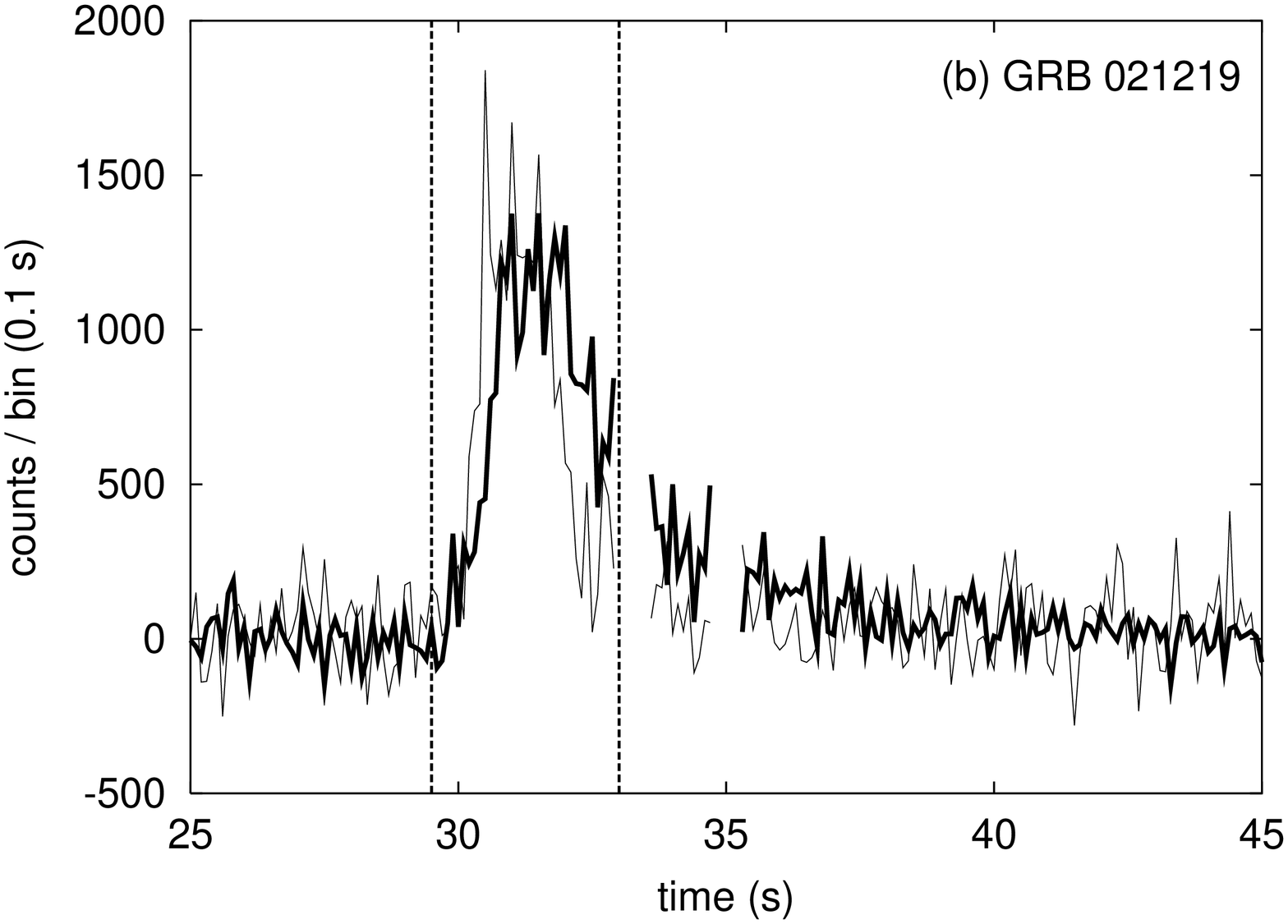}}
}
\mbox{
\subfigure{\label{030227}\includegraphics[height=0.35\textheight,width=0.18\textwidth,angle=270]{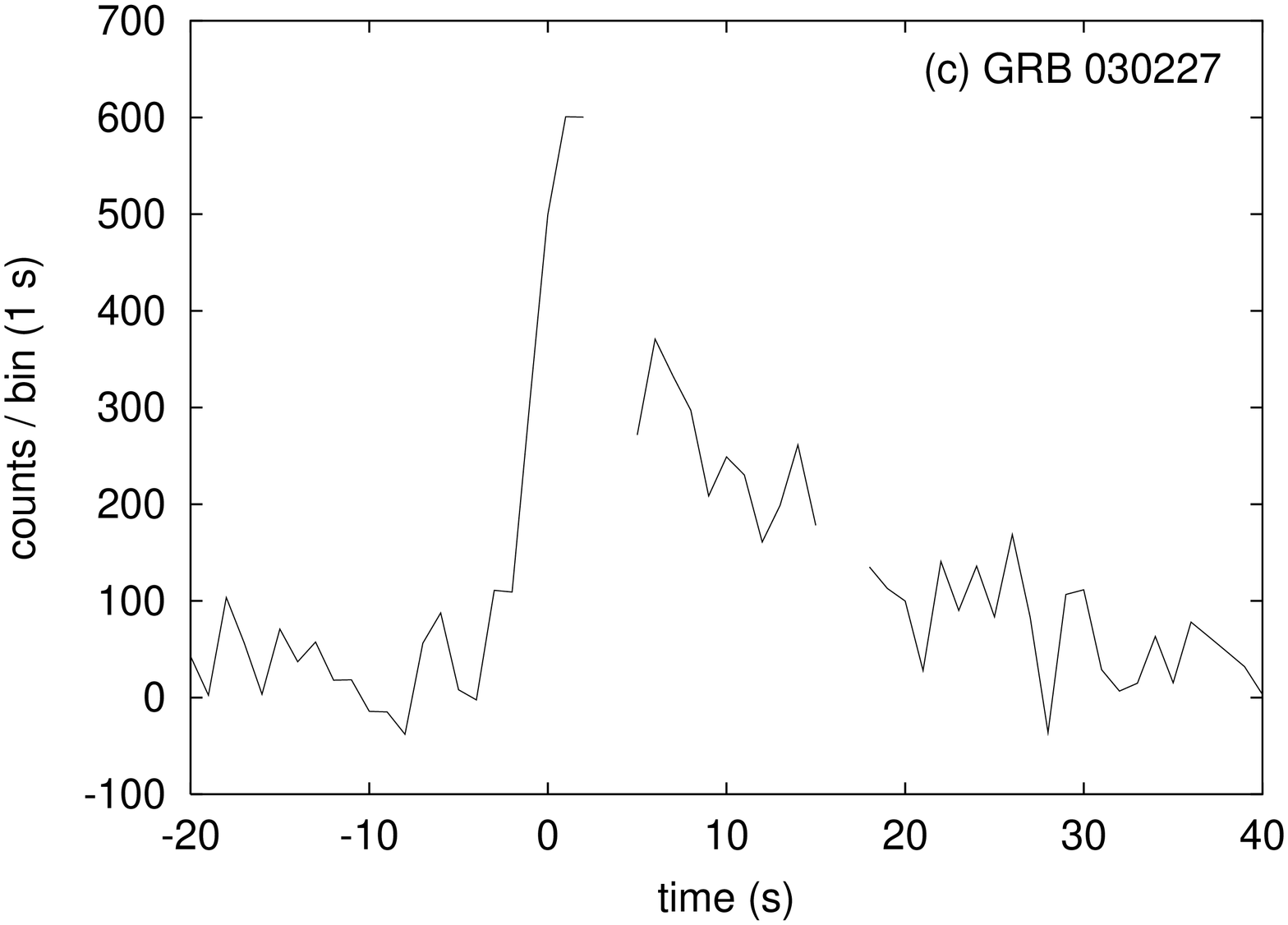}}
\subfigure{\includegraphics[height=0.35\textheight,width=0.18\textwidth,angle=270]{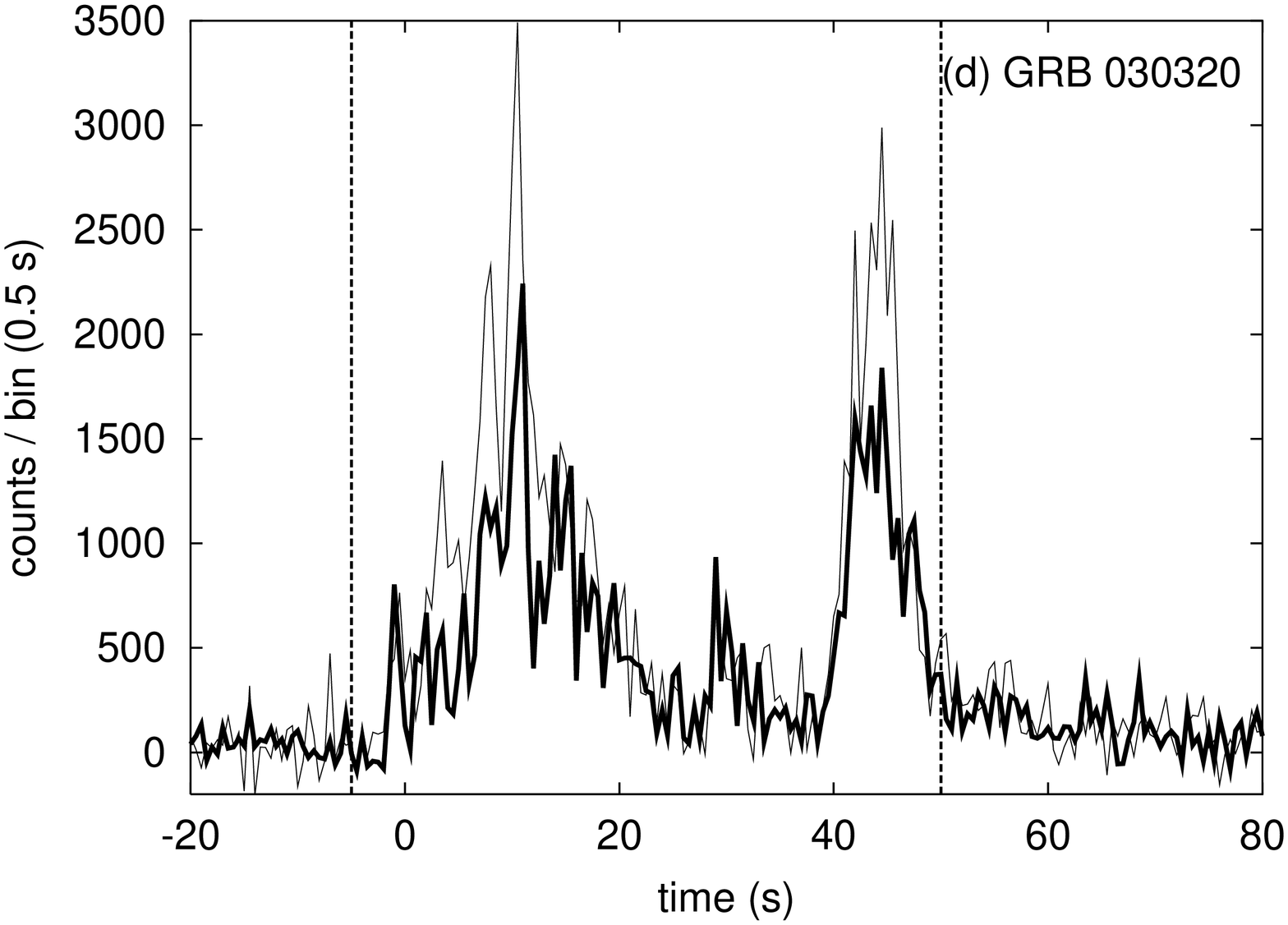}}
}
\mbox{
\subfigure{\includegraphics[height=0.35\textheight,width=0.18\textwidth,angle=270]{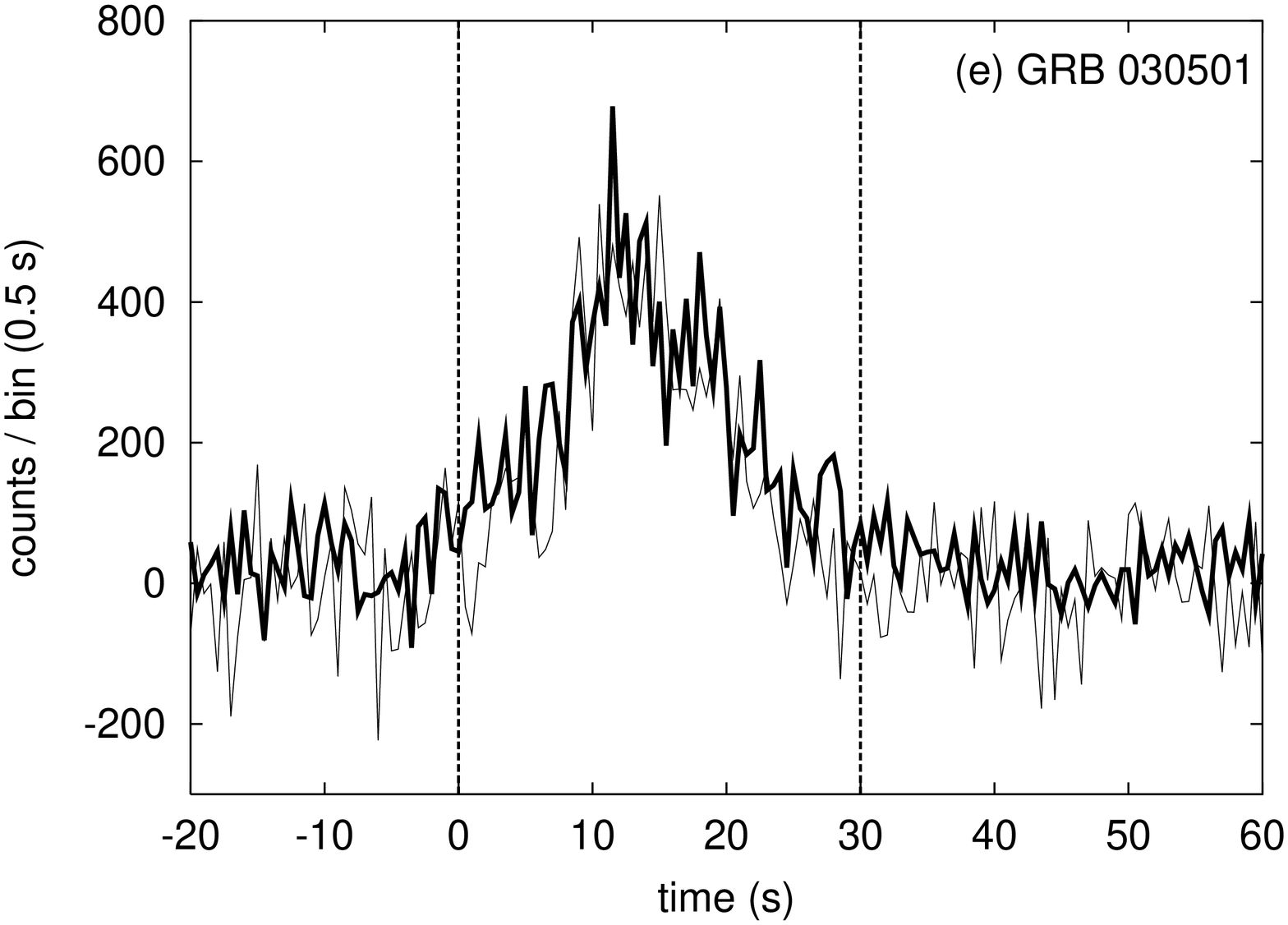}}
\subfigure{\includegraphics[height=0.35\textheight,width=0.18\textwidth,angle=270]{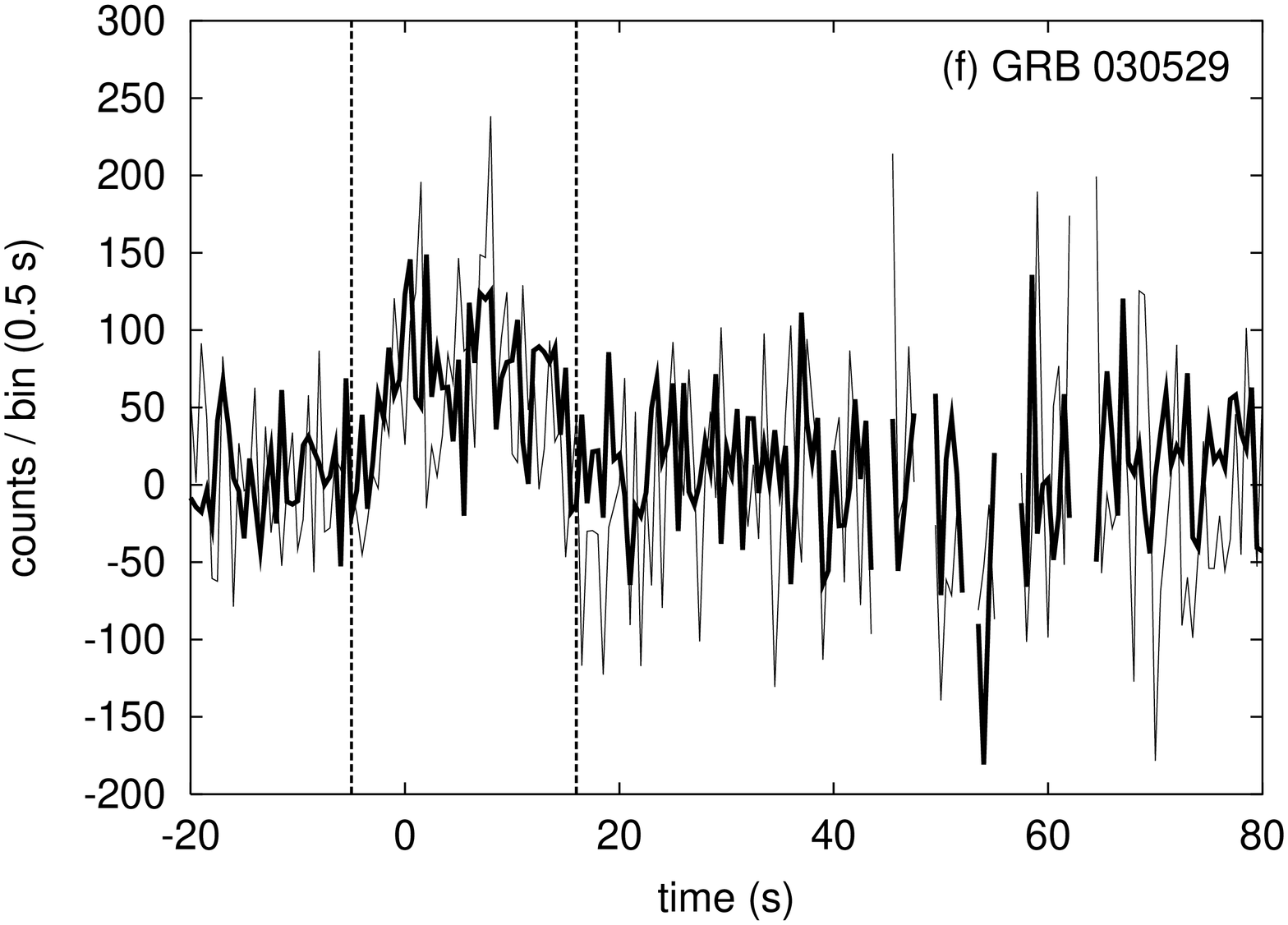}}
}
\mbox{
\subfigure{\includegraphics[height=0.35\textheight,width=0.18\textwidth,angle=270]{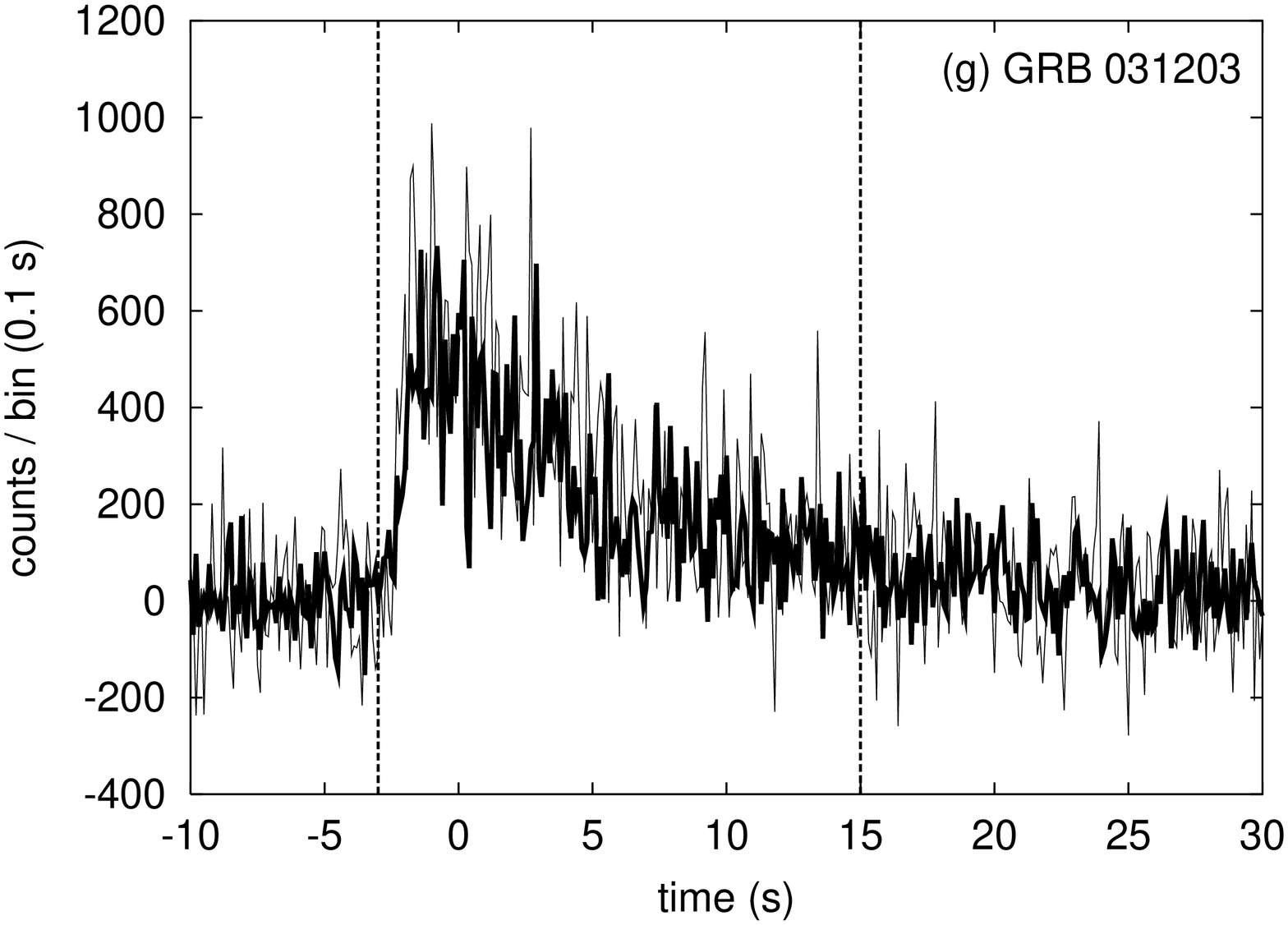}}
\subfigure{\includegraphics[height=0.35\textheight,width=0.18\textwidth,angle=270]{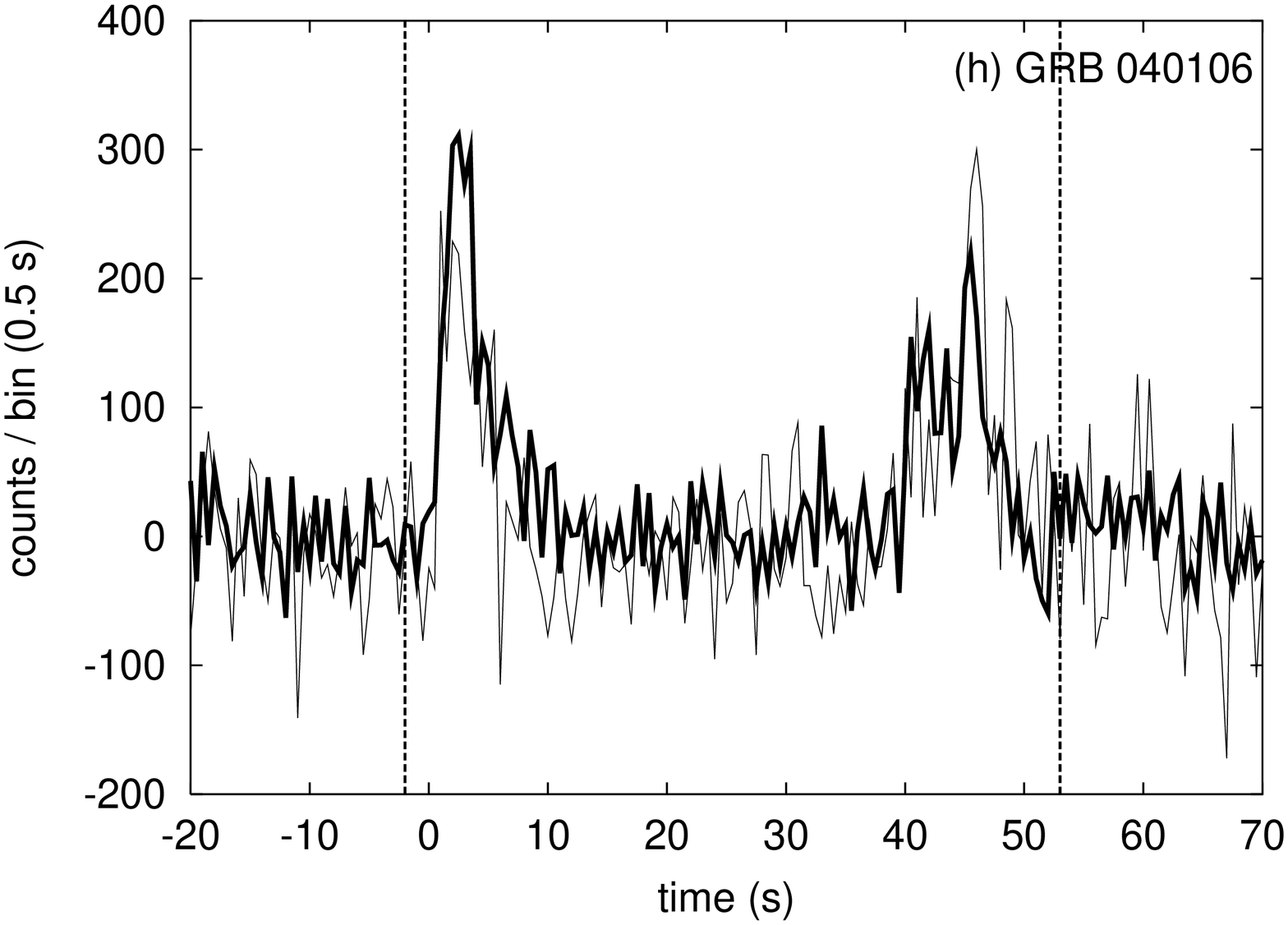}}
}
\mbox{
\subfigure{\includegraphics[height=0.35\textheight,width=0.18\textwidth,angle=270]{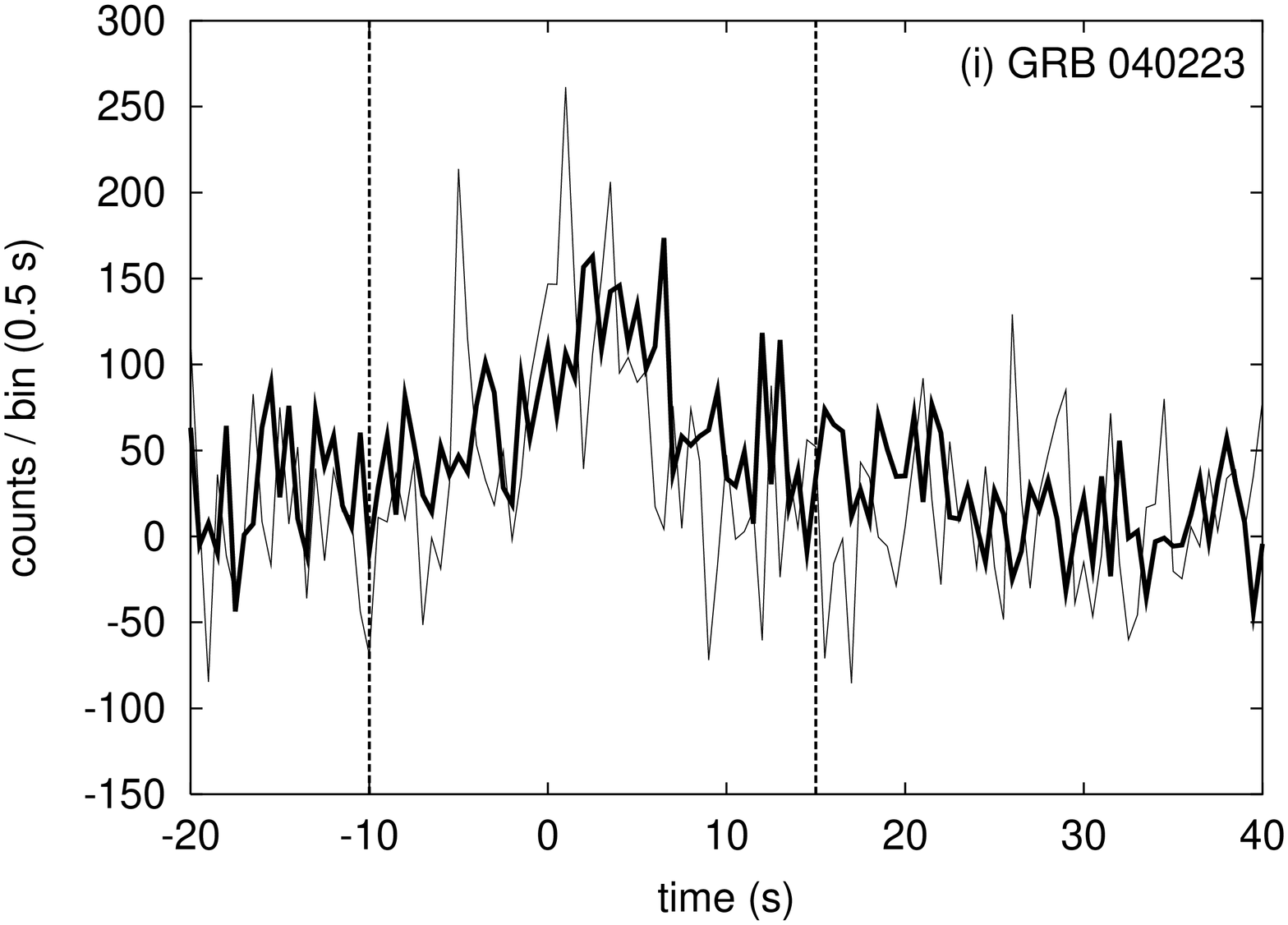}}
\subfigure{\includegraphics[height=0.35\textheight,width=0.18\textwidth,angle=270]{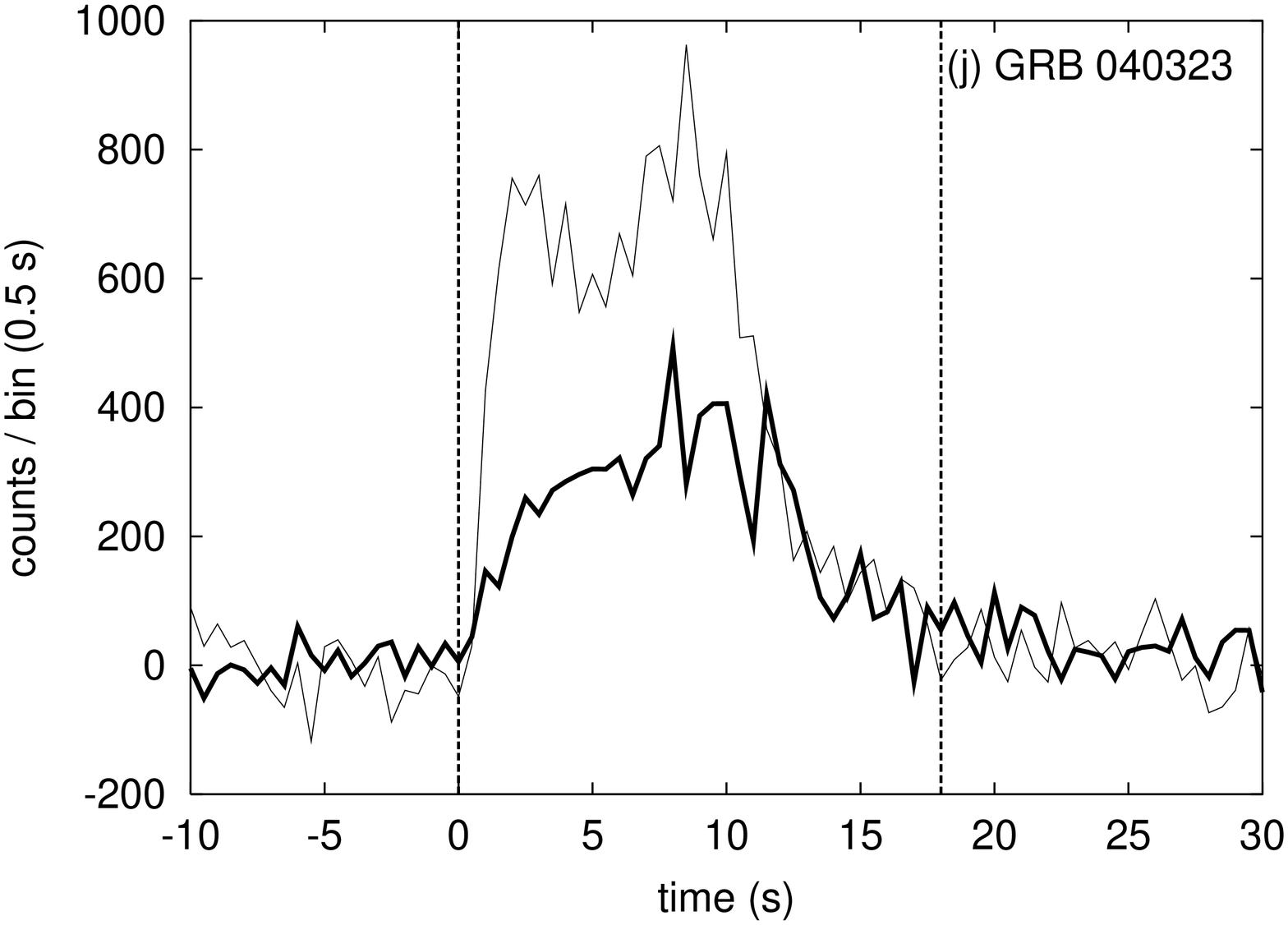}}
}
\mbox{
\subfigure{\includegraphics[height=0.35\textheight,width=0.18\textwidth,angle=270]{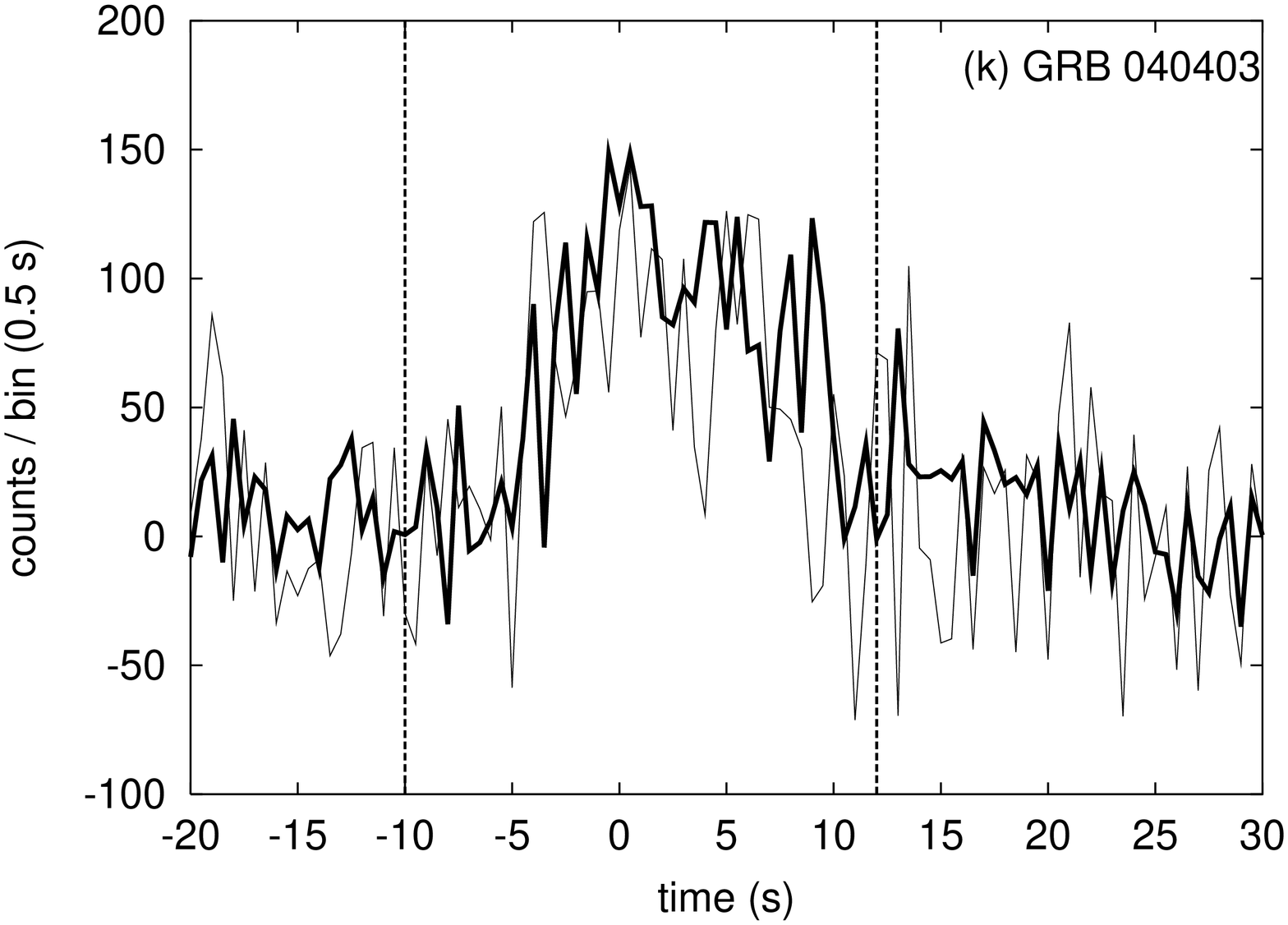}}
\subfigure{\includegraphics[height=0.35\textheight,width=0.18\textwidth,angle=270]{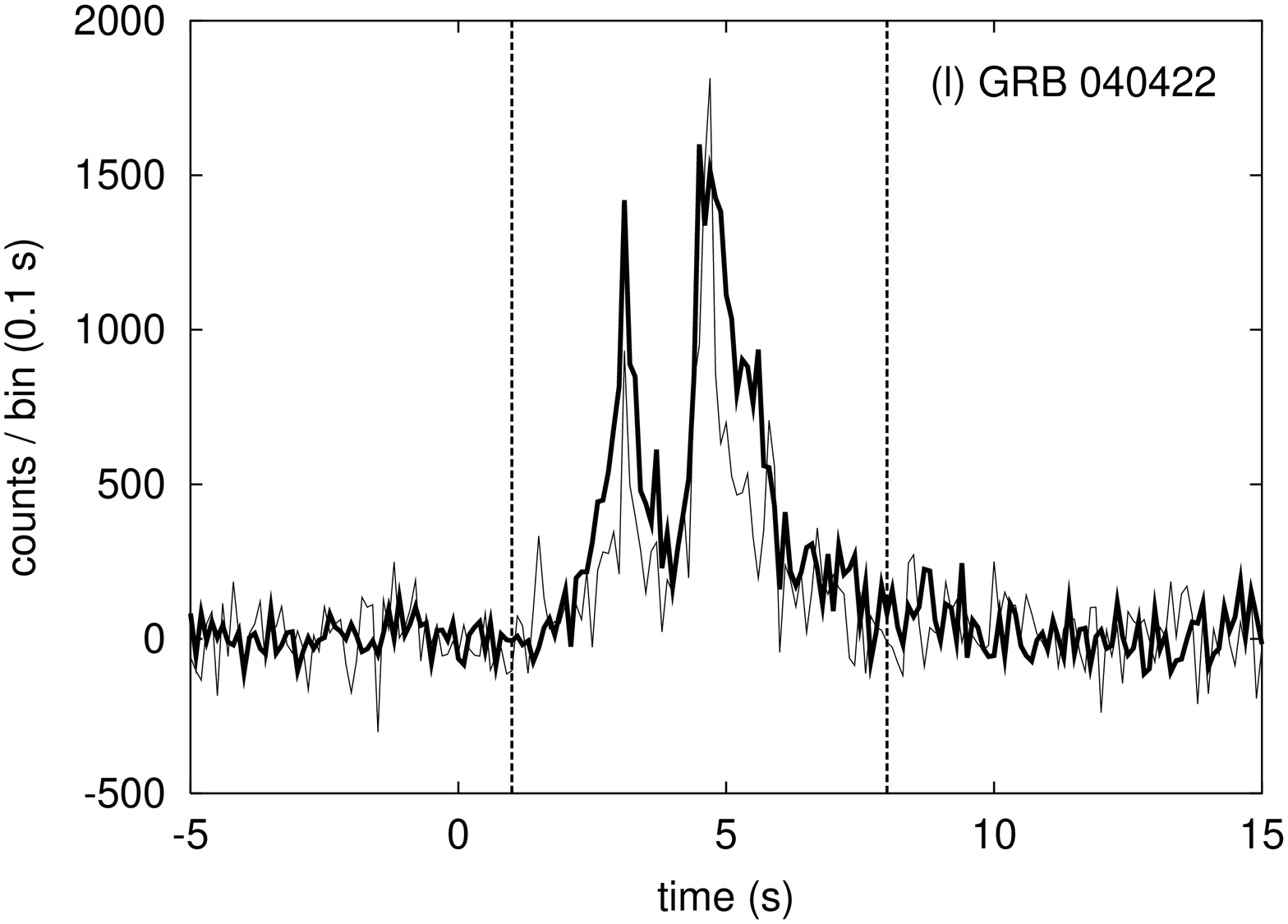}}
}
\end{figure*}

\begin{figure*}
  \centering
\caption{Lightcurves of GRBS observed with \textit{INTEGRAL} (continued).}
\label{fig:lcs2}
\mbox{
  \subfigure{\includegraphics[height=0.35\textheight, width=0.18\textwidth,angle=270]{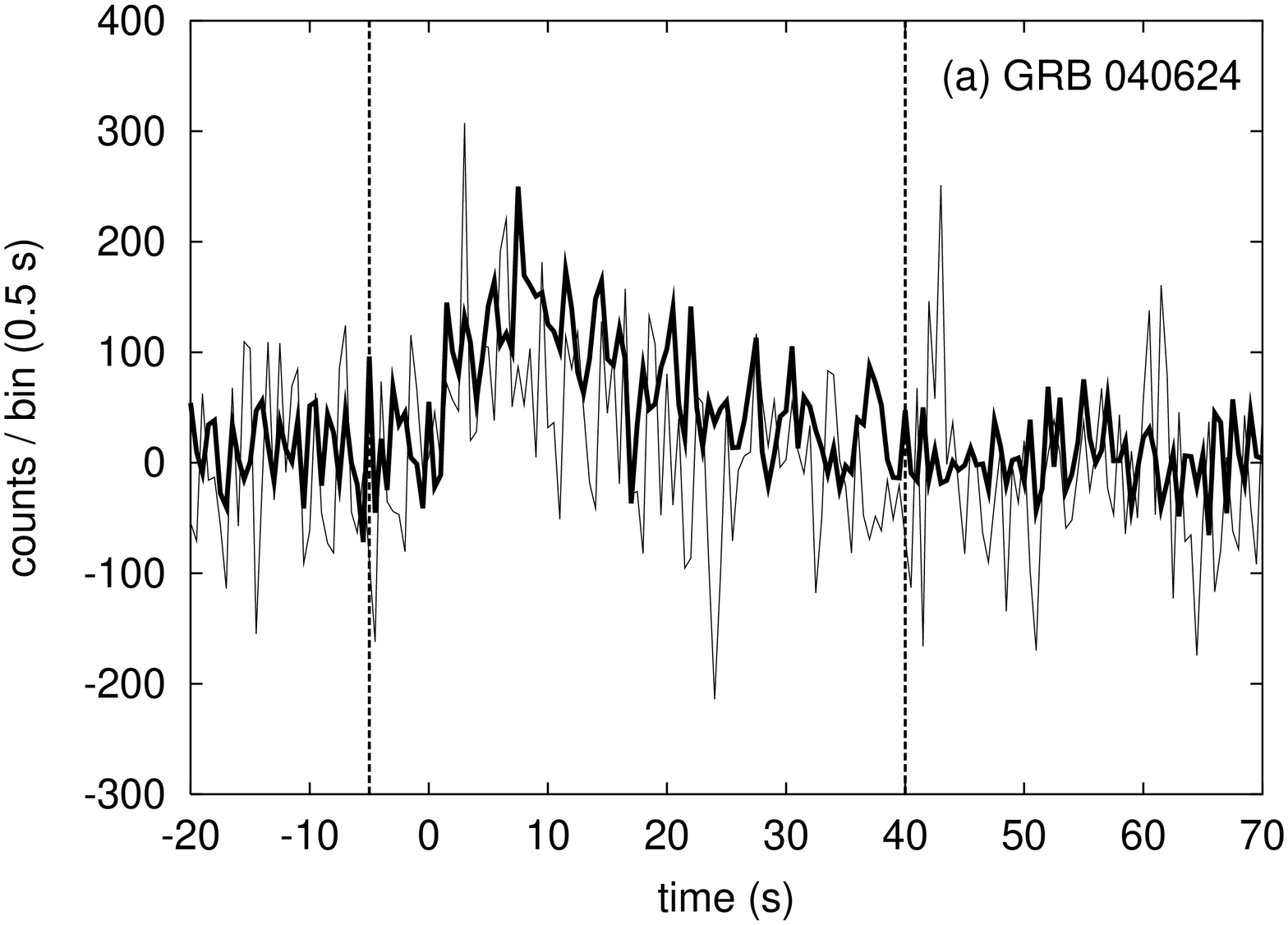}}
\subfigure{\includegraphics[height=0.35\textheight,width=0.18\textwidth,angle=270]{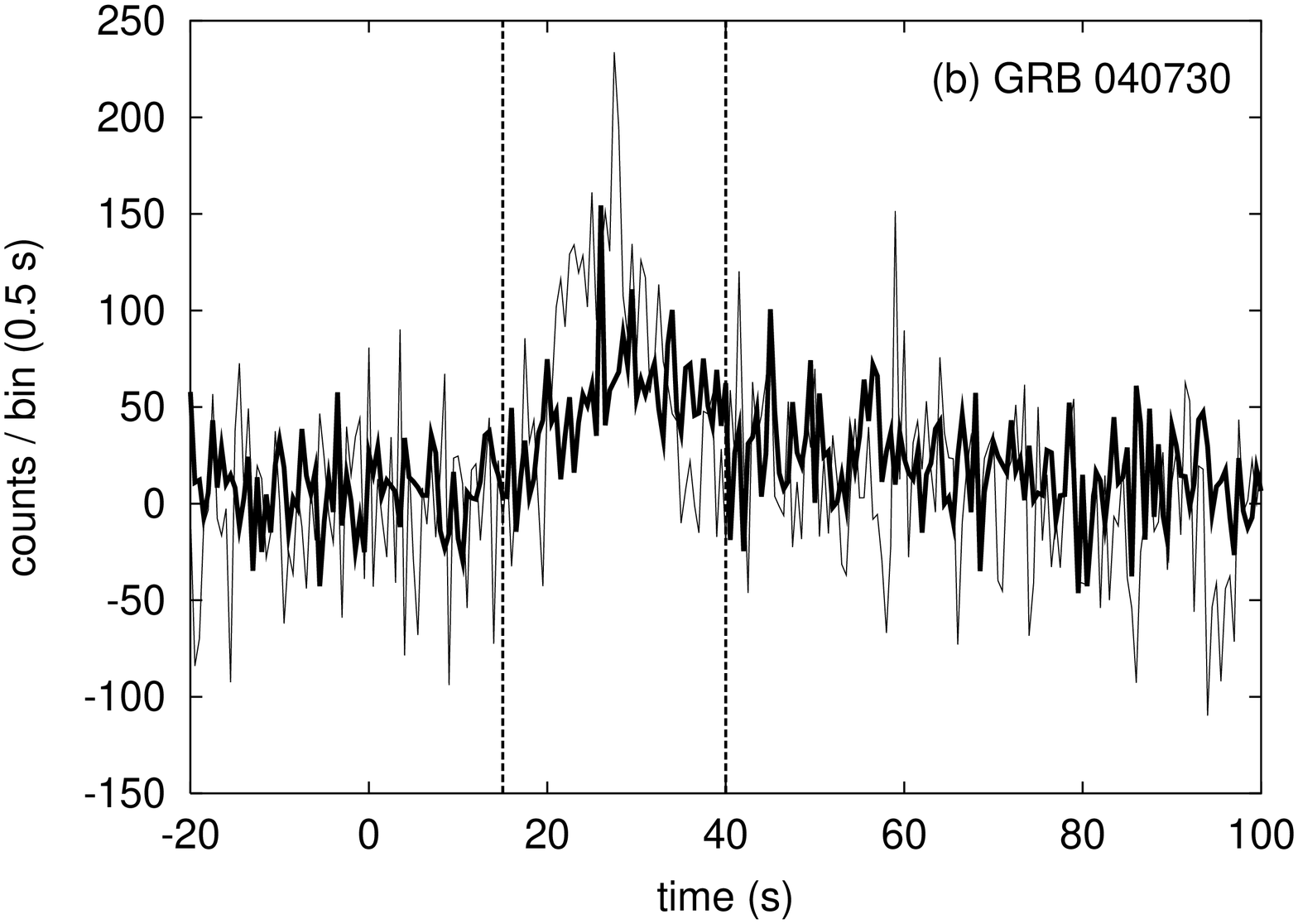}}
}
\mbox{
\subfigure{\includegraphics[height=0.35\textheight,width=0.18\textwidth,angle=270]{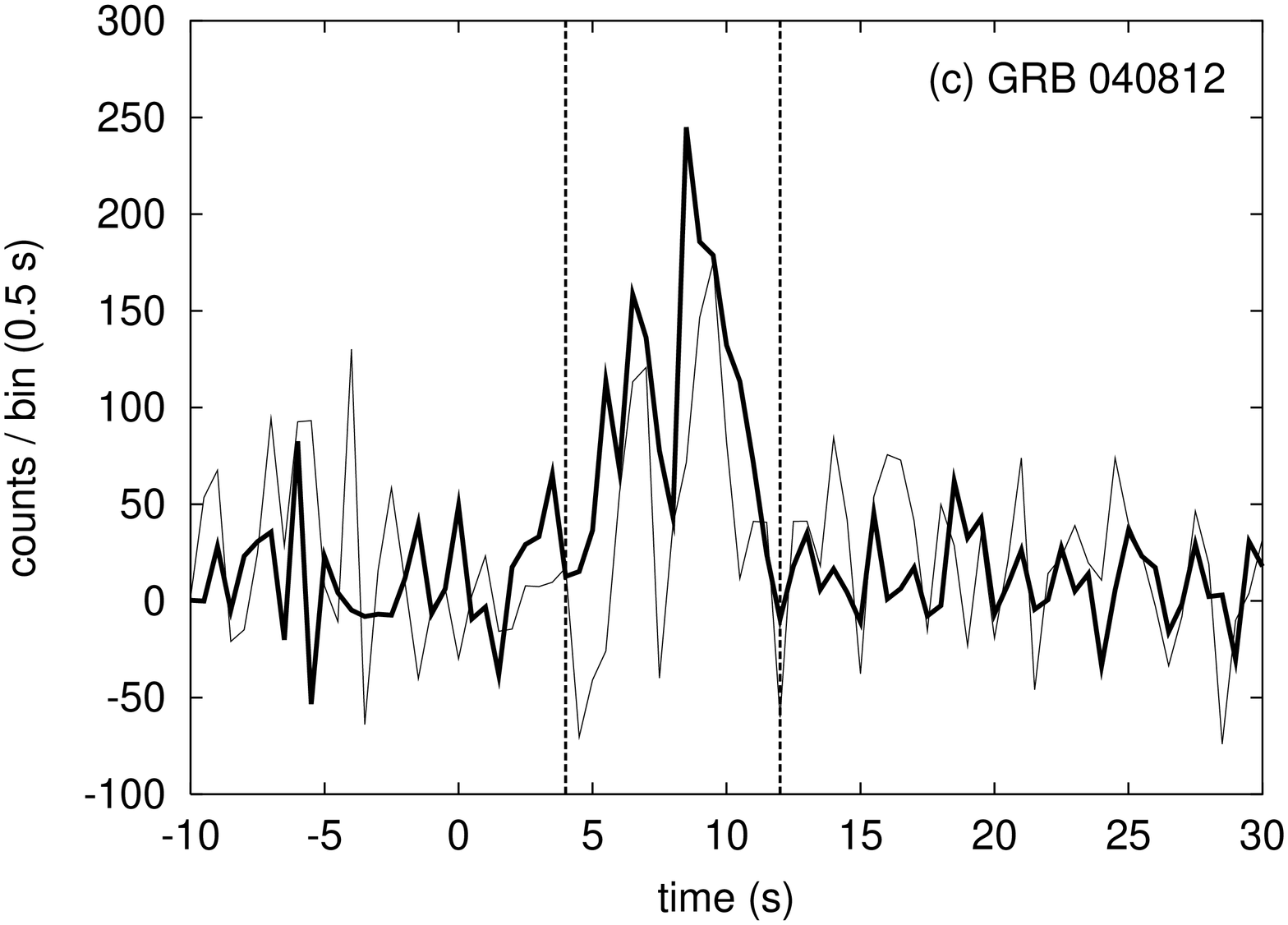}}
\subfigure{\includegraphics[height=0.35\textheight,width=0.18\textwidth,angle=270]{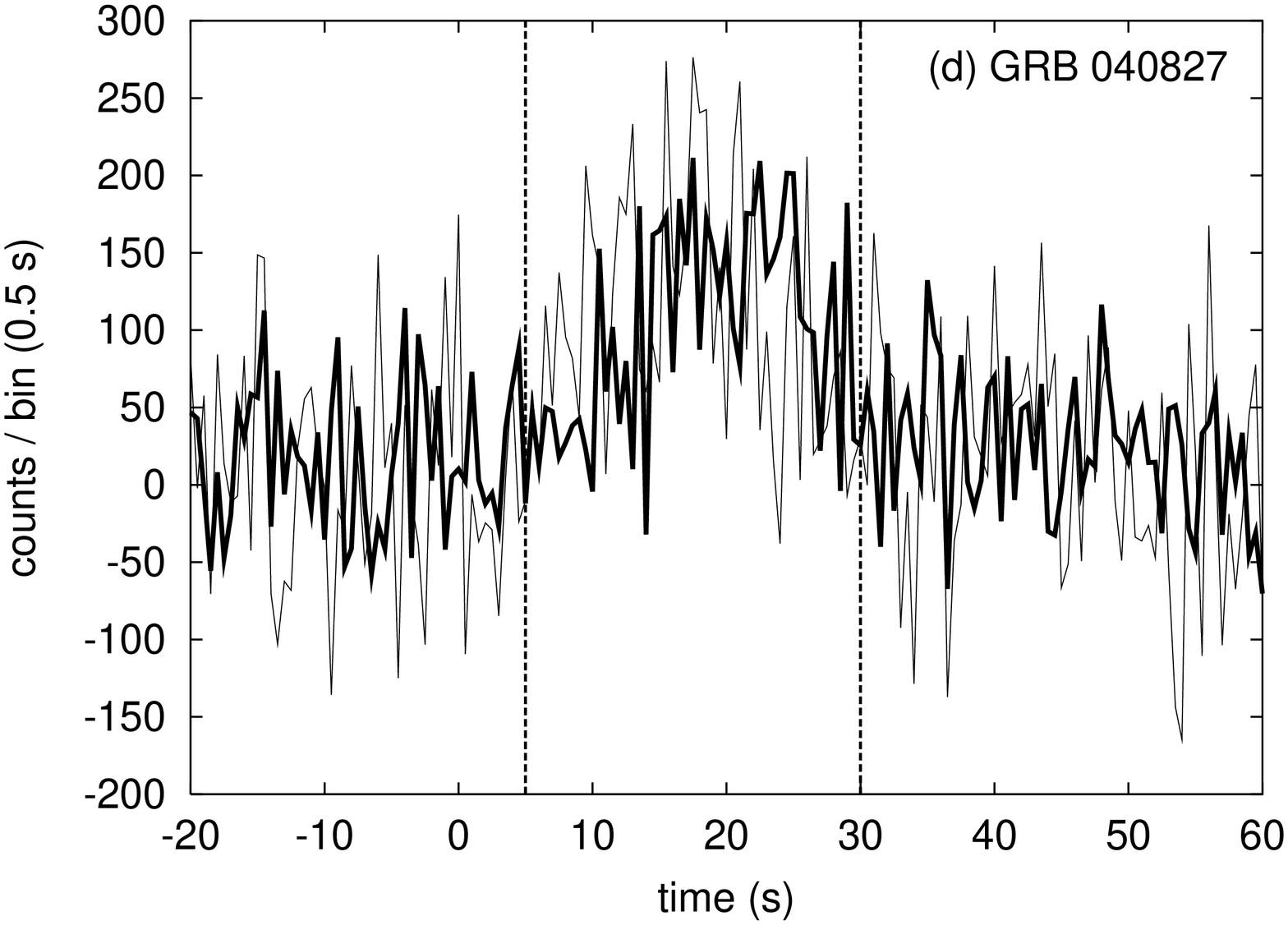}}
}
\mbox{
\subfigure{\includegraphics[height=0.35\textheight,width=0.18\textwidth,angle=270]{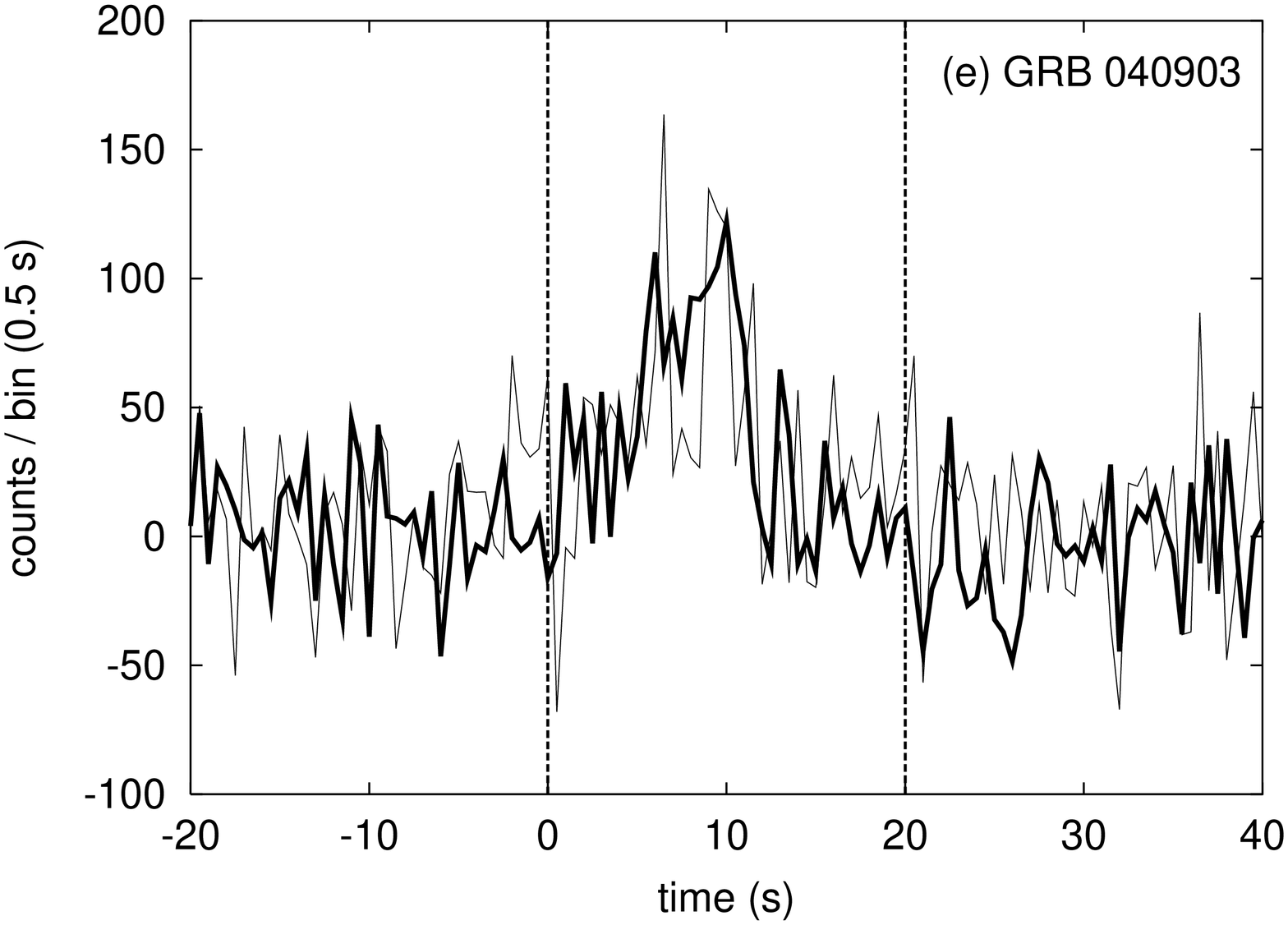}}
\subfigure{\includegraphics[height=0.35\textheight,width=0.18\textwidth,angle=270]{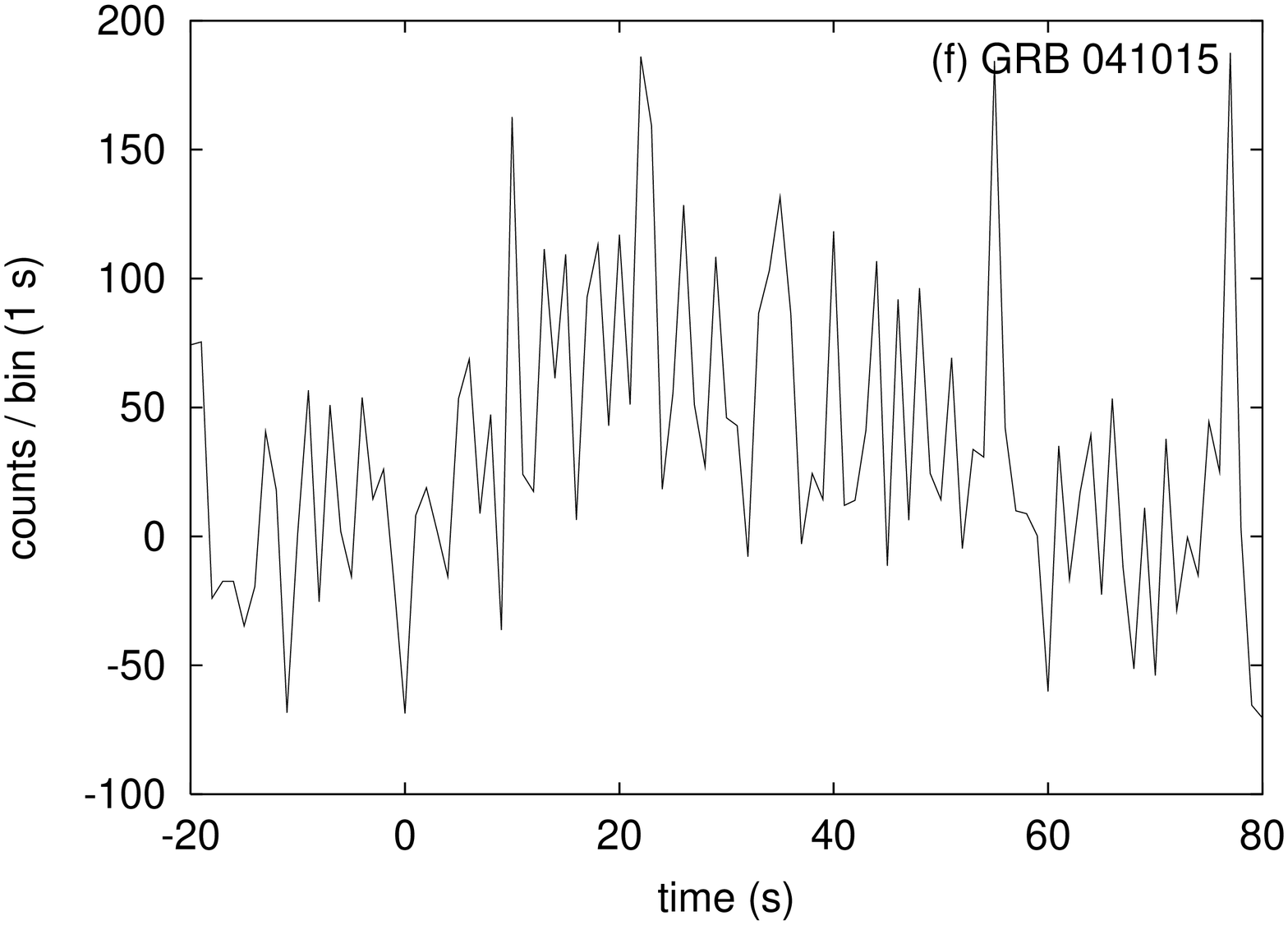}}
}
\mbox{
\subfigure{\includegraphics[height=0.35\textheight,width=0.18\textwidth,angle=270]{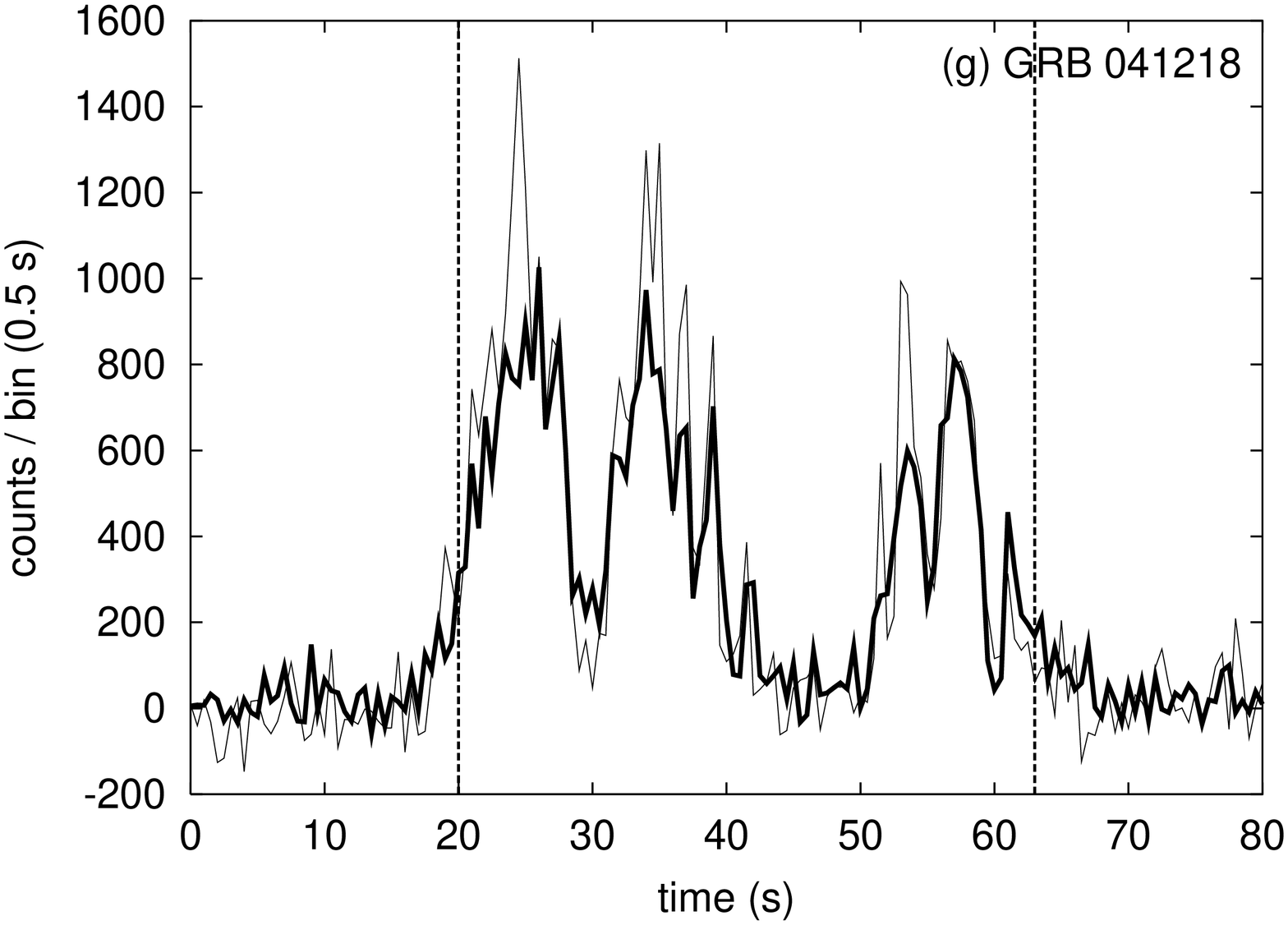}}
\subfigure{\includegraphics[height=0.35\textheight,width=0.18\textwidth,angle=270]{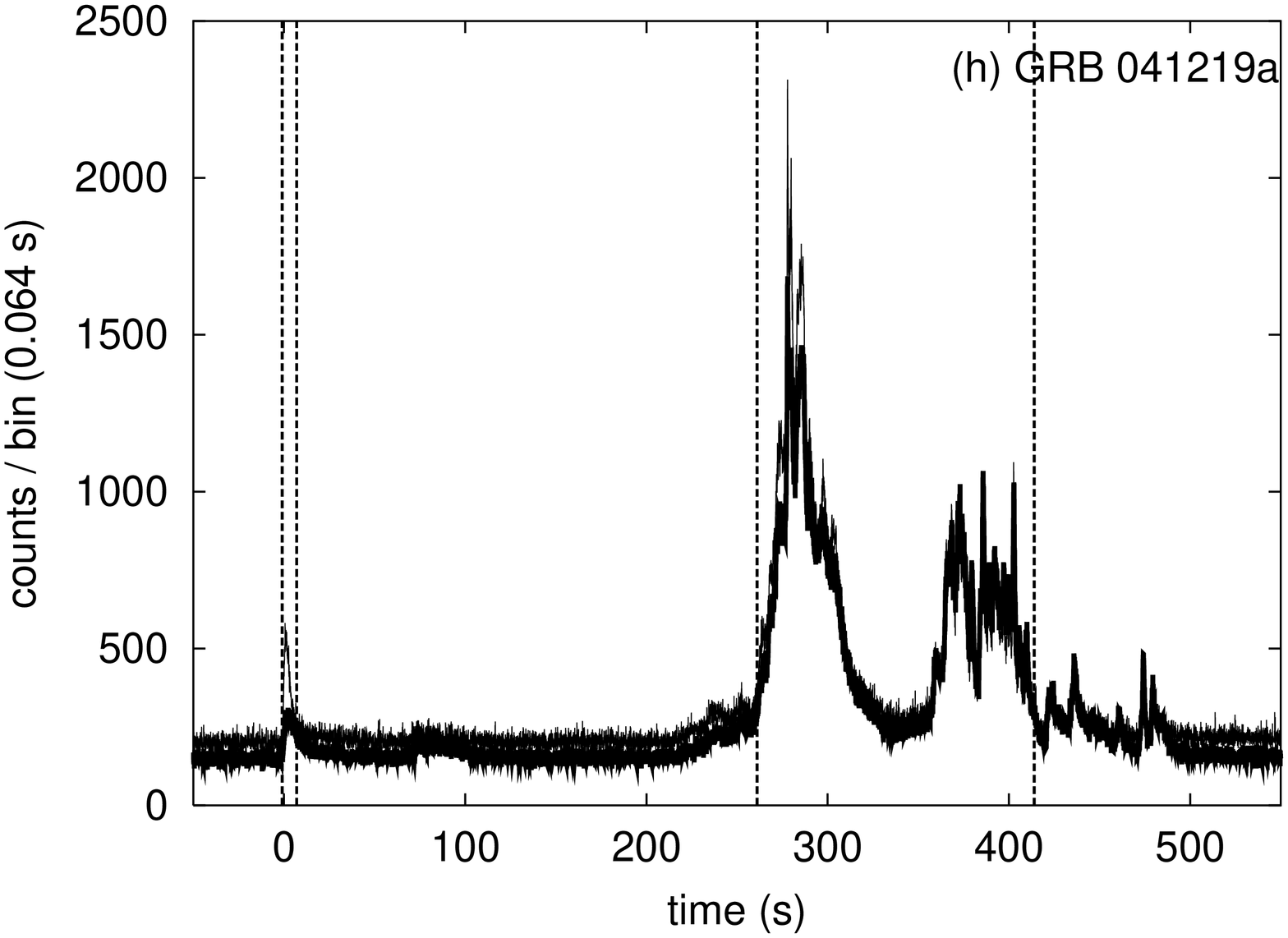}}
}
\mbox{
\subfigure{\includegraphics[height=0.35\textheight,width=0.18\textwidth,angle=270]{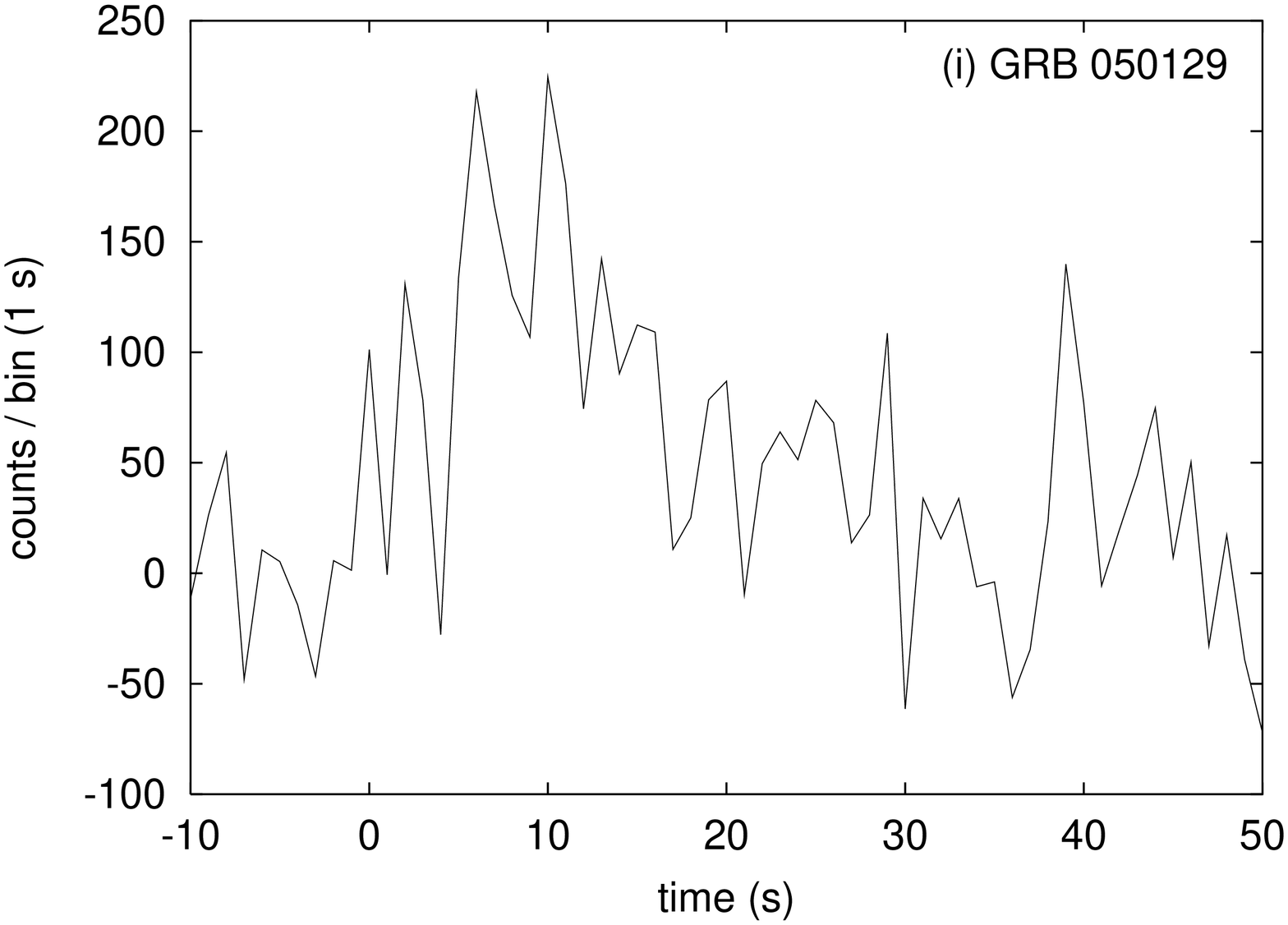}}
\subfigure{\includegraphics[height=0.35\textheight,width=0.18\textwidth,angle=270]{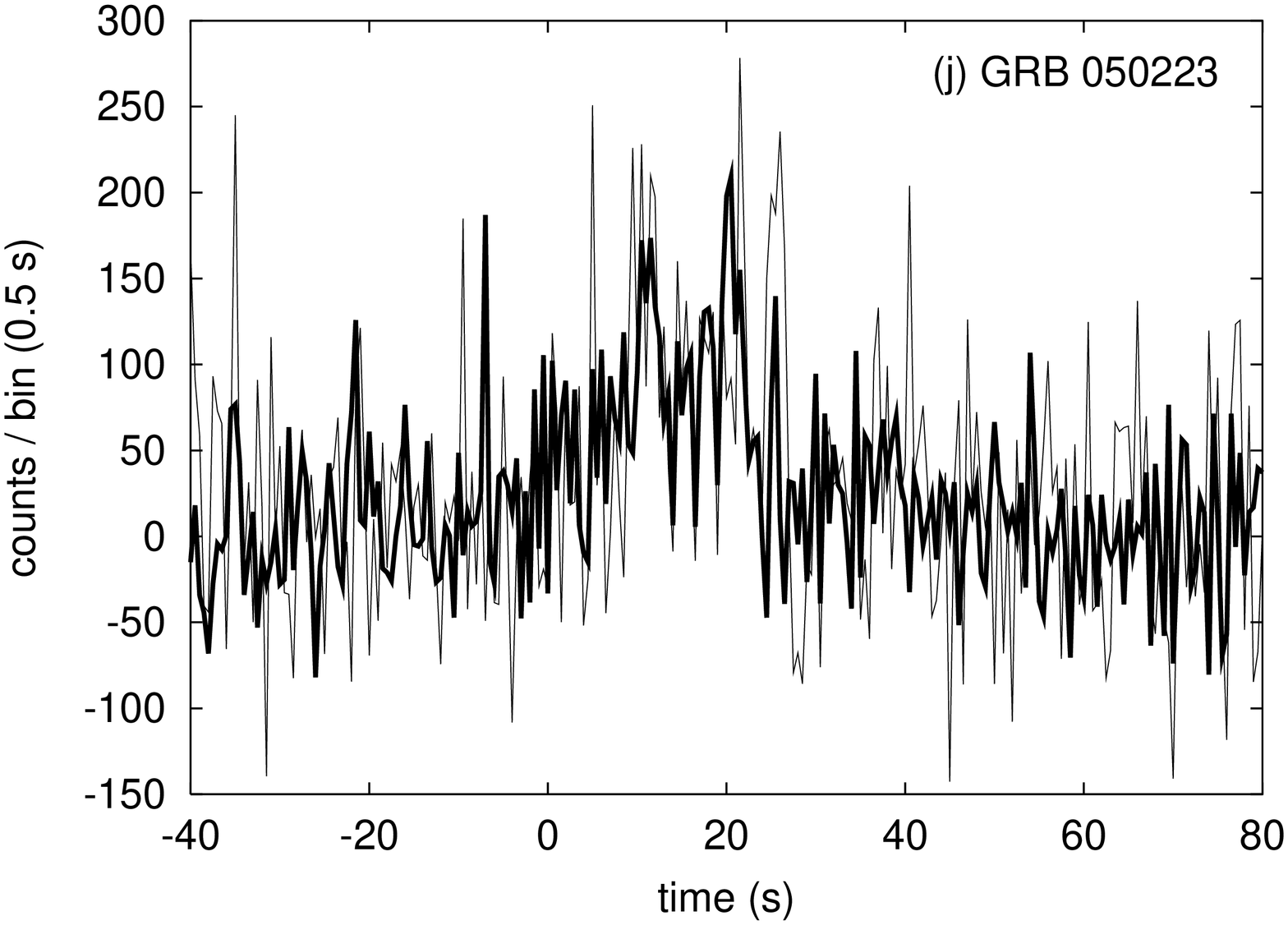}}
}
\mbox{
\subfigure{\includegraphics[height=0.35\textheight,width=0.18\textwidth,angle=270]{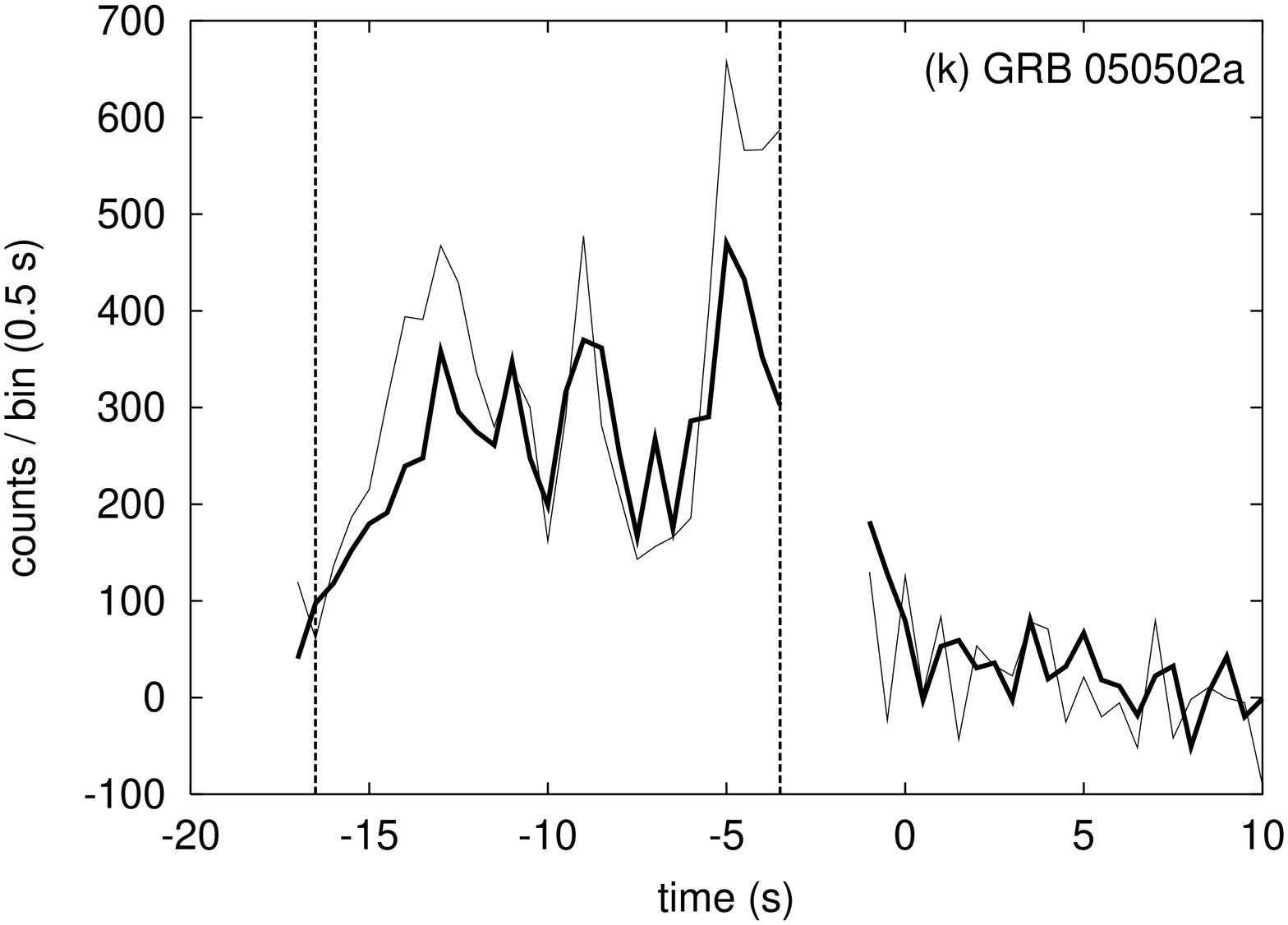}}
\subfigure{\includegraphics[height=0.35\textheight,width=0.18\textwidth,angle=270]{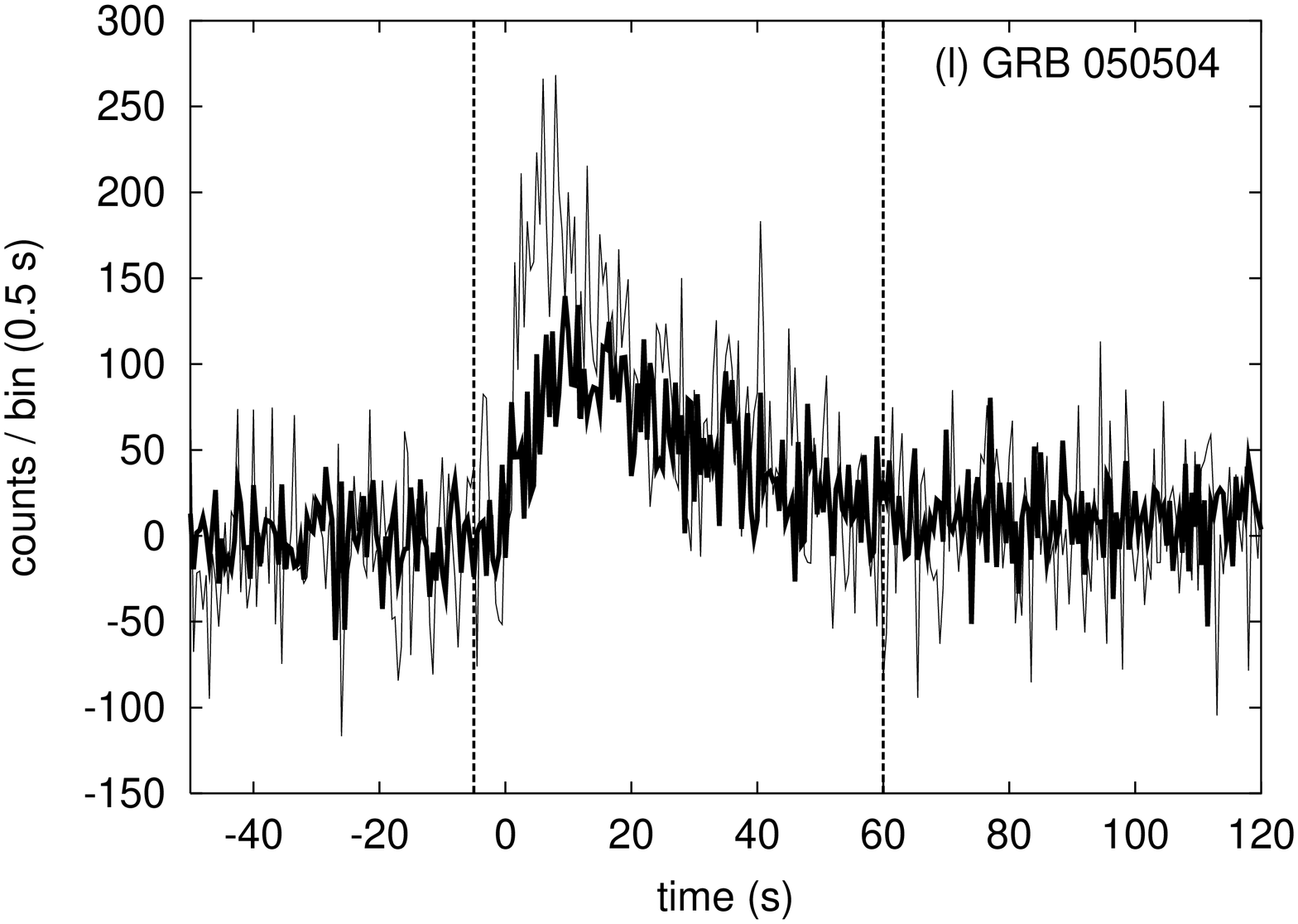}}
}
\end{figure*}

\begin{figure*}
\caption{Lightcurves of GRBS observed with \textit{INTEGRAL} (continued).}
\label{fig:lcs3}
\centering
\mbox{
\subfigure{\includegraphics[height=0.35\textheight, width=0.18\textwidth,angle=270]{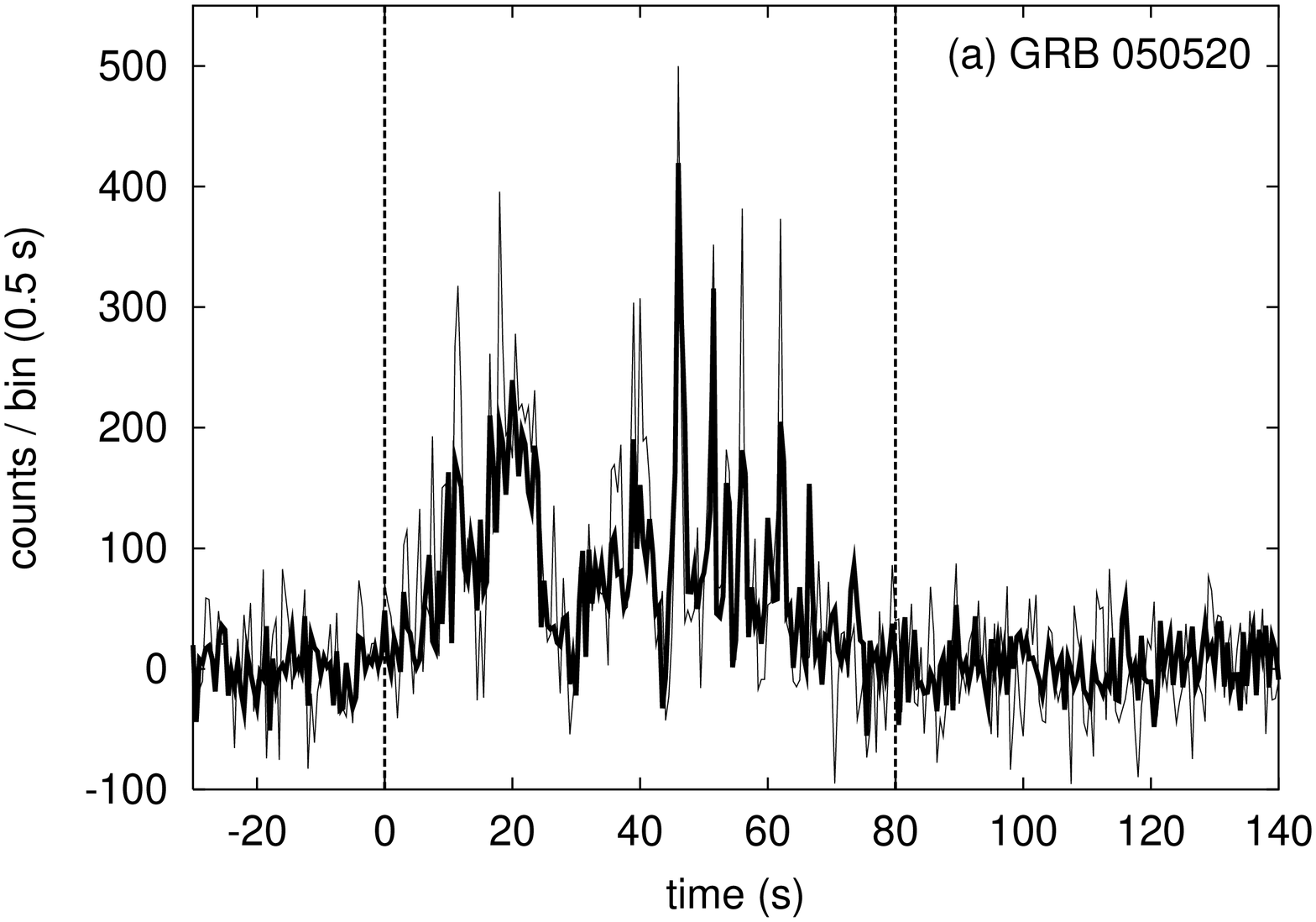}}
\subfigure{\includegraphics[height=0.35\textheight,width=0.18\textwidth,angle=270]{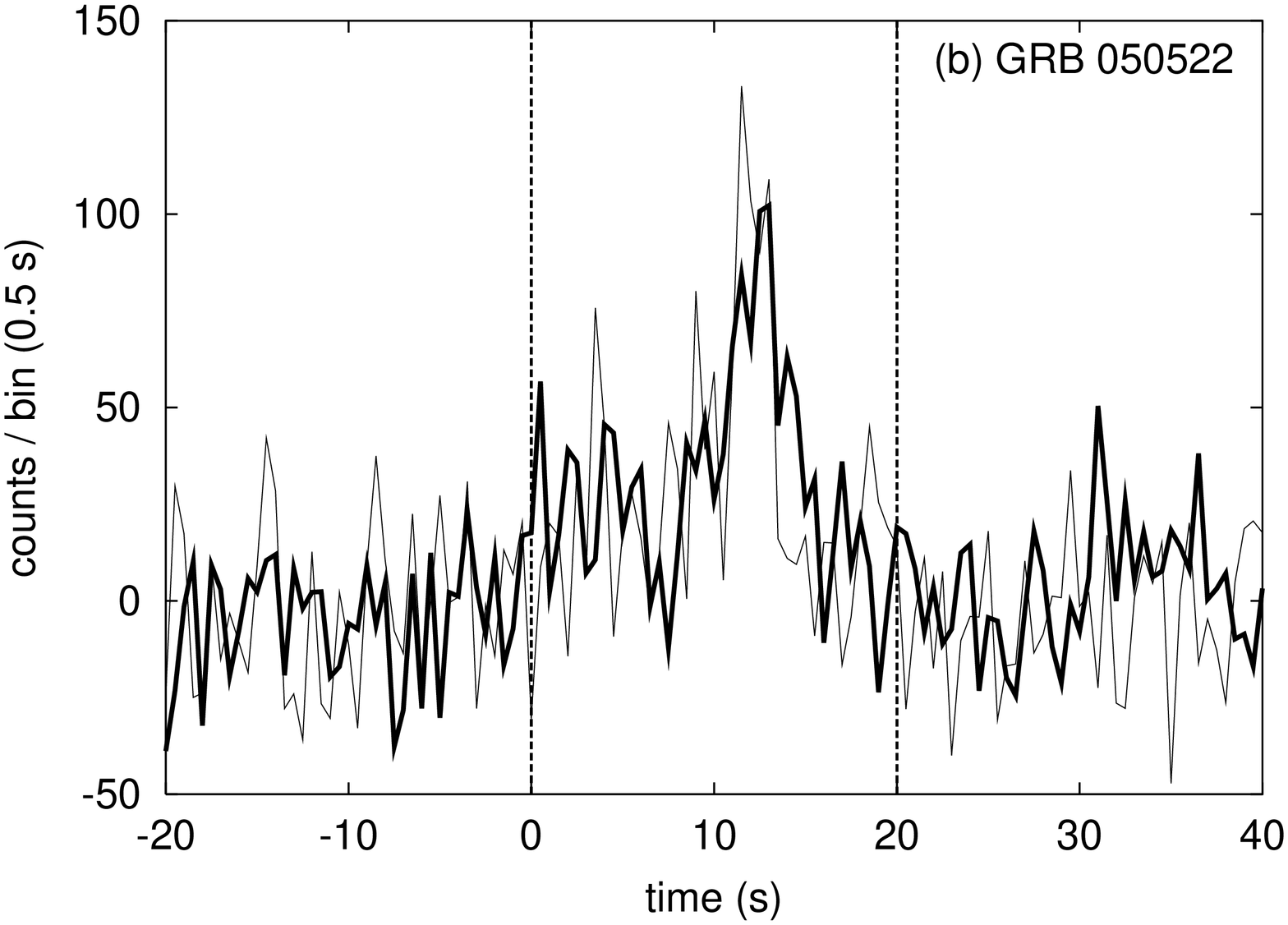}}
}
\mbox{
\subfigure{\includegraphics[height=0.35\textheight,width=0.18\textwidth,angle=270]{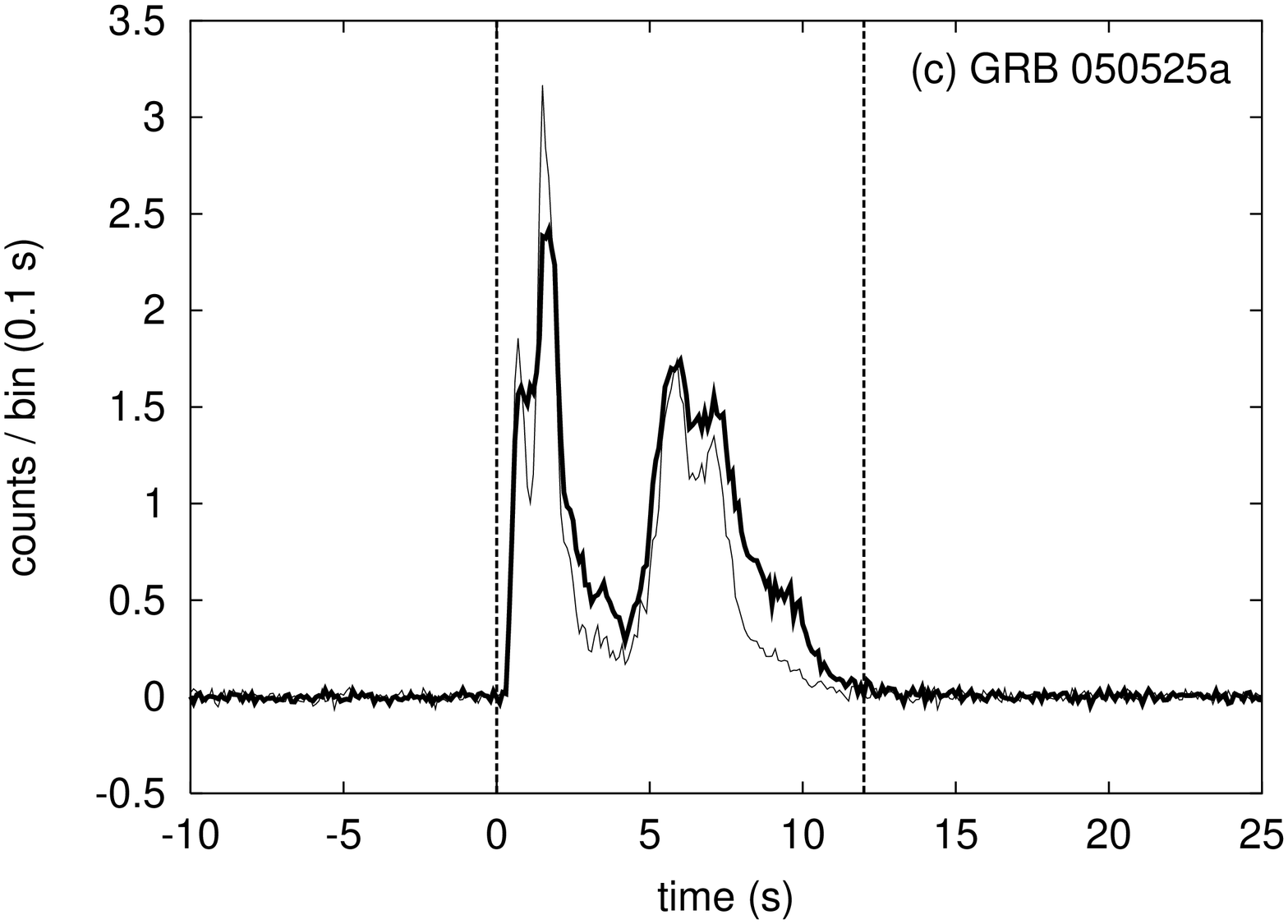}}
\subfigure{\includegraphics[height=0.35\textheight,width=0.18\textwidth,angle=270]{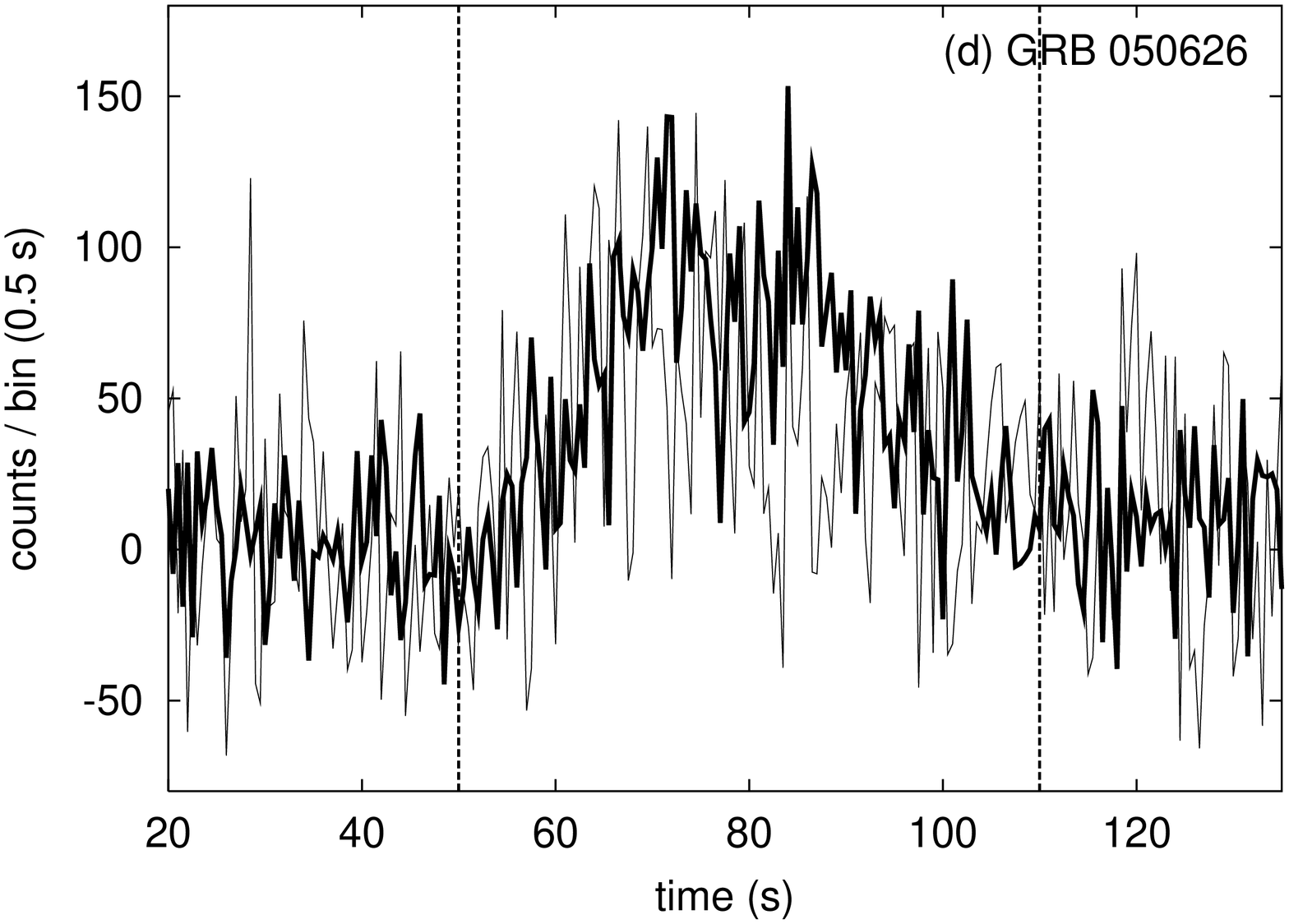}}
}
\mbox{
\subfigure{\includegraphics[height=0.35\textheight,width=0.18\textwidth,angle=270]{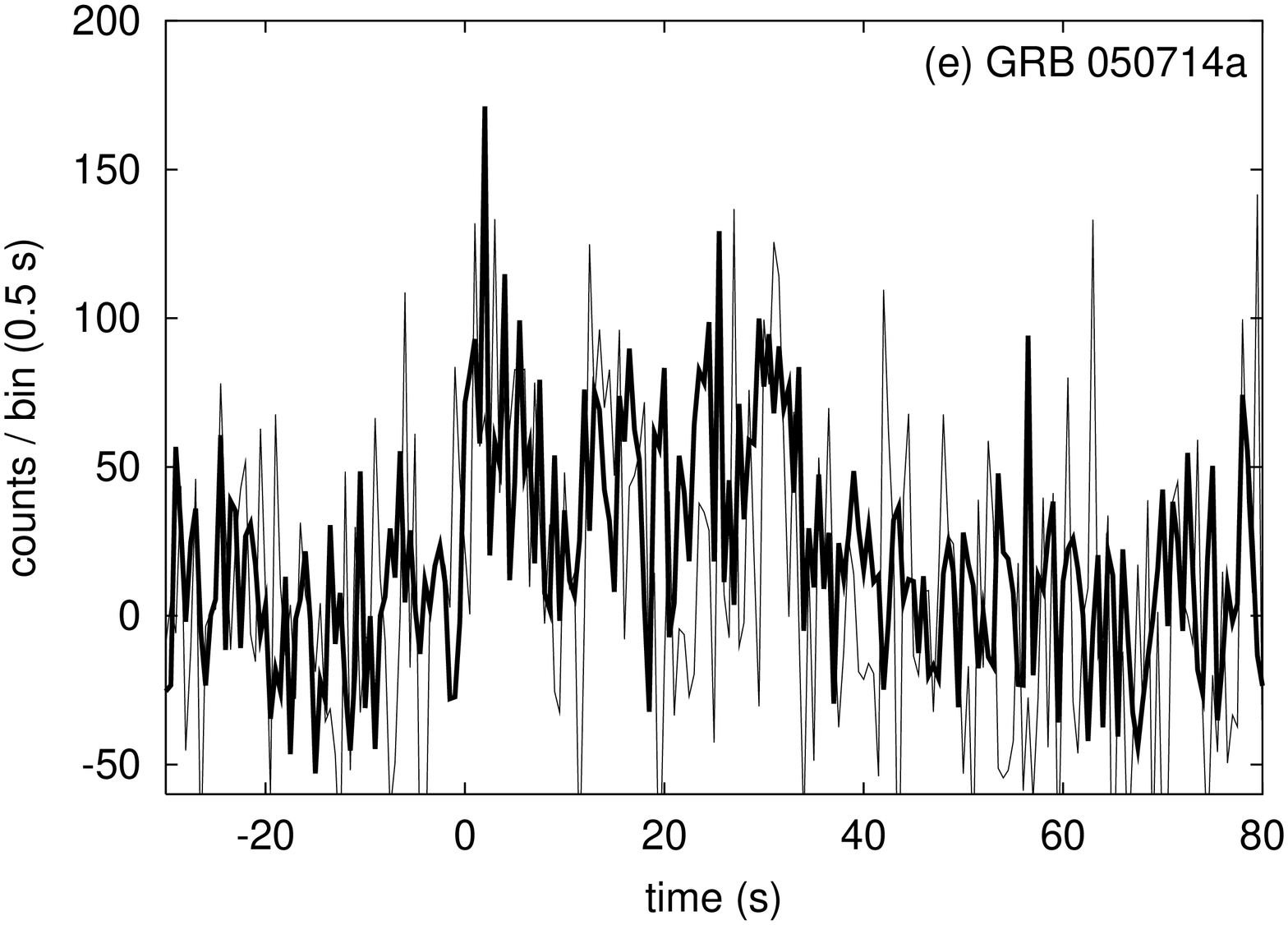}}
\subfigure{\includegraphics[height=0.35\textheight,width=0.18\textwidth,angle=270]{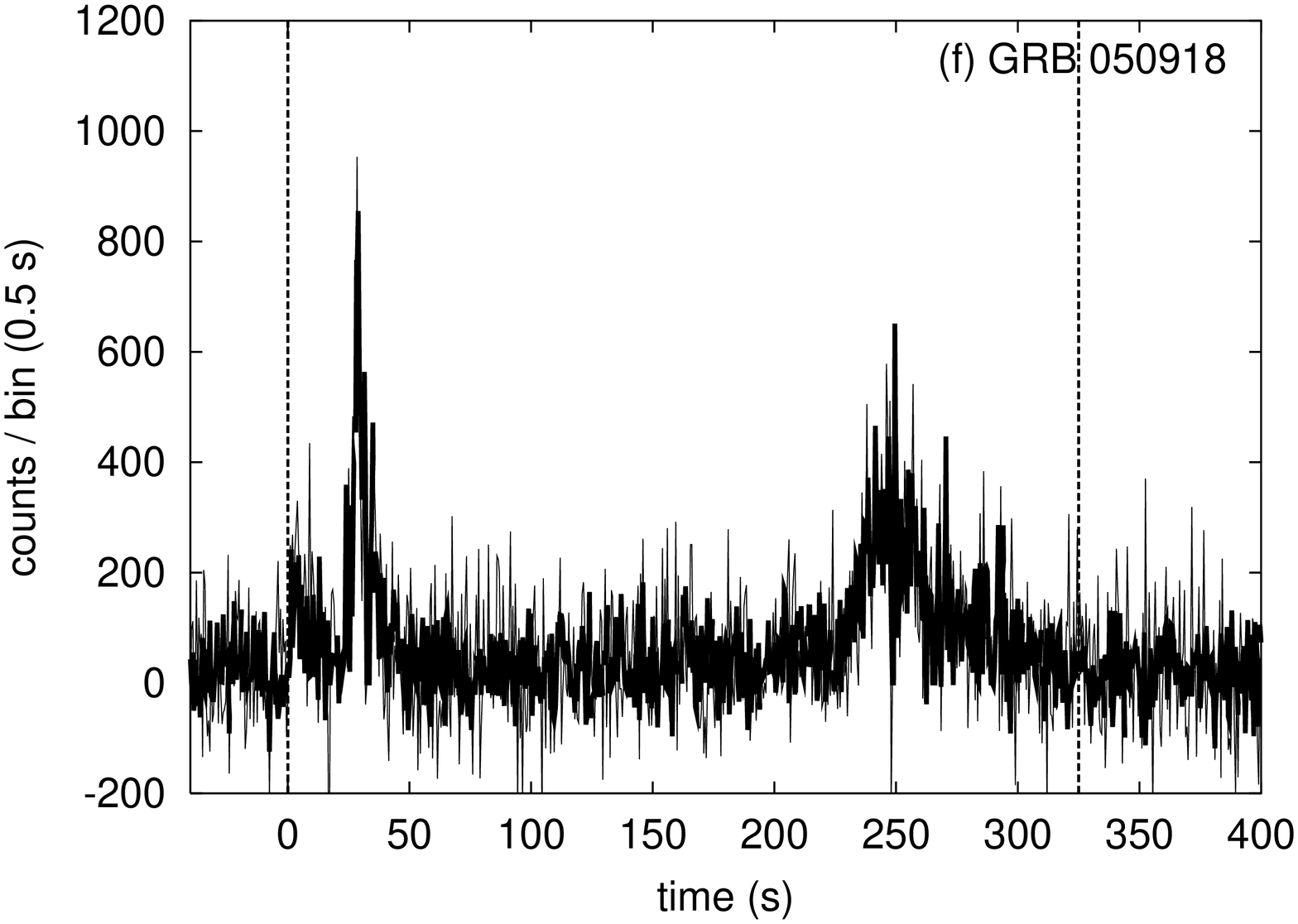}}
}
\mbox{
\subfigure{\includegraphics[height=0.35\textheight,width=0.18\textwidth,angle=270]{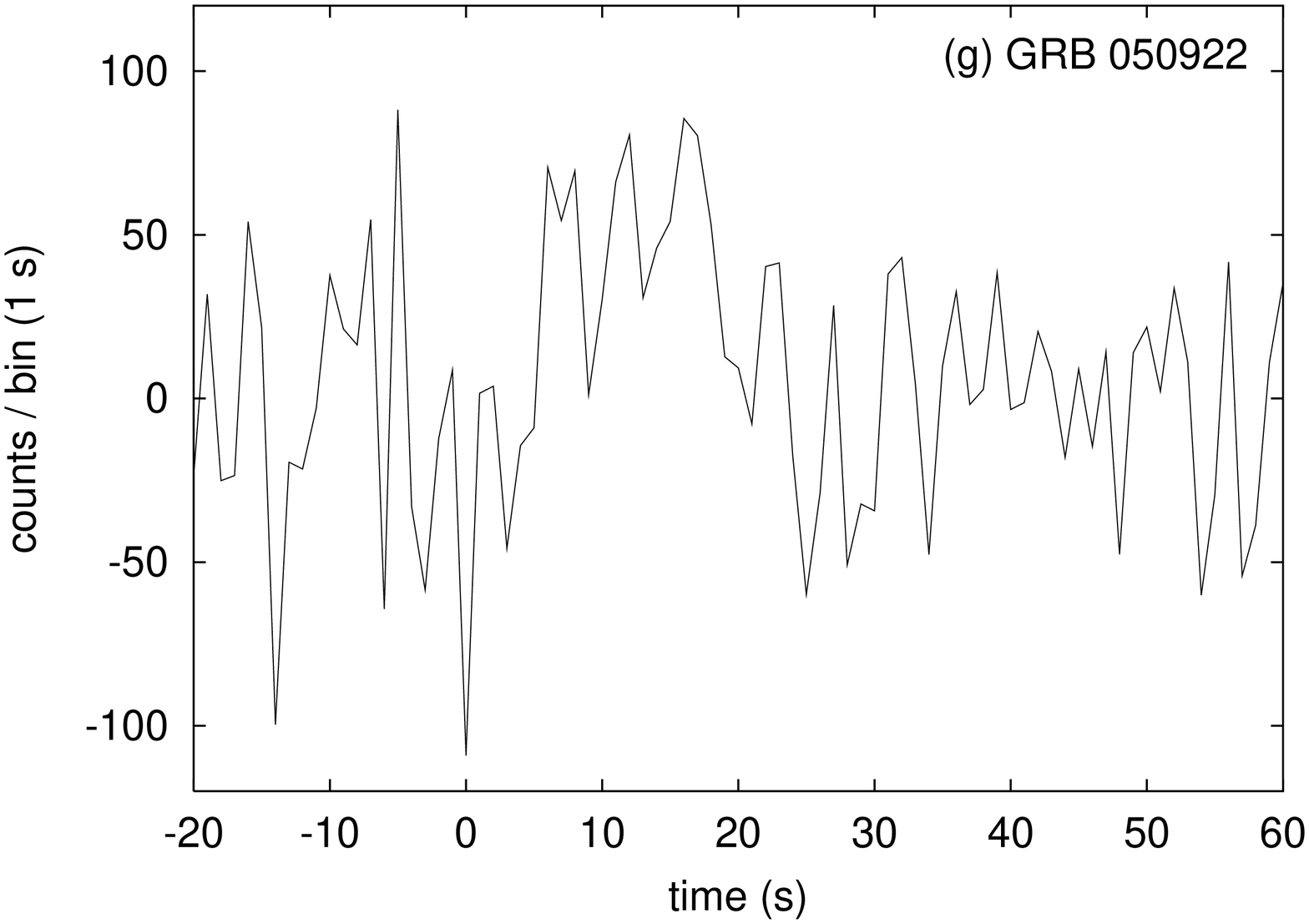}}
\subfigure{\includegraphics[height=0.35\textheight,width=0.18\textwidth,angle=270]{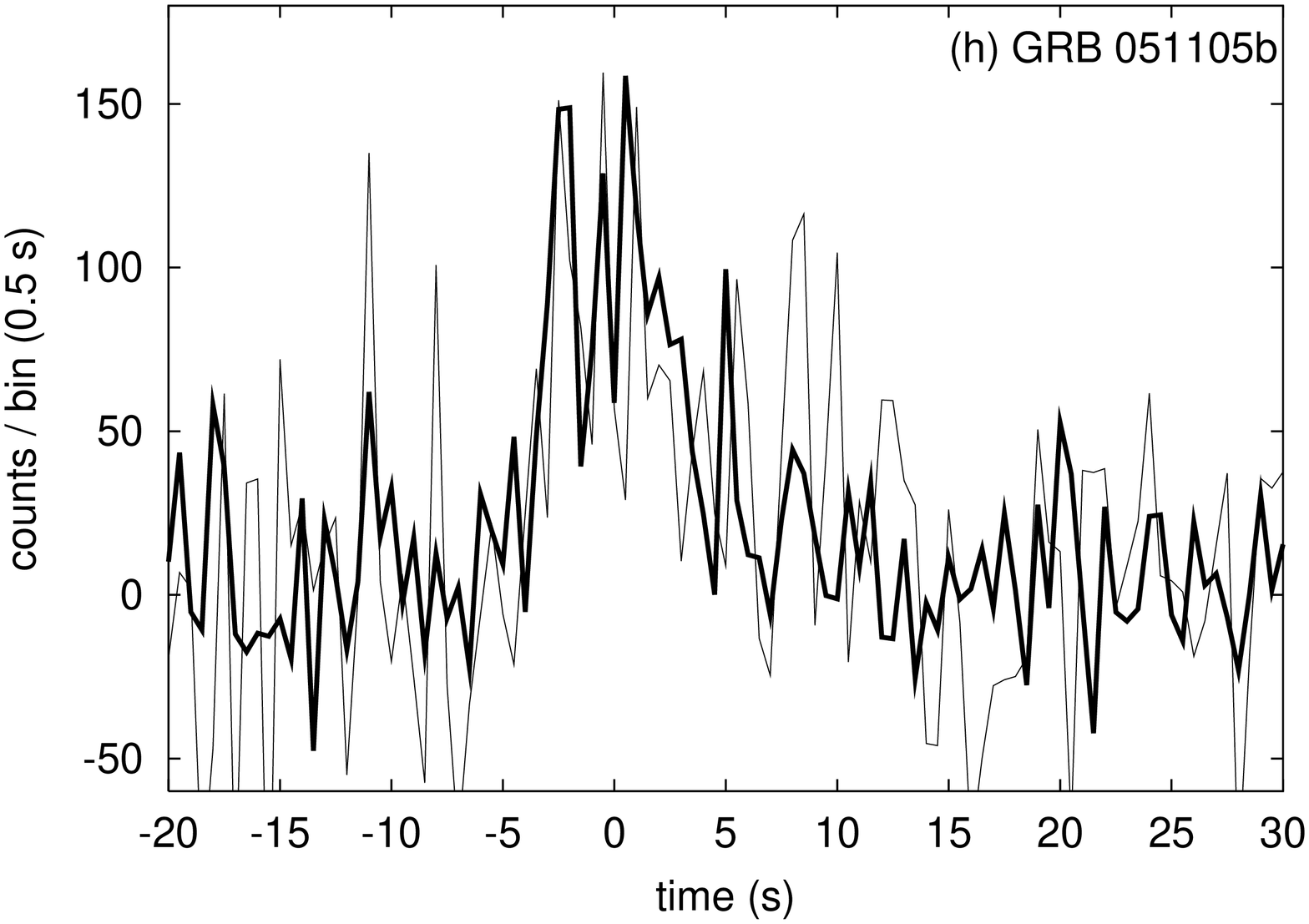}}
}
\mbox{
\subfigure{\includegraphics[height=0.35\textheight,width=0.18\textwidth,angle=270]{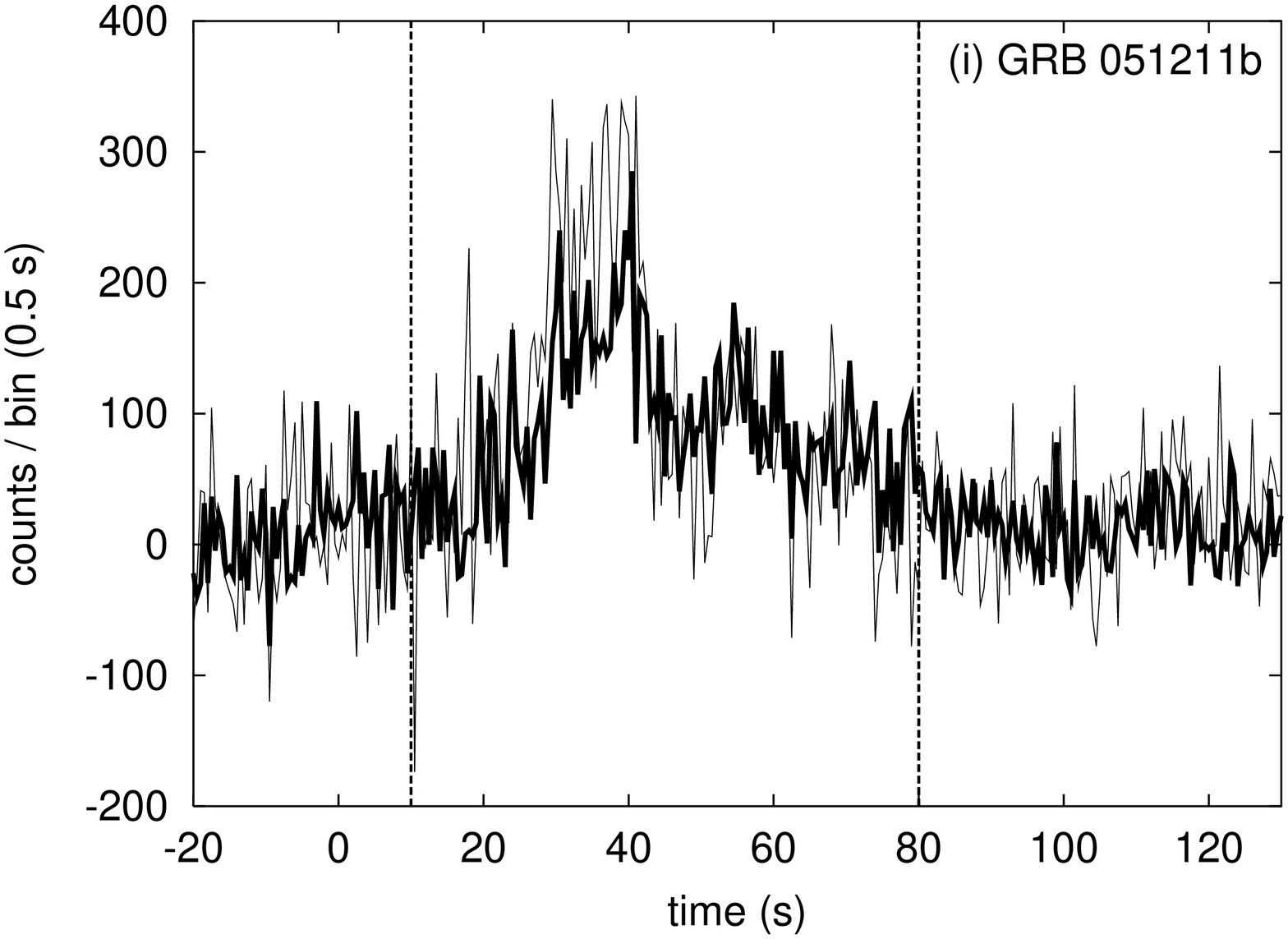}}
\subfigure{\includegraphics[height=0.35\textheight,width=0.18\textwidth,angle=270]{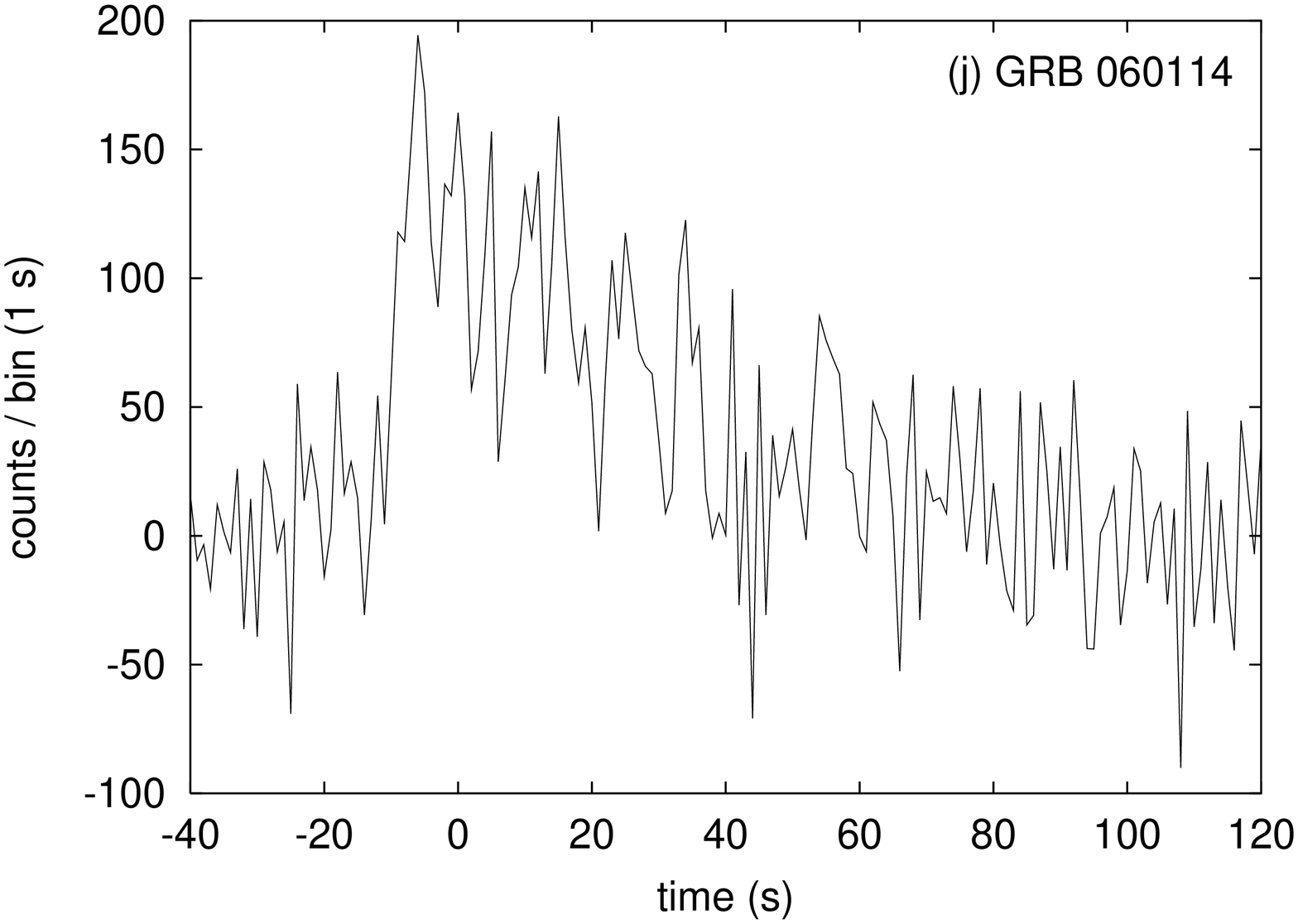}}
}
\mbox{
\subfigure{\includegraphics[height=0.35\textheight,width=0.18\textwidth,angle=270]{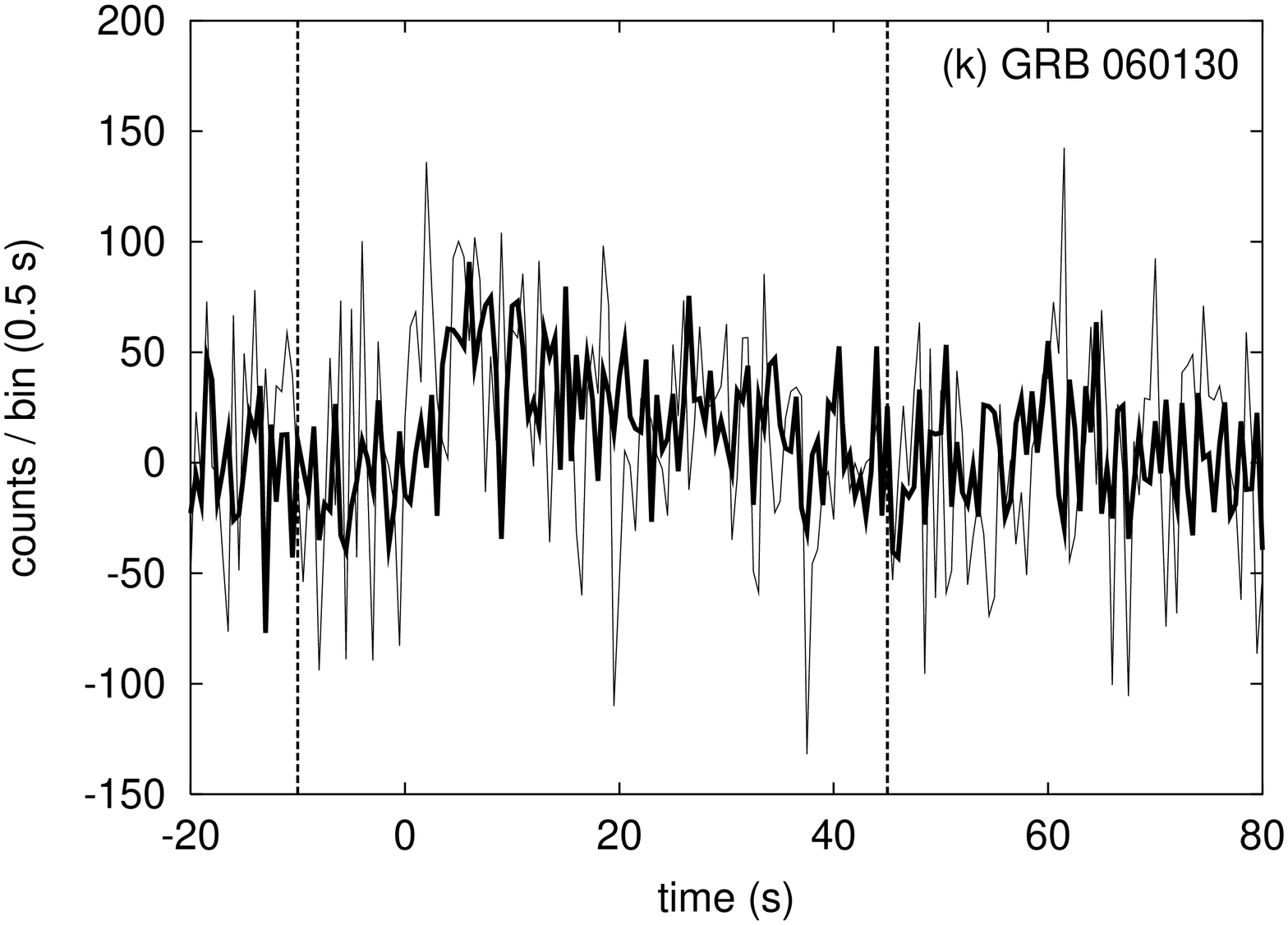}}
\subfigure{\includegraphics[height=0.35\textheight,width=0.18\textwidth,angle=270]{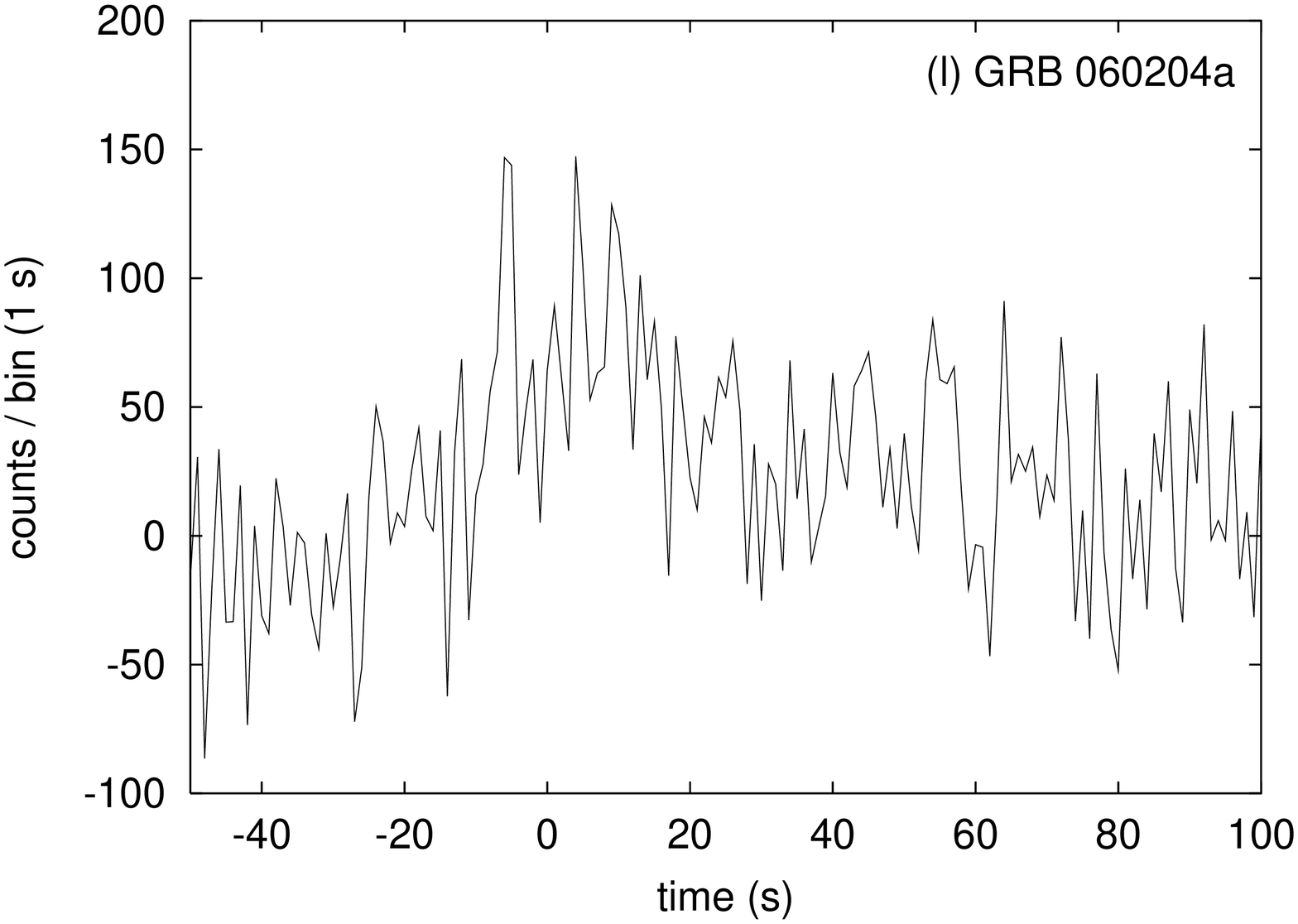}}
}
\end{figure*}

\begin{figure*}
\caption{Lightcurves of GRBS observed with \textit{INTEGRAL} (continued).}
\label{fig:lcs4}
\mbox{
\subfigure{\includegraphics[height=0.35\textheight, width=0.18\textwidth,angle=270]{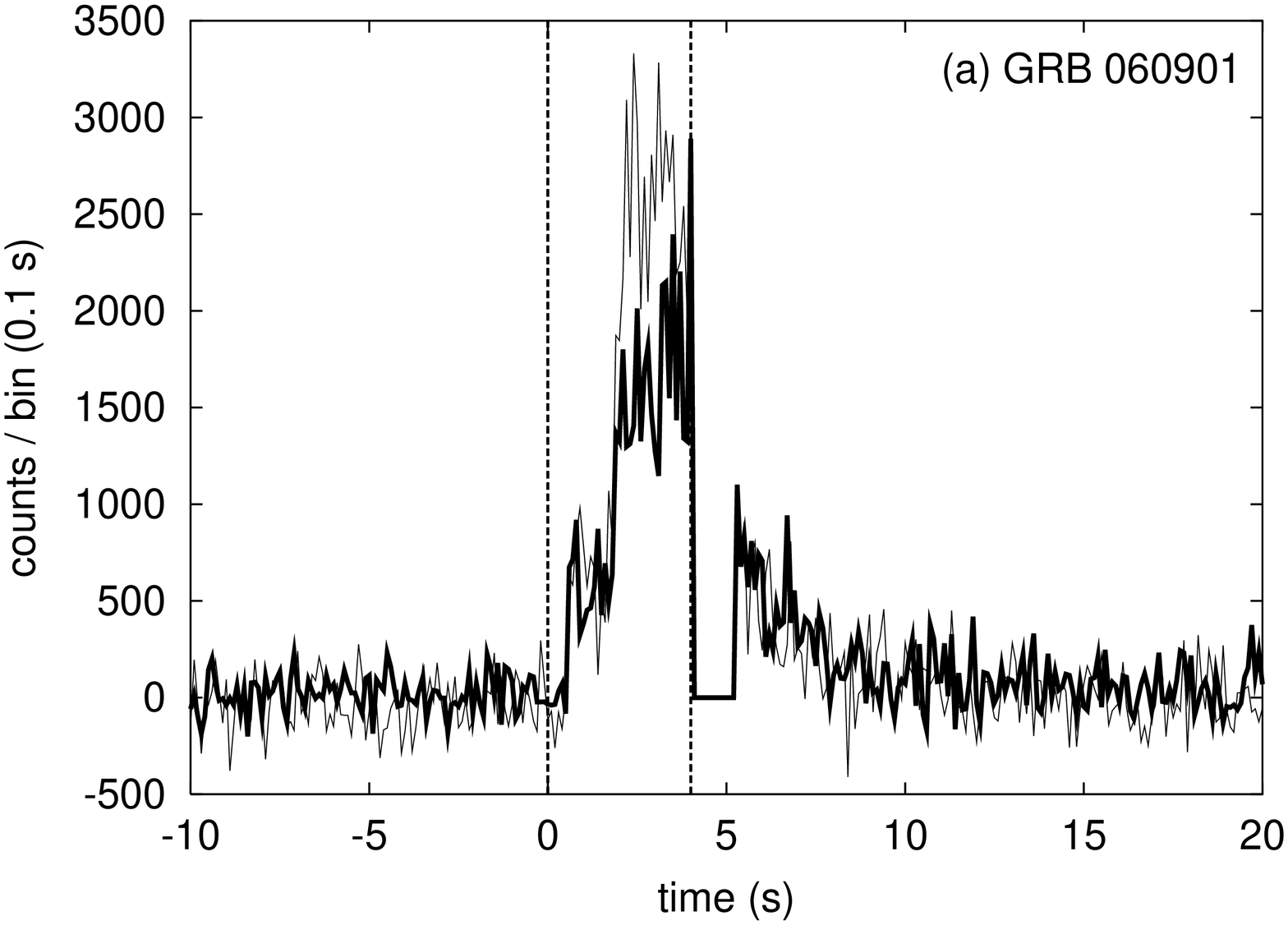}}
\subfigure{\includegraphics[height=0.35\textheight,width=0.18\textwidth,angle=270]{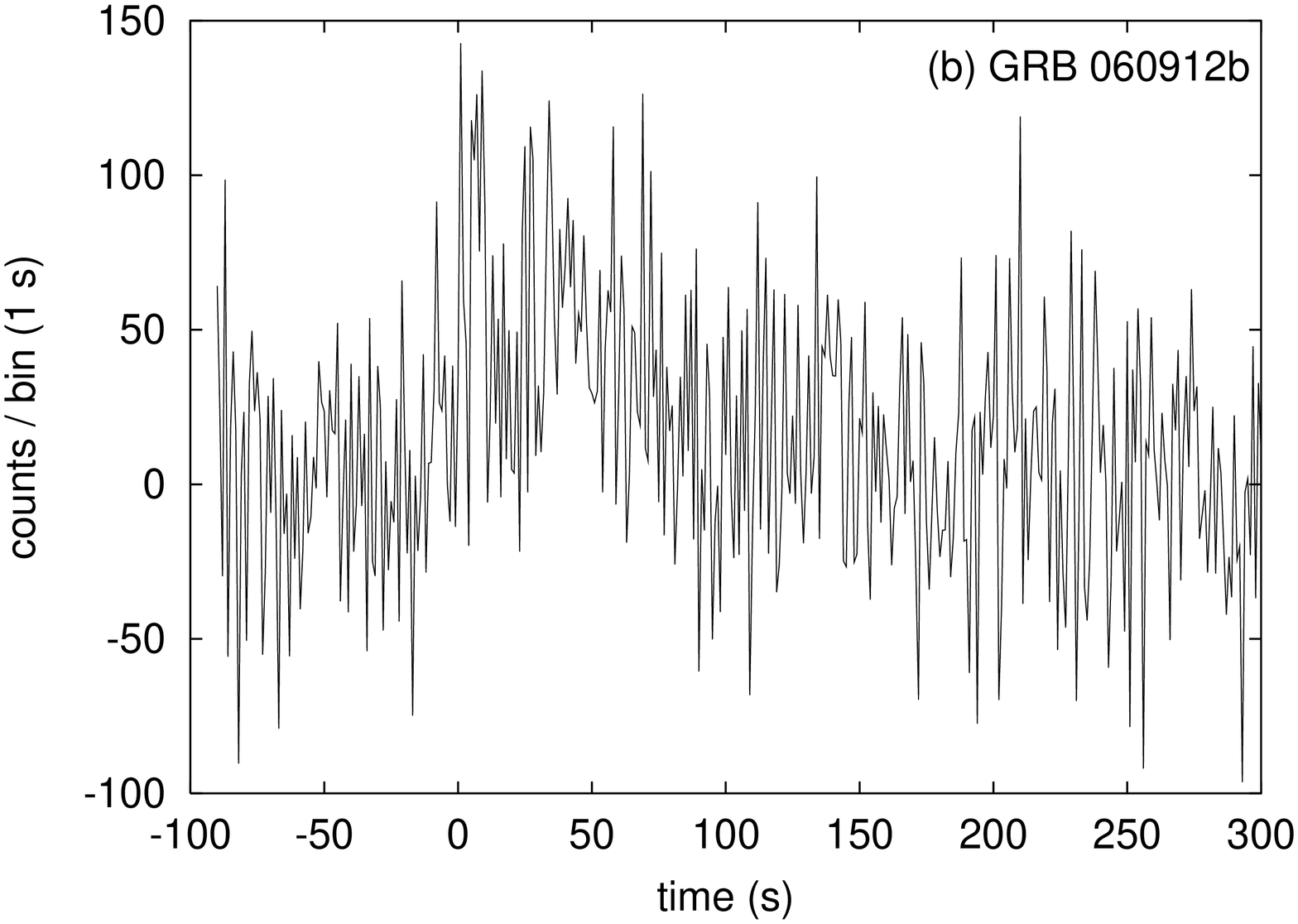}}
}
\mbox{
\subfigure{\includegraphics[height=0.35\textheight,width=0.18\textwidth,angle=270]{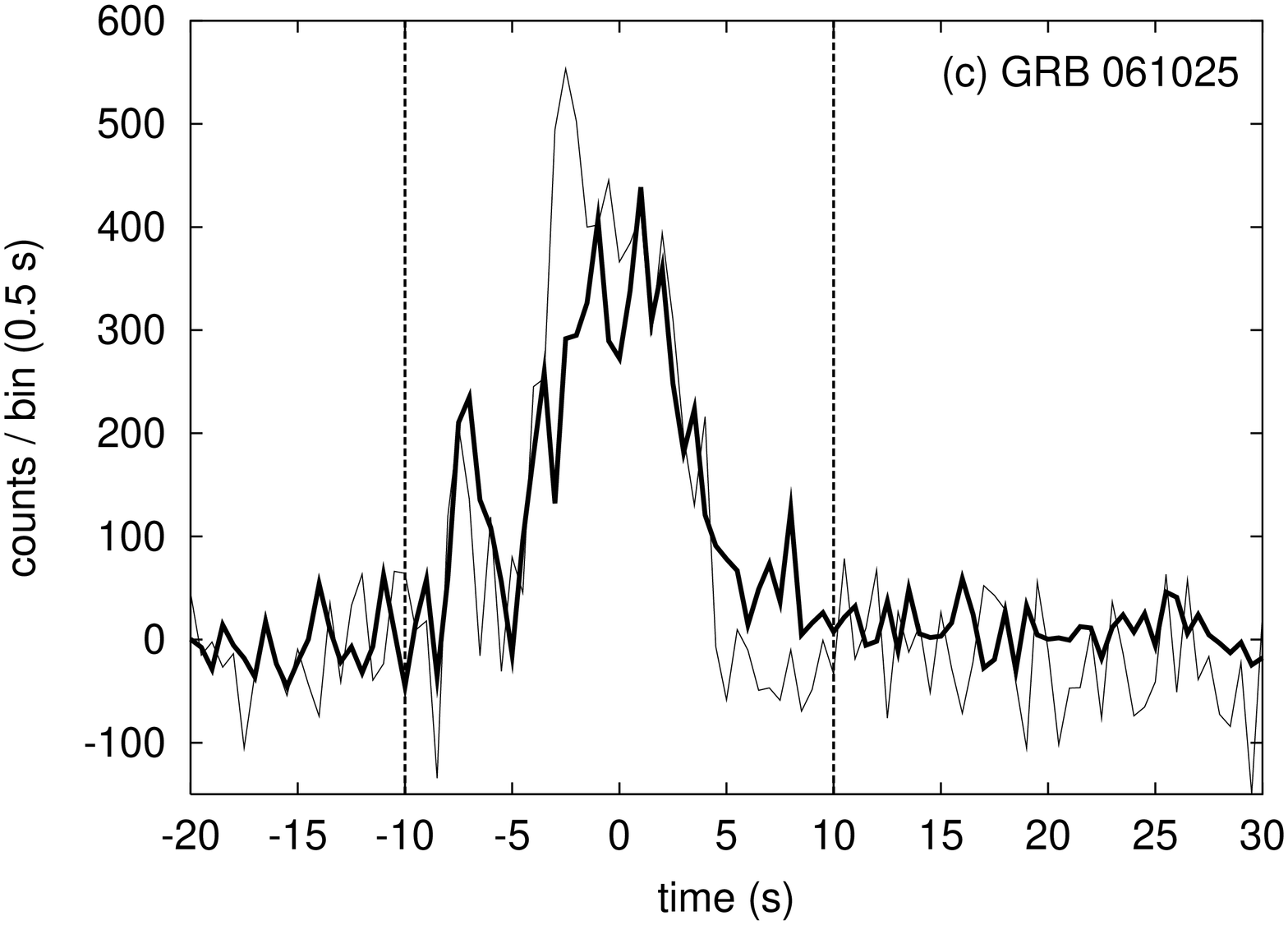}}
\subfigure{\includegraphics[height=0.35\textheight,width=0.18\textwidth,angle=270]{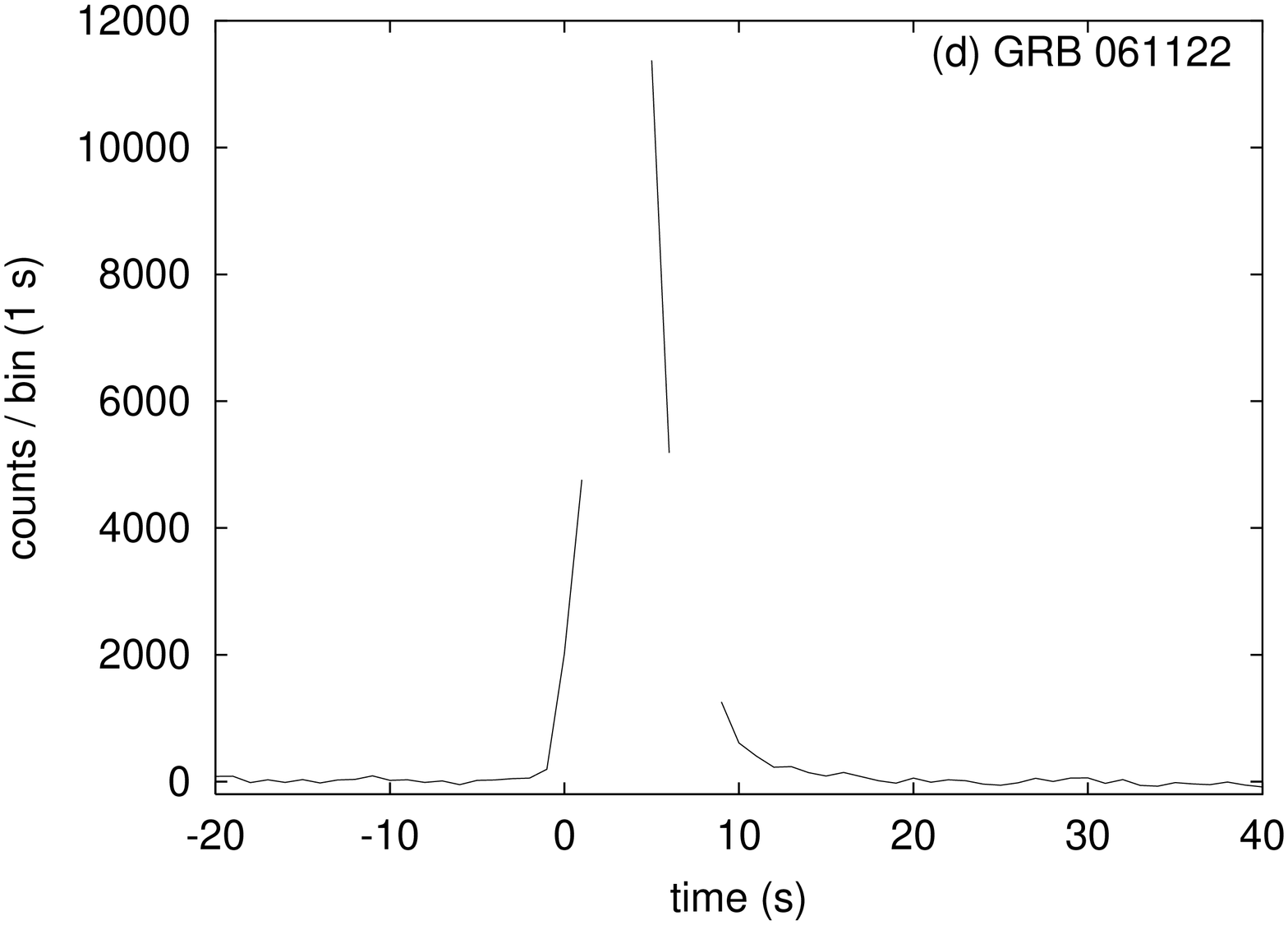}}
}
\mbox{
\subfigure{\includegraphics[height=0.35\textheight,width=0.18\textwidth,angle=270]{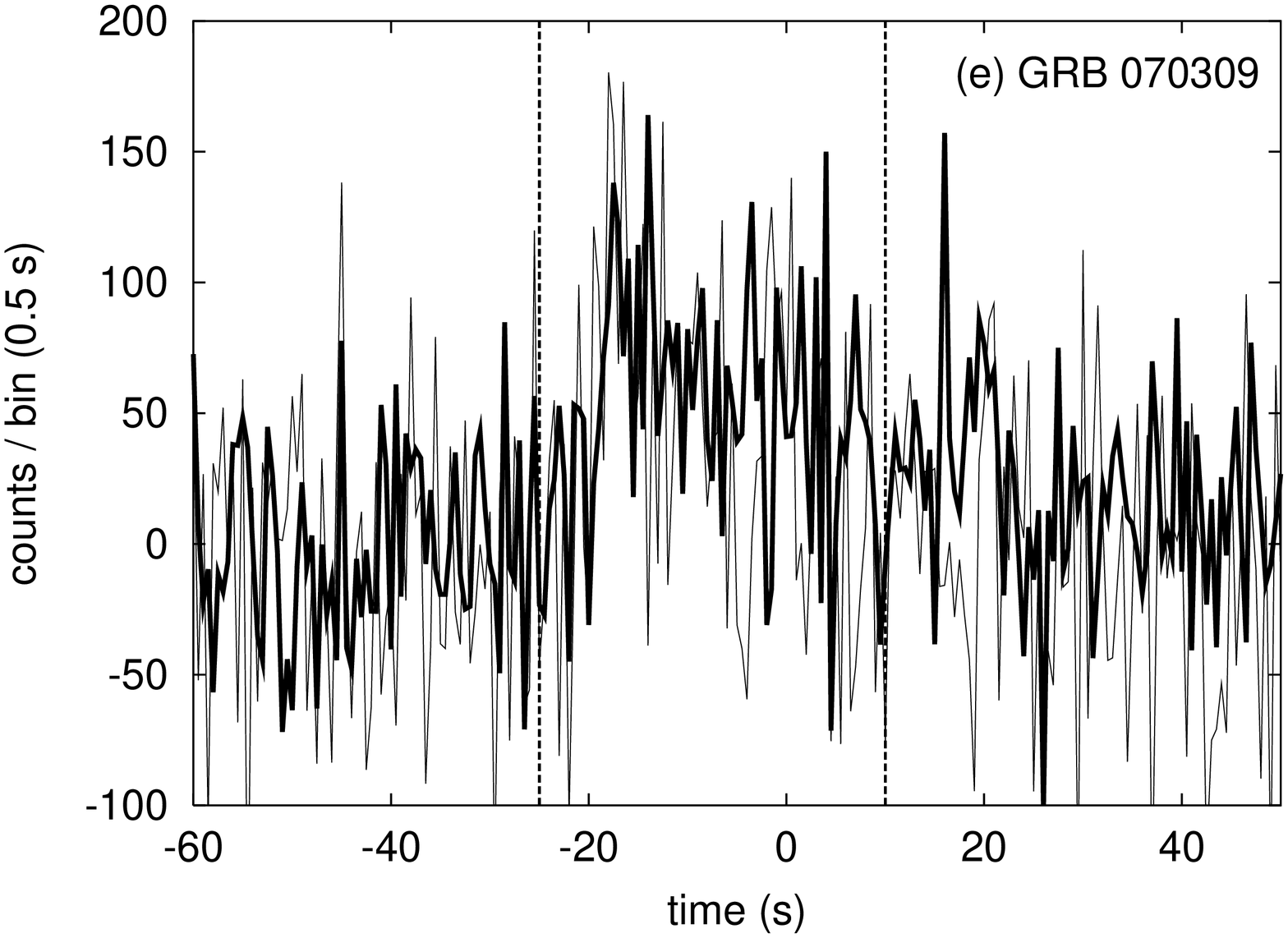}}
\subfigure{\includegraphics[height=0.35\textheight,width=0.18\textwidth,angle=270]{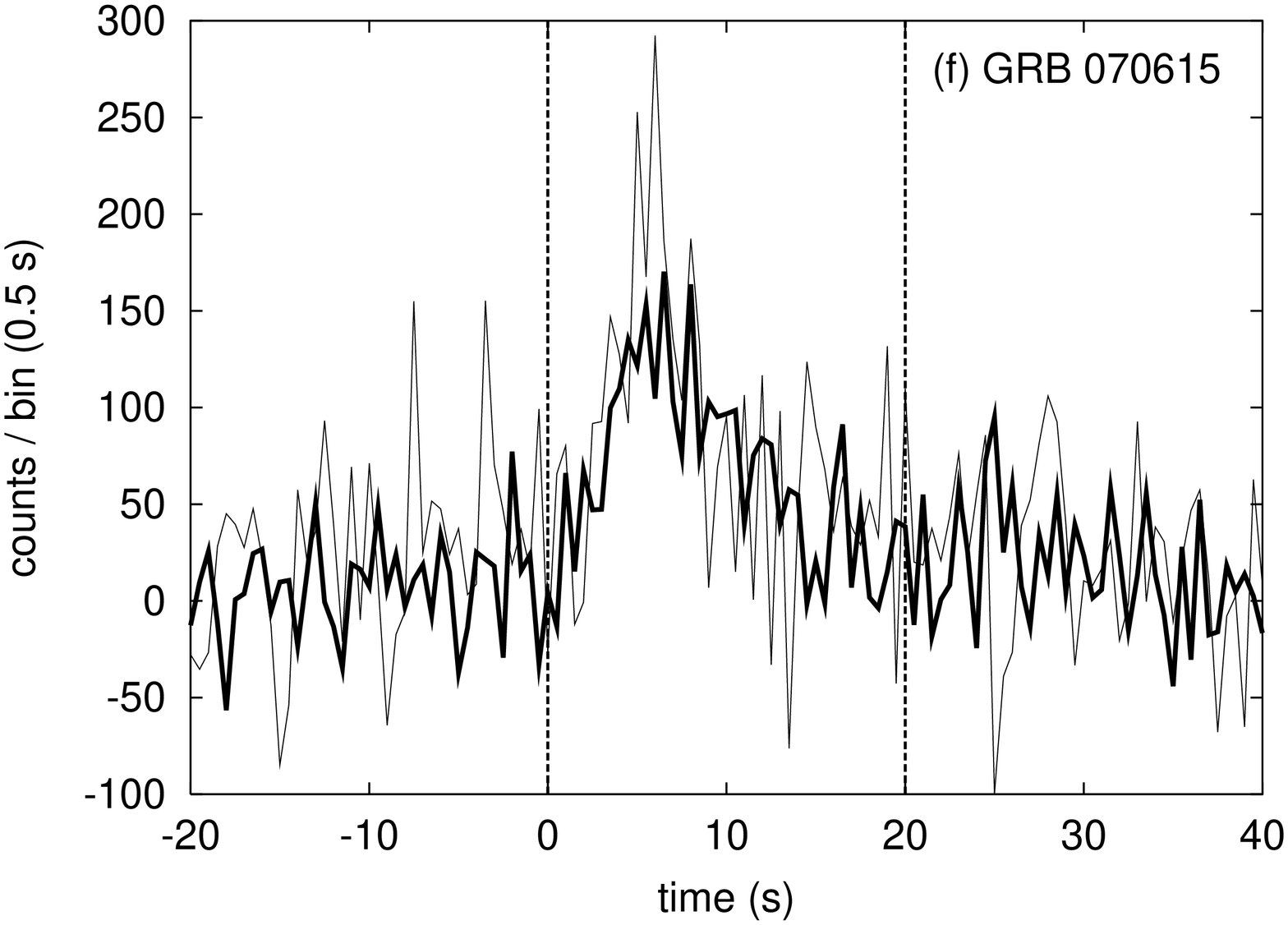}}
}
\subfigure{\includegraphics[height=0.35\textheight,width=0.18\textwidth,angle=270]{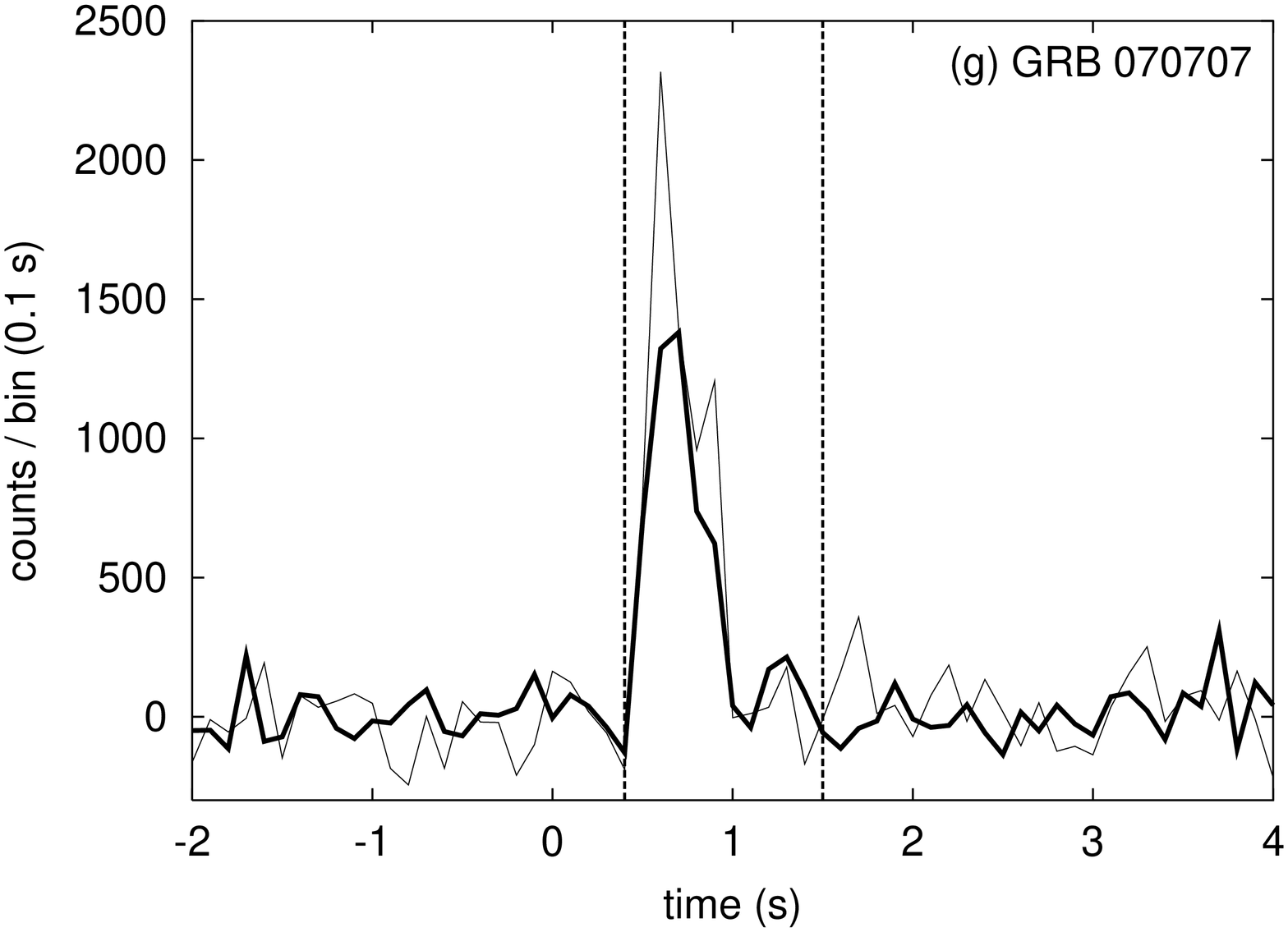}}
\end{figure*}

\end{document}